\documentclass[longauth,pdfencoding=auto,psdextra]{aa}

\usepackage{graphicx}
\usepackage{natbib}
\usepackage{scalerel}
\usepackage{comment}
\usepackage{orcidlink}

\usepackage[table]{xcolor}

\usepackage{txfonts}
\hypersetup{
    colorlinks=true,
    linkcolor=blue,
    filecolor=magenta,      
    urlcolor=blue,
    citecolor=blue
}
\makeatletter
\renewcommand*\aa@pageof{, page \thepage{} of \pageref*{LastPage}}
\makeatother

\usepackage[utf8]{inputenc}

\usepackage[switch, modulo]{lineno}
\linenumbers

\usepackage{euclid}

\usepackage[nolist,nohyperlinks]{acronym}

\begin{document}

\acrodef{PSF}{point-spread function}

\title{Euclid Data Release 1 (DR1)}
\subtitle{NISP-P point-spread function modelling pipelines}

\newcommand{\orcid}[1]{} 
\author{Euclid Collaboration: R.~da~Silva\orcid{0000-0003-4788-677X}\thanks{\email{ronaldo.oliveira@inaf.it}}\inst{\ref{aff1},\ref{aff2}}
\and F.~Faustini\orcid{0000-0001-6274-5145}\inst{\ref{aff1}}
\and G.~Polenta\orcid{0000-0003-4067-9196}\inst{\ref{aff2}}
\and A.~Alavi\orcid{0000-0002-8630-6435}\inst{\ref{aff3}}
\and A.~Alessio\inst{\ref{aff4}}
\and P.~N.~Appleton\orcid{0000-0002-7607-8766}\inst{\ref{aff3}}
\and P.~Awad\orcid{0000-0002-0428-849X}\inst{\ref{aff5}}
\and A.~Bonchi\orcid{0000-0002-2667-5482}\inst{\ref{aff2}}
\and L.~Bramante\inst{\ref{aff4}}
\and G.~Calderone\orcid{0000-0002-7738-5389}\inst{\ref{aff6}}
\and S.~Conseil\orcid{0000-0002-3657-4191}\inst{\ref{aff7}}
\and M.~Correnti\orcid{0000-0001-6464-3257}\inst{\ref{aff1},\ref{aff2}}
\and L.~V.~Costa\inst{\ref{aff4}}
\and I.~Das\orcid{0009-0007-7088-2044}\inst{\ref{aff3}}
\and M.~Frailis\orcid{0000-0002-7400-2135}\inst{\ref{aff6}}
\and Y.~Fu\orcid{0000-0002-0759-0504}\inst{\ref{aff5},\ref{aff8}}
\and T.~Gasparetto\orcid{0000-0002-7913-4866}\inst{\ref{aff1}}
\and W.~Gillard\orcid{0000-0003-4744-9748}\inst{\ref{aff9}}
\and S.~Hemmati\orcid{0000-0003-2226-5395}\inst{\ref{aff3}}
\and J.~Jacobson\inst{\ref{aff3}}
\and K.~Jahnke\orcid{0000-0003-3804-2137}\inst{\ref{aff10}}
\and B.~Kubik\orcid{0009-0006-5823-4880}\inst{\ref{aff7}}
\and X.~Liu\inst{\ref{aff3}}
\and C.~Macabiau\inst{\ref{aff7}}
\and P.~W.~Morris\orcid{0000-0002-5186-4381}\inst{\ref{aff11}}
\and K.~Paterson\orcid{0000-0001-8340-3486}\inst{\ref{aff10}}
\and M.~Radovich\orcid{0000-0002-3585-866X}\inst{\ref{aff12}}
\and M.~Schirmer\orcid{0000-0003-2568-9994}\inst{\ref{aff10}}
\and H.~I.~Teplitz\orcid{0000-0002-7064-5424}\inst{\ref{aff13}}
\and A.~Troja\orcid{0000-0003-0239-4595}\inst{\ref{aff6}}
\and B.~P.~Venemans\orcid{0000-0001-9024-8322}\inst{\ref{aff5}}
\and B.~Altieri\orcid{0000-0003-3936-0284}\inst{\ref{aff14}}
\and A.~Amara\inst{\ref{aff15}}
\and S.~Andreon\orcid{0000-0002-2041-8784}\inst{\ref{aff16}}
\and N.~Auricchio\orcid{0000-0003-4444-8651}\inst{\ref{aff17}}
\and C.~Baccigalupi\orcid{0000-0002-8211-1630}\inst{\ref{aff18},\ref{aff6},\ref{aff19},\ref{aff20}}
\and M.~Baldi\orcid{0000-0003-4145-1943}\inst{\ref{aff21},\ref{aff17},\ref{aff22}}
\and A.~Balestra\orcid{0000-0002-6967-261X}\inst{\ref{aff12}}
\and S.~Bardelli\orcid{0000-0002-8900-0298}\inst{\ref{aff17}}
\and P.~Battaglia\orcid{0000-0002-7337-5909}\inst{\ref{aff17}}
\and R.~Bender\orcid{0000-0001-7179-0626}\inst{\ref{aff23},\ref{aff24}}
\and A.~Biviano\orcid{0000-0002-0857-0732}\inst{\ref{aff6},\ref{aff18}}
\and E.~Branchini\orcid{0000-0002-0808-6908}\inst{\ref{aff25},\ref{aff26},\ref{aff16}}
\and M.~Brescia\orcid{0000-0001-9506-5680}\inst{\ref{aff27},\ref{aff28}}
\and S.~Camera\orcid{0000-0003-3399-3574}\inst{\ref{aff29},\ref{aff30},\ref{aff31}}
\and G.~Ca\~nas-Herrera\orcid{0000-0003-2796-2149}\inst{\ref{aff5}}
\and V.~Capobianco\orcid{0000-0002-3309-7692}\inst{\ref{aff31}}
\and C.~Carbone\orcid{0000-0003-0125-3563}\inst{\ref{aff32}}
\and J.~Carretero\orcid{0000-0002-3130-0204}\inst{\ref{aff33},\ref{aff34}}
\and M.~Castellano\orcid{0000-0001-9875-8263}\inst{\ref{aff1}}
\and G.~Castignani\orcid{0000-0001-6831-0687}\inst{\ref{aff17}}
\and S.~Cavuoti\orcid{0000-0002-3787-4196}\inst{\ref{aff28},\ref{aff35}}
\and A.~Cimatti\inst{\ref{aff36}}
\and C.~Colodro-Conde\inst{\ref{aff37}}
\and G.~Congedo\orcid{0000-0003-2508-0046}\inst{\ref{aff38}}
\and C.~J.~Conselice\orcid{0000-0003-1949-7638}\inst{\ref{aff39}}
\and L.~Conversi\orcid{0000-0002-6710-8476}\inst{\ref{aff40},\ref{aff14}}
\and Y.~Copin\orcid{0000-0002-5317-7518}\inst{\ref{aff7}}
\and A.~Costille\inst{\ref{aff41}}
\and F.~Courbin\orcid{0000-0003-0758-6510}\inst{\ref{aff42},\ref{aff43},\ref{aff44}}
\and H.~M.~Courtois\orcid{0000-0003-0509-1776}\inst{\ref{aff45}}
\and H.~Degaudenzi\orcid{0000-0002-5887-6799}\inst{\ref{aff46}}
\and G.~De~Lucia\orcid{0000-0002-6220-9104}\inst{\ref{aff6}}
\and H.~Dole\orcid{0000-0002-9767-3839}\inst{\ref{aff47}}
\and F.~Dubath\orcid{0000-0002-6533-2810}\inst{\ref{aff46}}
\and X.~Dupac\inst{\ref{aff14}}
\and S.~Dusini\orcid{0000-0002-1128-0664}\inst{\ref{aff48}}
\and A.~Ealet\orcid{0000-0003-3070-014X}\inst{\ref{aff7}}
\and S.~Escoffier\orcid{0000-0002-2847-7498}\inst{\ref{aff9}}
\and M.~Farina\orcid{0000-0002-3089-7846}\inst{\ref{aff49}}
\and R.~Farinelli\inst{\ref{aff17}}
\and S.~Ferriol\inst{\ref{aff7}}
\and S.~Fotopoulou\orcid{0000-0002-9686-254X}\inst{\ref{aff50}}
\and N.~Fourmanoit\orcid{0009-0005-6816-6925}\inst{\ref{aff9}}
\and E.~Franceschi\orcid{0000-0002-0585-6591}\inst{\ref{aff17}}
\and M.~Fumana\orcid{0000-0001-6787-5950}\inst{\ref{aff32}}
\and L.~Gabarra\orcid{0000-0002-8486-8856}\inst{\ref{aff51}}
\and S.~Galeotta\orcid{0000-0002-3748-5115}\inst{\ref{aff6}}
\and K.~George\orcid{0000-0002-1734-8455}\inst{\ref{aff52}}
\and B.~Gillis\orcid{0000-0002-4478-1270}\inst{\ref{aff38}}
\and C.~Giocoli\orcid{0000-0002-9590-7961}\inst{\ref{aff17},\ref{aff22}}
\and J.~Gracia-Carpio\orcid{0000-0003-4689-3134}\inst{\ref{aff23}}
\and B.~R.~Granett\orcid{0000-0003-2694-9284}\inst{\ref{aff16}}
\and A.~Grazian\orcid{0000-0002-5688-0663}\inst{\ref{aff12}}
\and F.~Grupp\inst{\ref{aff23},\ref{aff24}}
\and W.~G.~Hartley\orcid{0000-0002-4563-5101}\inst{\ref{aff46}}
\and S.~V.~H.~Haugan\orcid{0000-0001-9648-7260}\inst{\ref{aff53}}
\and H.~Hoekstra\orcid{0000-0002-0641-3231}\inst{\ref{aff5}}
\and W.~Holmes\orcid{0009-0007-8554-4646}\inst{\ref{aff54}}
\and I.~M.~Hook\orcid{0000-0002-2960-978X}\inst{\ref{aff55}}
\and F.~Hormuth\inst{\ref{aff56}}
\and A.~Hornstrup\orcid{0000-0002-3363-0936}\inst{\ref{aff57},\ref{aff58}}
\and M.~Jhabvala\inst{\ref{aff59}}
\and B.~Joachimi\orcid{0000-0001-7494-1303}\inst{\ref{aff60}}
\and S.~Kermiche\orcid{0000-0002-0302-5735}\inst{\ref{aff9}}
\and A.~Kiessling\orcid{0000-0002-2590-1273}\inst{\ref{aff54}}
\and R.~Kohley\inst{\ref{aff14}}
\and M.~Kunz\orcid{0000-0002-3052-7394}\inst{\ref{aff61}}
\and H.~Kurki-Suonio\orcid{0000-0002-4618-3063}\inst{\ref{aff62},\ref{aff63}}
\and A.~M.~C.~Le~Brun\orcid{0000-0002-0936-4594}\inst{\ref{aff64}}
\and D.~Le~Mignant\orcid{0000-0002-5339-5515}\inst{\ref{aff41}}
\and S.~Ligori\orcid{0000-0003-4172-4606}\inst{\ref{aff31}}
\and P.~B.~Lilje\orcid{0000-0003-4324-7794}\inst{\ref{aff53}}
\and V.~Lindholm\orcid{0000-0003-2317-5471}\inst{\ref{aff62},\ref{aff63}}
\and I.~Lloro\orcid{0000-0001-5966-1434}\inst{\ref{aff65}}
\and M.~Magliocchetti\orcid{0000-0001-9158-4838}\inst{\ref{aff49}}
\and G.~Mainetti\orcid{0000-0003-2384-2377}\inst{\ref{aff66}}
\and D.~Maino\inst{\ref{aff67},\ref{aff32},\ref{aff68}}
\and O.~Mansutti\orcid{0000-0001-5758-4658}\inst{\ref{aff6}}
\and S.~Marcin\inst{\ref{aff69}}
\and O.~Marggraf\orcid{0000-0001-7242-3852}\inst{\ref{aff70}}
\and M.~Martinelli\orcid{0000-0002-6943-7732}\inst{\ref{aff1},\ref{aff71}}
\and N.~Martinet\orcid{0000-0003-2786-7790}\inst{\ref{aff41}}
\and F.~Marulli\orcid{0000-0002-8850-0303}\inst{\ref{aff72},\ref{aff17},\ref{aff22}}
\and R.~J.~Massey\orcid{0000-0002-6085-3780}\inst{\ref{aff73}}
\and N.~Mauri\orcid{0000-0001-8196-1548}\inst{\ref{aff36},\ref{aff22}}
\and C.~J.~R.~McPartland\orcid{0000-0003-0639-025X}\inst{\ref{aff58},\ref{aff74}}
\and E.~Medinaceli\orcid{0000-0002-4040-7783}\inst{\ref{aff17}}
\and S.~Mei\orcid{0000-0002-2849-559X}\inst{\ref{aff75},\ref{aff76}}
\and M.~Melchior\inst{\ref{aff77}}
\and M.~Meneghetti\orcid{0000-0003-1225-7084}\inst{\ref{aff17},\ref{aff22}}
\and E.~Merlin\orcid{0000-0001-6870-8900}\inst{\ref{aff12}}
\and G.~Meylan\orcid{0000-0001-6503-0209}\inst{\ref{aff78}}
\and J.~J.~Mohr\orcid{0000-0002-6875-2087}\inst{\ref{aff52}}
\and P.~Monaco\orcid{0000-0003-2083-7564}\inst{\ref{aff79},\ref{aff6},\ref{aff19},\ref{aff18}}
\and A.~Mora\orcid{0000-0002-1922-8529}\inst{\ref{aff80}}
\and M.~Moresco\orcid{0000-0002-7616-7136}\inst{\ref{aff72},\ref{aff17}}
\and C.~Moretti\orcid{0000-0003-3314-8936}\inst{\ref{aff6},\ref{aff18},\ref{aff19}}
\and L.~Moscardini\orcid{0000-0002-3473-6716}\inst{\ref{aff72},\ref{aff17},\ref{aff22}}
\and R.~Nakajima\orcid{0009-0009-1213-7040}\inst{\ref{aff70}}
\and C.~Neissner\orcid{0000-0001-8524-4968}\inst{\ref{aff81},\ref{aff34}}
\and R.~C.~Nichol\orcid{0000-0003-0939-6518}\inst{\ref{aff15}}
\and S.-M.~Niemi\orcid{0009-0005-0247-0086}\inst{\ref{aff82}}
\and C.~Padilla\orcid{0000-0001-7951-0166}\inst{\ref{aff81}}
\and S.~Paltani\orcid{0000-0002-8108-9179}\inst{\ref{aff46}}
\and F.~Pasian\orcid{0000-0002-4869-3227}\inst{\ref{aff6}}
\and W.~J.~Percival\orcid{0000-0002-0644-5727}\inst{\ref{aff83},\ref{aff84},\ref{aff85}}
\and V.~Pettorino\orcid{0000-0002-4203-9320}\inst{\ref{aff82}}
\and A.~Pezzotta\orcid{0000-0003-0726-2268}\inst{\ref{aff16}}
\and S.~Pires\orcid{0000-0002-0249-2104}\inst{\ref{aff86}}
\and M.~Poncet\inst{\ref{aff87}}
\and L.~A.~Popa\inst{\ref{aff88}}
\and G.~D.~Racca\orcid{0000-0002-9883-8981}\inst{\ref{aff5},\ref{aff82}}
\and F.~Raison\orcid{0000-0002-7819-6918}\inst{\ref{aff23}}
\and R.~Rebolo\orcid{0000-0003-3767-7085}\inst{\ref{aff37},\ref{aff89},\ref{aff90}}
\and A.~Renzi\orcid{0000-0001-9856-1970}\inst{\ref{aff91},\ref{aff48},\ref{aff17}}
\and J.~Rhodes\orcid{0000-0002-4485-8549}\inst{\ref{aff54}}
\and G.~Riccio\inst{\ref{aff28}}
\and I.~Risso\orcid{0000-0003-2525-7761}\inst{\ref{aff25},\ref{aff26},\ref{aff16}}
\and F.~Rizzo\orcid{0000-0002-9407-585X}\inst{\ref{aff6}}
\and E.~Romelli\orcid{0000-0003-3069-9222}\inst{\ref{aff6}}
\and M.~Roncarelli\orcid{0000-0001-9587-7822}\inst{\ref{aff17}}
\and C.~Rosset\orcid{0000-0003-0286-2192}\inst{\ref{aff75}}
\and B.~Rusholme\orcid{0000-0001-7648-4142}\inst{\ref{aff3}}
\and R.~Saglia\orcid{0000-0003-0378-7032}\inst{\ref{aff24},\ref{aff23}}
\and Z.~Sakr\orcid{0000-0002-4823-3757}\inst{\ref{aff92},\ref{aff93},\ref{aff94}}
\and A.~G.~S\'anchez\orcid{0000-0003-1198-831X}\inst{\ref{aff23}}
\and D.~Sapone\orcid{0000-0001-7089-4503}\inst{\ref{aff95}}
\and B.~Sartoris\orcid{0000-0003-1337-5269}\inst{\ref{aff23},\ref{aff6}}
\and P.~Schneider\orcid{0000-0001-8561-2679}\inst{\ref{aff70}}
\and T.~Schrabback\orcid{0000-0002-6987-7834}\inst{\ref{aff96}}
\and A.~Secroun\orcid{0000-0003-0505-3710}\inst{\ref{aff9}}
\and E.~Sihvola\orcid{0000-0003-1804-7715}\inst{\ref{aff97}}
\and P.~Simon\inst{\ref{aff70}}
\and C.~Sirignano\orcid{0000-0002-0995-7146}\inst{\ref{aff91},\ref{aff48}}
\and G.~Sirri\orcid{0000-0003-2626-2853}\inst{\ref{aff22}}
\and L.~Stanco\orcid{0000-0002-9706-5104}\inst{\ref{aff48}}
\and P.~Tallada-Cresp\'{i}\orcid{0000-0002-1336-8328}\inst{\ref{aff33},\ref{aff34}}
\and D.~Tavagnacco\orcid{0000-0001-7475-9894}\inst{\ref{aff6}}
\and A.~N.~Taylor\inst{\ref{aff38}}
\and I.~Tereno\orcid{0000-0002-4537-6218}\inst{\ref{aff98},\ref{aff99}}
\and S.~Toft\orcid{0000-0003-3631-7176}\inst{\ref{aff100},\ref{aff74}}
\and R.~Toledo-Moreo\orcid{0000-0002-2997-4859}\inst{\ref{aff101},\ref{aff102}}
\and F.~Torradeflot\orcid{0000-0003-1160-1517}\inst{\ref{aff34},\ref{aff33}}
\and I.~Tutusaus\orcid{0000-0002-3199-0399}\inst{\ref{aff103},\ref{aff104},\ref{aff93}}
\and L.~Valenziano\orcid{0000-0002-1170-0104}\inst{\ref{aff17},\ref{aff105}}
\and J.~Valiviita\orcid{0000-0001-6225-3693}\inst{\ref{aff62},\ref{aff63}}
\and T.~Vassallo\orcid{0000-0001-6512-6358}\inst{\ref{aff6},\ref{aff52}}
\and A.~Veropalumbo\orcid{0000-0003-2387-1194}\inst{\ref{aff16},\ref{aff26},\ref{aff25}}
\and Y.~Wang\orcid{0000-0002-4749-2984}\inst{\ref{aff3}}
\and J.~Weller\orcid{0000-0002-8282-2010}\inst{\ref{aff24},\ref{aff23}}
\and A.~Zacchei\orcid{0000-0003-0396-1192}\inst{\ref{aff6},\ref{aff18}}
\and G.~Zamorani\orcid{0000-0002-2318-301X}\inst{\ref{aff17}}
\and F.~M.~Zerbi\orcid{0000-0002-9996-973X}\inst{\ref{aff16}}
\and E.~Zucca\orcid{0000-0002-5845-8132}\inst{\ref{aff17}}
\and V.~Allevato\orcid{0000-0001-7232-5152}\inst{\ref{aff28}}
\and M.~Ballardini\orcid{0000-0003-4481-3559}\inst{\ref{aff106},\ref{aff107},\ref{aff17}}
\and P.~Bergamini\orcid{0000-0003-1383-9414}\inst{\ref{aff17}}
\and S.~Borgani\orcid{0000-0001-6151-6439}\inst{\ref{aff79},\ref{aff18},\ref{aff6},\ref{aff19},\ref{aff108}}
\and E.~Bozzo\orcid{0000-0002-8201-1525}\inst{\ref{aff46}}
\and C.~Burigana\orcid{0000-0002-3005-5796}\inst{\ref{aff109},\ref{aff105}}
\and R.~Cabanac\orcid{0000-0001-6679-2600}\inst{\ref{aff93}}
\and M.~Calabrese\orcid{0000-0002-2637-2422}\inst{\ref{aff110},\ref{aff32}}
\and A.~Cappi\inst{\ref{aff111},\ref{aff17}}
\and F.~Caro\orcid{0009-0003-1053-0507}\inst{\ref{aff1}}
\and T.~Castro\orcid{0000-0002-6292-3228}\inst{\ref{aff112},\ref{aff6},\ref{aff18}}
\and J.~A.~Escartin~Vigo\inst{\ref{aff23}}
\and J.~Garc\'ia-Bellido\orcid{0000-0002-9370-8360}\inst{\ref{aff92}}
\and V.~Gautard\inst{\ref{aff113}}
\and A.~Gregorio\orcid{0000-0003-4028-8785}\inst{\ref{aff79},\ref{aff6},\ref{aff19}}
\and C.~M.~Gutierrez\orcid{0000-0001-7854-783X}\inst{\ref{aff37},\ref{aff90}}
\and A.~Loureiro\orcid{0000-0002-4371-0876}\inst{\ref{aff114},\ref{aff115}}
\and J.~Macias-Perez\orcid{0000-0002-5385-2763}\inst{\ref{aff116}}
\and R.~Maoli\orcid{0000-0002-6065-3025}\inst{\ref{aff117},\ref{aff1}}
\and R.~B.~Metcalf\orcid{0000-0003-3167-2574}\inst{\ref{aff72},\ref{aff17}}
\and M.~P\"ontinen\orcid{0000-0001-5442-2530}\inst{\ref{aff62}}
\and E.~Sarpa\orcid{0000-0002-1256-655X}\inst{\ref{aff6}}
\and V.~Scottez\orcid{0009-0008-3864-940X}\inst{\ref{aff118},\ref{aff119}}
\and M.~Sereno\orcid{0000-0003-0302-0325}\inst{\ref{aff17},\ref{aff22}}
\and M.~Tenti\orcid{0000-0002-4254-5901}\inst{\ref{aff22}}
\and M.~Tucci\inst{\ref{aff46}}
\and M.~Viel\orcid{0000-0002-2642-5707}\inst{\ref{aff18},\ref{aff6},\ref{aff20},\ref{aff19},\ref{aff108}}
\and M.~Wiesmann\orcid{0009-0000-8199-5860}\inst{\ref{aff53}}
\and J.~A.~Acevedo~Barroso\orcid{0000-0002-9654-1711}\inst{\ref{aff54}}
\and Y.~Akrami\orcid{0000-0002-2407-7956}\inst{\ref{aff92},\ref{aff120}}
\and I.~T.~Andika\orcid{0000-0001-6102-9526}\inst{\ref{aff24}}
\and G.~Angora\orcid{0000-0002-0316-6562}\inst{\ref{aff28},\ref{aff106}}
\and S.~Anselmi\orcid{0000-0002-3579-9583}\inst{\ref{aff48},\ref{aff91},\ref{aff121}}
\and M.~Archidiacono\orcid{0000-0003-4952-9012}\inst{\ref{aff67},\ref{aff68}}
\and G.~Aric\`o\orcid{0000-0002-2802-2928}\inst{\ref{aff22}}
\and F.~Atrio-Barandela\orcid{0000-0002-2130-2513}\inst{\ref{aff122}}
\and M.~Baes\orcid{0000-0002-3930-2757}\inst{\ref{aff123}}
\and L.~Bazzanini\orcid{0000-0003-0727-0137}\inst{\ref{aff106},\ref{aff17}}
\and D.~Bertacca\orcid{0000-0002-2490-7139}\inst{\ref{aff91},\ref{aff12},\ref{aff48}}
\and M.~Bethermin\orcid{0000-0002-3915-2015}\inst{\ref{aff124}}
\and F.~Beutler\orcid{0000-0003-0467-5438}\inst{\ref{aff38}}
\and L.~Blot\orcid{0000-0002-9622-7167}\inst{\ref{aff125},\ref{aff64}}
\and M.~L.~Brown\orcid{0000-0002-0370-8077}\inst{\ref{aff39}}
\and S.~Bruton\orcid{0000-0002-6503-5218}\inst{\ref{aff11}}
\and A.~Calabro\orcid{0000-0003-2536-1614}\inst{\ref{aff1}}
\and B.~Camacho~Quevedo\orcid{0000-0002-8789-4232}\inst{\ref{aff18},\ref{aff20},\ref{aff6}}
\and P.~Carrilho\orcid{0000-0003-1339-0194}\inst{\ref{aff126}}
\and C.~S.~Carvalho\inst{\ref{aff99}}
\and Y.~Charles\orcid{0009-0000-3636-834X}\inst{\ref{aff41}}
\and F.~Cogato\orcid{0000-0003-4632-6113}\inst{\ref{aff72},\ref{aff17}}
\and T.~E.~Collett\orcid{0000-0001-5564-3140}\inst{\ref{aff127}}
\and A.~R.~Cooray\orcid{0000-0002-3892-0190}\inst{\ref{aff128}}
\and P.~Corcho-Caballero\orcid{0000-0001-6327-7080}\inst{\ref{aff8}}
\and B.~Csizi\orcid{0000-0003-3227-6581}\inst{\ref{aff96}}
\and O.~Cucciati\orcid{0000-0002-9336-7551}\inst{\ref{aff17}}
\and J.~E.~Davies\orcid{0000-0002-5079-9098}\inst{\ref{aff10}}
\and S.~Davini\orcid{0000-0003-3269-1718}\inst{\ref{aff26}}
\and T.~de~Boer\orcid{0000-0001-5486-2747}\inst{\ref{aff129}}
\and F.~De~Paolis\orcid{0000-0001-6460-7563}\inst{\ref{aff130},\ref{aff131},\ref{aff132}}
\and G.~Desprez\orcid{0000-0001-8325-1742}\inst{\ref{aff8}}
\and A.~D\'iaz-S\'anchez\orcid{0000-0003-0748-4768}\inst{\ref{aff133}}
\and S.~Di~Domizio\orcid{0000-0003-2863-5895}\inst{\ref{aff25},\ref{aff26}}
\and J.~M.~Diego\orcid{0000-0001-9065-3926}\inst{\ref{aff134}}
\and V.~Duret\orcid{0009-0009-0383-4960}\inst{\ref{aff9}}
\and M.~Y.~Elkhashab\orcid{0000-0001-9306-2603}\inst{\ref{aff6},\ref{aff19},\ref{aff79},\ref{aff18}}
\and A.~Enia\orcid{0000-0002-0200-2857}\inst{\ref{aff113},\ref{aff17}}
\and Y.~Fang\orcid{0000-0002-0334-6950}\inst{\ref{aff24}}
\and A.~Farina\orcid{0009-0000-3420-929X}\inst{\ref{aff16},\ref{aff26}}
\and A.~M.~N.~Ferguson\orcid{0000-0001-7934-1278}\inst{\ref{aff38}}
\and A.~Finoguenov\orcid{0000-0002-4606-5403}\inst{\ref{aff62}}
\and A.~Franco\orcid{0000-0002-4761-366X}\inst{\ref{aff130},\ref{aff131},\ref{aff132}}
\and K.~Ganga\orcid{0000-0001-8159-8208}\inst{\ref{aff75}}
\and E.~Gaztanaga\orcid{0000-0001-9632-0815}\inst{\ref{aff103},\ref{aff104},\ref{aff127}}
\and Z.~Ghaffari\orcid{0000-0002-6467-8078}\inst{\ref{aff6},\ref{aff18}}
\and F.~Giacomini\orcid{0000-0002-3129-2814}\inst{\ref{aff22}}
\and F.~Gianotti\orcid{0000-0003-4666-119X}\inst{\ref{aff17}}
\and G.~Gozaliasl\orcid{0000-0002-0236-919X}\inst{\ref{aff135},\ref{aff62}}
\and A.~Gruppuso\orcid{0000-0001-9272-5292}\inst{\ref{aff17},\ref{aff22}}
\and M.~Guidi\orcid{0000-0001-9408-1101}\inst{\ref{aff21},\ref{aff17}}
\and A.~Hall\orcid{0000-0002-3139-8651}\inst{\ref{aff38}}
\and N.~A.~Hatch\orcid{0000-0001-5600-0534}\inst{\ref{aff136}}
\and C.~Hern\'andez-Monteagudo\orcid{0000-0001-5471-9166}\inst{\ref{aff90},\ref{aff37}}
\and T.~Heuberger\orcid{0009-0009-3474-8467}\inst{\ref{aff69}}
\and H.~Hildebrandt\orcid{0000-0002-9814-3338}\inst{\ref{aff137}}
\and J.~Hjorth\orcid{0000-0002-4571-2306}\inst{\ref{aff138}}
\and M.~Jauzac\orcid{0000-0003-1974-8732}\inst{\ref{aff139},\ref{aff73},\ref{aff140},\ref{aff141},\ref{aff93}}
\and J.~J.~E.~Kajava\orcid{0000-0002-3010-8333}\inst{\ref{aff142},\ref{aff143},\ref{aff144}}
\and Y.~Kang\orcid{0009-0000-8588-7250}\inst{\ref{aff96}}
\and V.~Kansal\orcid{0000-0002-4008-6078}\inst{\ref{aff145},\ref{aff146}}
\and D.~Karagiannis\orcid{0000-0002-4927-0816}\inst{\ref{aff106},\ref{aff147}}
\and J.~Kim\orcid{0000-0003-2776-2761}\inst{\ref{aff51}}
\and C.~C.~Kirkpatrick\inst{\ref{aff97}}
\and A.~Kov\'acs\orcid{0000-0002-5825-579X}\inst{\ref{aff148},\ref{aff149}}
\and I.~Kova{\v{c}}i{\'{c}}\orcid{0000-0001-6751-3263}\inst{\ref{aff123}}
\and K.~Koyama\orcid{0000-0001-6727-6915}\inst{\ref{aff127}}
\and S.~Kruk\orcid{0000-0001-8010-8879}\inst{\ref{aff14}}
\and M.~C.~Lam\orcid{0000-0002-9347-2298}\inst{\ref{aff38}}
\and M.~Lattanzi\orcid{0000-0003-1059-2532}\inst{\ref{aff107}}
\and J.~Le~Graet\orcid{0000-0001-6523-7971}\inst{\ref{aff9}}
\and G.~Leroy\orcid{0009-0004-2523-4425}\inst{\ref{aff139},\ref{aff73}}
\and J.~Lesgourgues\orcid{0000-0001-7627-353X}\inst{\ref{aff150}}
\and T.~I.~Liaudat\orcid{0000-0002-9104-314X}\inst{\ref{aff151}}
\and L.~Linke\orcid{0000-0002-2622-8113}\inst{\ref{aff96}}
\and S.~J.~Liu\orcid{0000-0001-7680-2139}\inst{\ref{aff49}}
\and G.~Maggio\orcid{0000-0003-4020-4836}\inst{\ref{aff6}}
\and F.~Mannucci\orcid{0000-0002-4803-2381}\inst{\ref{aff152}}
\and F.~R.~Marleau\orcid{0000-0002-1442-2947}\inst{\ref{aff96}}
\and C.~J.~A.~P.~Martins\orcid{0000-0002-4886-9261}\inst{\ref{aff153},\ref{aff154}}
\and M.~Miluzio\inst{\ref{aff14},\ref{aff155}}
\and G.~Morgante\inst{\ref{aff17}}
\and S.~Nadathur\orcid{0000-0001-9070-3102}\inst{\ref{aff127}}
\and K.~Naidoo\orcid{0000-0002-9182-1802}\inst{\ref{aff127},\ref{aff10}}
\and P.~Natoli\orcid{0000-0003-0126-9100}\inst{\ref{aff106},\ref{aff107}}
\and A.~Navarro-Alsina\orcid{0000-0002-3173-2592}\inst{\ref{aff70}}
\and S.~Nesseris\orcid{0000-0002-0567-0324}\inst{\ref{aff92}}
\and L.~Nicastro\orcid{0000-0001-8534-6788}\inst{\ref{aff17}}
\and F.~Oppizzi\orcid{0000-0003-3904-8370}\inst{\ref{aff26},\ref{aff48},\ref{aff91}}
\and F.~Pace\orcid{0000-0001-8039-0480}\inst{\ref{aff29},\ref{aff30},\ref{aff31},\ref{aff156}}
\and D.~Paoletti\orcid{0000-0003-4761-6147}\inst{\ref{aff17},\ref{aff105}}
\and F.~Passalacqua\orcid{0000-0002-8606-4093}\inst{\ref{aff48}}
\and L.~Patrizii\inst{\ref{aff22}}
\and C.~Pattison\orcid{0000-0003-3272-2617}\inst{\ref{aff127}}
\and D.~Potter\orcid{0000-0002-0757-5195}\inst{\ref{aff157}}
\and G.~W.~Pratt\inst{\ref{aff86}}
\and S.~Quai\orcid{0000-0002-0449-8163}\inst{\ref{aff72},\ref{aff17}}
\and G.~Rodighiero\orcid{0000-0002-9415-2296}\inst{\ref{aff91},\ref{aff12}}
\and W.~Roster\orcid{0000-0002-9149-6528}\inst{\ref{aff23}}
\and S.~Sacquegna\orcid{0000-0002-8433-6630}\inst{\ref{aff158}}
\and M.~Sahl\'en\orcid{0000-0003-0973-4804}\inst{\ref{aff159}}
\and D.~B.~Sanders\orcid{0000-0002-1233-9998}\inst{\ref{aff129}}
\and A.~Schneider\orcid{0000-0001-7055-8104}\inst{\ref{aff157}}
\and D.~Sciotti\orcid{0009-0008-4519-2620}\inst{\ref{aff1},\ref{aff71}}
\and E.~Sellentin\orcid{0009-0002-2655-3458}\inst{\ref{aff160},\ref{aff5}}
\and S.~Serjeant\orcid{0000-0002-0517-7943}\inst{\ref{aff161}}
\and F.~Shankar\orcid{0000-0001-8973-5051}\inst{\ref{aff162}}
\and A.~Shulevski\orcid{0000-0002-1827-0469}\inst{\ref{aff163},\ref{aff8},\ref{aff5},\ref{aff164}}
\and L.~C.~Smith\orcid{0000-0002-3259-2771}\inst{\ref{aff165}}
\and I.~Szapudi\orcid{0000-0003-2274-0301}\inst{\ref{aff129}}
\and M.~Talia\orcid{0000-0003-4352-2063}\inst{\ref{aff72},\ref{aff17}}
\and K.~Tanidis\orcid{0000-0001-9843-5130}\inst{\ref{aff166}}
\and C.~Tao\orcid{0000-0001-7961-8177}\inst{\ref{aff9}}
\and F.~Tarsitano\orcid{0000-0002-5919-0238}\inst{\ref{aff167},\ref{aff168},\ref{aff46}}
\and G.~Testera\orcid{0000-0003-2970-766X}\inst{\ref{aff26}}
\and R.~Teyssier\orcid{0000-0001-7689-0933}\inst{\ref{aff169}}
\and S.~Tosi\orcid{0000-0002-7275-9193}\inst{\ref{aff25},\ref{aff16},\ref{aff26}}
\and C.~Uhlemann\orcid{0000-0001-7831-1579}\inst{\ref{aff170},\ref{aff171}}
\and M.~Urbano\orcid{0000-0001-5640-0650}\inst{\ref{aff136},\ref{aff124}}
\and C.~Valieri\inst{\ref{aff22}}
\and A.~Venhola\orcid{0000-0001-6071-4564}\inst{\ref{aff172}}
\and D.~Vergani\orcid{0000-0003-0898-2216}\inst{\ref{aff17}}
\and P.~Vielzeuf\orcid{0000-0003-2035-9339}\inst{\ref{aff9}}
\and E.~Vilenius\orcid{0000-0002-6184-7681}\inst{\ref{aff173}}
\and S.~Vinciguerra\orcid{0009-0005-4018-3184}\inst{\ref{aff41}}
\and M.~von~Wietersheim-Kramsta\orcid{0000-0003-4986-5091}\inst{\ref{aff73},\ref{aff139}}
\and L.~Wang\orcid{0000-0002-6736-9158}\inst{\ref{aff174},\ref{aff8}}
\and A.~H.~Wright\orcid{0000-0001-7363-7932}\inst{\ref{aff137}}
\and D.~Scott\orcid{0000-0002-6878-9840}\inst{\ref{aff175}}}
										   
\institute{INAF-Osservatorio Astronomico di Roma, Via Frascati 33, 00078 Monteporzio Catone, Italy\label{aff1}
\and
Space Science Data Center, Italian Space Agency, via del Politecnico snc, 00133 Roma, Italy\label{aff2}
\and
Caltech/IPAC, 1200 E. California Blvd., Pasadena, CA 91125, USA\label{aff3}
\and
Aerospace Logistics Technology Engineering Company, Corso Marche 79, 10146 Torino, Italy\label{aff4}
\and
Leiden Observatory, Leiden University, Einsteinweg 55, 2333 CC Leiden, The Netherlands\label{aff5}
\and
INAF-Osservatorio Astronomico di Trieste, Via G. B. Tiepolo 11, 34143 Trieste, Italy\label{aff6}
\and
Universit\'e Lyon 1, CNRS/IN2P3, IP2I Lyon, Villeurbanne, France\label{aff7}
\and
Kapteyn Astronomical Institute, University of Groningen, PO Box 800, 9700 AV Groningen, The Netherlands\label{aff8}
\and
Aix-Marseille Universit\'e, CNRS/IN2P3, CPPM, Marseille, France\label{aff9}
\and
Max-Planck-Institut f\"ur Astronomie, K\"onigstuhl 17, 69117 Heidelberg, Germany\label{aff10}
\and
California Institute of Technology, 1200 E California Blvd, Pasadena, CA 91125, USA\label{aff11}
\and
INAF-Osservatorio Astronomico di Padova, Via dell'Osservatorio 5, 35122 Padova, Italy\label{aff12}
\and
Infrared Processing and Analysis Center, California Institute of Technology, Pasadena, CA 91125, USA\label{aff13}
\and
ESAC/ESA, Camino Bajo del Castillo, s/n., Urb. Villafranca del Castillo, 28692 Villanueva de la Ca\~nada, Madrid, Spain\label{aff14}
\and
School of Mathematics and Physics, University of Surrey, Guildford, Surrey, GU2 7XH, UK\label{aff15}
\and
INAF-Osservatorio Astronomico di Brera, Via Brera 28, 20122 Milano, Italy\label{aff16}
\and
INAF-Osservatorio di Astrofisica e Scienza dello Spazio di Bologna, Via Piero Gobetti 93/3, 40129 Bologna, Italy\label{aff17}
\and
IFPU, Institute for Fundamental Physics of the Universe, via Beirut 2, 34151 Trieste, Italy\label{aff18}
\and
INFN, Sezione di Trieste, Via Valerio 2, 34127 Trieste TS, Italy\label{aff19}
\and
SISSA, International School for Advanced Studies, Via Bonomea 265, 34136 Trieste TS, Italy\label{aff20}
\and
Dipartimento di Fisica e Astronomia, Universit\`a di Bologna, Via Gobetti 93/2, 40129 Bologna, Italy\label{aff21}
\and
INFN-Sezione di Bologna, Viale Berti Pichat 6/2, 40127 Bologna, Italy\label{aff22}
\and
Max Planck Institute for Extraterrestrial Physics, Giessenbachstr. 1, 85748 Garching, Germany\label{aff23}
\and
Universit\"ats-Sternwarte M\"unchen, Fakult\"at f\"ur Physik, Ludwig-Maximilians-Universit\"at M\"unchen, Scheinerstr.~1, 81679 M\"unchen, Germany\label{aff24}
\and
Dipartimento di Fisica, Universit\`a di Genova, Via Dodecaneso 33, 16146, Genova, Italy\label{aff25}
\and
INFN-Sezione di Genova, Via Dodecaneso 33, 16146, Genova, Italy\label{aff26}
\and
Department of Physics "E. Pancini", University Federico II, Via Cinthia 6, 80126, Napoli, Italy\label{aff27}
\and
INAF-Osservatorio Astronomico di Capodimonte, Via Moiariello 16, 80131 Napoli, Italy\label{aff28}
\and
Dipartimento di Fisica, Universit\`a degli Studi di Torino, Via P. Giuria 1, 10125 Torino, Italy\label{aff29}
\and
INFN-Sezione di Torino, Via P. Giuria 1, 10125 Torino, Italy\label{aff30}
\and
INAF-Osservatorio Astrofisico di Torino, Via Osservatorio 20, 10025 Pino Torinese (TO), Italy\label{aff31}
\and
INAF-IASF Milano, Via Alfonso Corti 12, 20133 Milano, Italy\label{aff32}
\and
Centro de Investigaciones Energ\'eticas, Medioambientales y Tecnol\'ogicas (CIEMAT), Avenida Complutense 40, 28040 Madrid, Spain\label{aff33}
\and
Port d'Informaci\'{o} Cient\'{i}fica, Campus UAB, C. Albareda s/n, 08193 Bellaterra (Barcelona), Spain\label{aff34}
\and
INFN -- Sezione di Napoli, Via Cinthia 6, 80126, Napoli, Italy\label{aff35}
\and
Dipartimento di Fisica e Astronomia "Augusto Righi" - Alma Mater Studiorum Universit\`a di Bologna, Viale Berti Pichat 6/2, 40127 Bologna, Italy\label{aff36}
\and
Instituto de Astrof\'{\i}sica de Canarias, E-38205 La Laguna, Tenerife, Spain\label{aff37}
\and
Institute for Astronomy, University of Edinburgh, Royal Observatory, Blackford Hill, Edinburgh EH9 3HJ, UK\label{aff38}
\and
Jodrell Bank Centre for Astrophysics, Department of Physics and Astronomy, University of Manchester, Oxford Road, Manchester M13 9PL, UK\label{aff39}
\and
European Space Agency/ESRIN, Largo Galileo Galilei 1, 00044 Frascati, Roma, Italy\label{aff40}
\and
Aix-Marseille Universit\'e, CNRS, CNES, LAM, Marseille, France\label{aff41}
\and
Institut de Ci\`{e}ncies del Cosmos (ICCUB), Universitat de Barcelona (IEEC-UB), Mart\'{i} i Franqu\`{e}s 1, 08028 Barcelona, Spain\label{aff42}
\and
Instituci\'o Catalana de Recerca i Estudis Avan\c{c}ats (ICREA), Passeig de Llu\'{\i}s Companys 23, 08010 Barcelona, Spain\label{aff43}
\and
Institut de Ciencies de l'Espai (IEEC-CSIC), Campus UAB, Carrer de Can Magrans, s/n Cerdanyola del Vall\'es, 08193 Barcelona, Spain\label{aff44}
\and
Universit\'e Lyon 1, CNRS/IN2P3, IUF, IP2I Lyon, Villeurbanne, France \label{aff45}
\and
Department of Astronomy, University of Geneva, ch. d'Ecogia 16, 1290 Versoix, Switzerland\label{aff46}
\and
Universit\'e Paris-Saclay, CNRS, Institut d'astrophysique spatiale, 91405, Orsay, France\label{aff47}
\and
INFN-Padova, Via Marzolo 8, 35131 Padova, Italy\label{aff48}
\and
INAF-Istituto di Astrofisica e Planetologia Spaziali, via del Fosso del Cavaliere, 100, 00100 Roma, Italy\label{aff49}
\and
School of Physics, HH Wills Physics Laboratory, University of Bristol, Tyndall Avenue, Bristol, BS8 1TL, UK\label{aff50}
\and
Department of Physics, University of Oxford, Keble Road, Oxford OX1 3RH, UK\label{aff51}
\and
University Observatory, LMU Faculty of Physics, Scheinerstr.~1, 81679 Munich, Germany\label{aff52}
\and
Institute of Theoretical Astrophysics, University of Oslo, P.O. Box 1029 Blindern, 0315 Oslo, Norway\label{aff53}
\and
Jet Propulsion Laboratory, California Institute of Technology, 4800 Oak Grove Drive, Pasadena, CA, 91109, USA\label{aff54}
\and
School of Physics and Astronomy, Lancaster University, Lancaster, LA1 4YB, UK\label{aff55}
\and
Felix Hormuth Engineering, Goethestr. 17, 69181 Leimen, Germany\label{aff56}
\and
Technical University of Denmark, Elektrovej 327, 2800 Kgs. Lyngby, Denmark\label{aff57}
\and
Cosmic Dawn Center (DAWN), Denmark\label{aff58}
\and
NASA Goddard Space Flight Center, Greenbelt, MD 20771, USA\label{aff59}
\and
Department of Physics and Astronomy, University College London, Gower Street, London WC1E 6BT, UK\label{aff60}
\and
Universit\'e de Gen\`eve, D\'epartement de Physique Th\'eorique and Centre for Astroparticle Physics, 24 quai Ernest-Ansermet, CH-1211 Gen\`eve 4, Switzerland\label{aff61}
\and
Department of Physics, P.O. Box 64, University of Helsinki, 00014 Helsinki, Finland\label{aff62}
\and
Helsinki Institute of Physics, Gustaf H{\"a}llstr{\"o}min katu 2, University of Helsinki, 00014 Helsinki, Finland\label{aff63}
\and
Laboratoire d'etude de l'Univers et des phenomenes eXtremes, Observatoire de Paris, Universit\'e PSL, Sorbonne Universit\'e, CNRS, 92190 Meudon, France\label{aff64}
\and
SKAO, Jodrell Bank, Lower Withington, Macclesfield SK11 9FT, UK\label{aff65}
\and
Centre de Calcul de l'IN2P3/CNRS, 21 avenue Pierre de Coubertin 69627 Villeurbanne Cedex, France\label{aff66}
\and
Dipartimento di Fisica "Aldo Pontremoli", Universit\`a degli Studi di Milano, Via Celoria 16, 20133 Milano, Italy\label{aff67}
\and
INFN-Sezione di Milano, Via Celoria 16, 20133 Milano, Italy\label{aff68}
\and
University of Applied Sciences and Arts of Northwestern Switzerland, School of Computer Science, 5210 Windisch, Switzerland\label{aff69}
\and
Universit\"at Bonn, Argelander-Institut f\"ur Astronomie, Auf dem H\"ugel 71, 53121 Bonn, Germany\label{aff70}
\and
INFN-Sezione di Roma, Piazzale Aldo Moro, 2 - c/o Dipartimento di Fisica, Edificio G. Marconi, 00185 Roma, Italy\label{aff71}
\and
Dipartimento di Fisica e Astronomia "Augusto Righi" - Alma Mater Studiorum Universit\`a di Bologna, via Piero Gobetti 93/2, 40129 Bologna, Italy\label{aff72}
\and
Department of Physics, Institute for Computational Cosmology, Durham University, South Road, Durham, DH1 3LE, UK\label{aff73}
\and
Niels Bohr Institute, University of Copenhagen, Jagtvej 128, 2200 Copenhagen, Denmark\label{aff74}
\and
Universit\'e Paris Cit\'e, CNRS, Astroparticule et Cosmologie, 75013 Paris, France\label{aff75}
\and
CNRS-UCB International Research Laboratory, Centre Pierre Bin\'etruy, IRL2007, CPB-IN2P3, Berkeley, USA\label{aff76}
\and
University of Applied Sciences and Arts of Northwestern Switzerland, School of Engineering, 5210 Windisch, Switzerland\label{aff77}
\and
Institute of Physics, Laboratory of Astrophysics, Ecole Polytechnique F\'ed\'erale de Lausanne (EPFL), Observatoire de Sauverny, 1290 Versoix, Switzerland\label{aff78}
\and
Dipartimento di Fisica - Sezione di Astronomia, Universit\`a di Trieste, Via Tiepolo 11, 34131 Trieste, Italy\label{aff79}
\and
Telespazio UK S.L. for European Space Agency (ESA), Camino bajo del Castillo, s/n, Urbanizacion Villafranca del Castillo, Villanueva de la Ca\~nada, 28692 Madrid, Spain\label{aff80}
\and
Institut de F\'{i}sica d'Altes Energies (IFAE), The Barcelona Institute of Science and Technology, Campus UAB, 08193 Bellaterra (Barcelona), Spain\label{aff81}
\and
European Space Agency/ESTEC, Keplerlaan 1, 2201 AZ Noordwijk, The Netherlands\label{aff82}
\and
Waterloo Centre for Astrophysics, University of Waterloo, Waterloo, Ontario N2L 3G1, Canada\label{aff83}
\and
Department of Physics and Astronomy, University of Waterloo, Waterloo, Ontario N2L 3G1, Canada\label{aff84}
\and
Perimeter Institute for Theoretical Physics, Waterloo, Ontario N2L 2Y5, Canada\label{aff85}
\and
Universit\'e Paris-Saclay, Universit\'e Paris Cit\'e, CEA, CNRS, AIM, 91191, Gif-sur-Yvette, France\label{aff86}
\and
Centre National d'Etudes Spatiales -- Centre spatial de Toulouse, 18 avenue Edouard Belin, 31401 Toulouse Cedex 9, France\label{aff87}
\and
Institute of Space Science, Str. Atomistilor, nr. 409 M\u{a}gurele, Ilfov, 077125, Romania\label{aff88}
\and
Consejo Superior de Investigaciones Cientificas, Calle Serrano 117, 28006 Madrid, Spain\label{aff89}
\and
Universidad de La Laguna, Dpto. Astrof\'\i sica, E-38206 La Laguna, Tenerife, Spain\label{aff90}
\and
Dipartimento di Fisica e Astronomia "G. Galilei", Universit\`a di Padova, Via Marzolo 8, 35131 Padova, Italy\label{aff91}
\and
Instituto de F\'isica Te\'orica UAM-CSIC, Campus de Cantoblanco, 28049 Madrid, Spain\label{aff92}
\and
Institut de Recherche en Astrophysique et Plan\'etologie (IRAP), Universit\'e de Toulouse, CNRS, UPS, CNES, 14 Av. Edouard Belin, 31400 Toulouse, France\label{aff93}
\and
Universit\'e St Joseph; Faculty of Sciences, Beirut, Lebanon\label{aff94}
\and
Departamento de F\'isica, FCFM, Universidad de Chile, Blanco Encalada 2008, Santiago, Chile\label{aff95}
\and
Universit\"at Innsbruck, Institut f\"ur Astro- und Teilchenphysik, Technikerstr. 25/8, 6020 Innsbruck, Austria\label{aff96}
\and
Department of Physics and Helsinki Institute of Physics, Gustaf H\"allstr\"omin katu 2, University of Helsinki, 00014 Helsinki, Finland\label{aff97}
\and
Departamento de F\'isica, Faculdade de Ci\^encias, Universidade de Lisboa, Edif\'icio C8, Campo Grande, PT1749-016 Lisboa, Portugal\label{aff98}
\and
Instituto de Astrof\'isica e Ci\^encias do Espa\c{c}o, Faculdade de Ci\^encias, Universidade de Lisboa, Tapada da Ajuda, 1349-018 Lisboa, Portugal\label{aff99}
\and
Cosmic Dawn Center (DAWN)\label{aff100}
\and
Universidad Polit\'ecnica de Cartagena, Departamento de Electr\'onica y Tecnolog\'ia de Computadoras,  Plaza del Hospital 1, 30202 Cartagena, Spain\label{aff101}
\and
European University of Technology EUt+, European Union\label{aff102}
\and
Institute of Space Sciences (ICE, CSIC), Campus UAB, Carrer de Can Magrans, s/n, 08193 Barcelona, Spain\label{aff103}
\and
Institut d'Estudis Espacials de Catalunya (IEEC),  Edifici RDIT, Campus UPC, 08860 Castelldefels, Barcelona, Spain\label{aff104}
\and
INFN-Bologna, Via Irnerio 46, 40126 Bologna, Italy\label{aff105}
\and
Dipartimento di Fisica e Scienze della Terra, Universit\`a degli Studi di Ferrara, Via Giuseppe Saragat 1, 44122 Ferrara, Italy\label{aff106}
\and
Istituto Nazionale di Fisica Nucleare, Sezione di Ferrara, Via Giuseppe Saragat 1, 44122 Ferrara, Italy\label{aff107}
\and
ICSC - Centro Nazionale di Ricerca in High Performance Computing, Big Data e Quantum Computing, Via Magnanelli 2, Bologna, Italy\label{aff108}
\and
INAF, Istituto di Radioastronomia, Via Piero Gobetti 101, 40129 Bologna, Italy\label{aff109}
\and
Astronomical Observatory of the Autonomous Region of the Aosta Valley (OAVdA), Loc. Lignan 39, I-11020, Nus (Aosta Valley), Italy\label{aff110}
\and
Universit\'e C\^{o}te d'Azur, Observatoire de la C\^{o}te d'Azur, CNRS, Laboratoire Lagrange, Bd de l'Observatoire, CS 34229, 06304 Nice cedex 4, France\label{aff111}
\and
Department of Mathematical Physics, Institute of Physics, University of S\~ao Paulo, R. do Mat\~ao 1371, 05508-090, S\~ao Paulo, SP, Brazil\label{aff112}
\and
CEA Saclay, DFR/IRFU, Service d'Astrophysique, Bat. 709, 91191 Gif-sur-Yvette, France\label{aff113}
\and
Oskar Klein Centre for Cosmoparticle Physics, Department of Physics, Stockholm University, Stockholm, SE-106 91, Sweden\label{aff114}
\and
Astrophysics Group, Blackett Laboratory, Imperial College London, London SW7 2AZ, UK\label{aff115}
\and
Univ. Grenoble Alpes, CNRS, Grenoble INP, LPSC-IN2P3, 53, Avenue des Martyrs, 38000 Grenoble, France\label{aff116}
\and
Dipartimento di Fisica, Sapienza Universit\`a di Roma, Piazzale Aldo Moro 2, 00185 Roma, Italy\label{aff117}
\and
Institut d'Astrophysique de Paris, 98bis Boulevard Arago, 75014, Paris, France\label{aff118}
\and
ICL, Junia, Universit\'e Catholique de Lille, LITL, 59000 Lille, France\label{aff119}
\and
CERCA/ISO, Department of Physics, Case Western Reserve University, 10900 Euclid Avenue, Cleveland, OH 44106, USA\label{aff120}
\and
Laboratoire Univers et Th\'eorie, Observatoire de Paris, Universit\'e PSL, Universit\'e Paris Cit\'e, CNRS, 92190 Meudon, France\label{aff121}
\and
Departamento de F{\'\i}sica Fundamental. Universidad de Salamanca. Plaza de la Merced s/n. 37008 Salamanca, Spain\label{aff122}
\and
Universiteit Gent, Department of Physics and Astronomy, Proeftuinstraat 86 N3, 9000 Ghent, Belgium
\label{aff123}
\and
Universit\'e de Strasbourg, CNRS, Observatoire astronomique de Strasbourg, UMR 7550, 67000 Strasbourg, France\label{aff124}
\and
Center for Data-Driven Discovery, Kavli IPMU (WPI), UTIAS, The University of Tokyo, Kashiwa, Chiba 277-8583, Japan\label{aff125}
\and
Department of Physics, Astronomy and Mathematics, University of Hertfordshire, College Lane, Hatfield AL10 9AB, UK\label{aff126}
\and
Institute of Cosmology and Gravitation, University of Portsmouth, Portsmouth PO1 3FX, UK\label{aff127}
\and
Department of Physics \& Astronomy, University of California Irvine, Irvine CA 92697, USA\label{aff128}
\and
Institute for Astronomy, University of Hawaii, 2680 Woodlawn Drive, Honolulu, HI 96822, USA\label{aff129}
\and
Department of Mathematics and Physics E. De Giorgi, University of Salento, Via per Arnesano, CP-I93, 73100, Lecce, Italy\label{aff130}
\and
INFN, Sezione di Lecce, Via per Arnesano, CP-193, 73100, Lecce, Italy\label{aff131}
\and
INAF-Sezione di Lecce, c/o Dipartimento Matematica e Fisica, Via per Arnesano, 73100, Lecce, Italy\label{aff132}
\and
Departamento F\'isica Aplicada, Universidad Polit\'ecnica de Cartagena, Campus Muralla del Mar, 30202 Cartagena, Murcia, Spain\label{aff133}
\and
Instituto de F\'isica de Cantabria, Edificio Juan Jord\'a, Avenida de los Castros, 39005 Santander, Spain\label{aff134}
\and
Department of Computer Science, Aalto University, PO Box 15400, Espoo, FI-00 076, Finland\label{aff135}
\and
School of Physics and Astronomy, University of Nottingham, University Park, Nottingham NG7 2RD, UK\label{aff136}
\and
Ruhr University Bochum, Faculty of Physics and Astronomy, Astronomical Institute (AIRUB), German Centre for Cosmological Lensing (GCCL), 44780 Bochum, Germany\label{aff137}
\and
DARK, Niels Bohr Institute, University of Copenhagen, Jagtvej 155, 2200 Copenhagen, Denmark\label{aff138}
\and
Department of Physics, Centre for Extragalactic Astronomy, Durham University, South Road, Durham, DH1 3LE, UK\label{aff139}
\and
Astrophysics Research Centre, University of KwaZulu-Natal, Westville Campus, Durban 4041, South Africa\label{aff140}
\and
School of Mathematics, Statistics \& Computer Science, University of KwaZulu-Natal, Westville Campus, Durban 4041, South Africa\label{aff141}
\and
Department of Physics and Astronomy, Vesilinnantie 5, University of Turku, 20014 Turku, Finland\label{aff142}
\and
Finnish Centre for Astronomy with ESO (FINCA), Quantum, Vesilinnantie 5, University of Turku, 20014 Turku, Finland\label{aff143}
\and
Serco for European Space Agency (ESA), Camino bajo del Castillo, s/n, Urbanizacion Villafranca del Castillo, Villanueva de la Ca\~nada, 28692 Madrid, Spain\label{aff144}
\and
ARC Centre of Excellence for Dark Matter Particle Physics, Melbourne, Australia\label{aff145}
\and
Centre for Astrophysics \& Supercomputing, Swinburne University of Technology,  Hawthorn, Victoria 3122, Australia\label{aff146}
\and
Department of Physics and Astronomy, University of the Western Cape, Bellville, Cape Town, 7535, South Africa\label{aff147}
\and
MTA-CSFK Lend\"ulet Large-Scale Structure Research Group, Konkoly-Thege Mikl\'os \'ut 15-17, H-1121 Budapest, Hungary\label{aff148}
\and
Konkoly Observatory, HUN-REN CSFK, MTA Centre of Excellence, Budapest, Konkoly Thege Mikl\'os {\'u}t 15-17. H-1121, Hungary\label{aff149}
\and
Institute for Theoretical Particle Physics and Cosmology (TTK), RWTH Aachen University, 52056 Aachen, Germany\label{aff150}
\and
IRFU, CEA, Universit\'e Paris-Saclay 91191 Gif-sur-Yvette Cedex, France\label{aff151}
\and
INAF-Osservatorio Astrofisico di Arcetri, Largo E. Fermi 5, 50125, Firenze, Italy\label{aff152}
\and
Centro de Astrof\'{\i}sica da Universidade do Porto, Rua das Estrelas, 4150-762 Porto, Portugal\label{aff153}
\and
Instituto de Astrof\'isica e Ci\^encias do Espa\c{c}o, Universidade do Porto, CAUP, Rua das Estrelas, PT4150-762 Porto, Portugal\label{aff154}
\and
HE Space for European Space Agency (ESA), Camino bajo del Castillo, s/n, Urbanizacion Villafranca del Castillo, Villanueva de la Ca\~nada, 28692 Madrid, Spain\label{aff155}
\and
Instituto de Astrof\'isica e Ci\^encias do Espa\c{c}o, Faculdade de Ci\^encias, Universidade de Lisboa, Campo Grande, 1749-016 Lisboa, Portugal\label{aff156}
\and
Department of Astrophysics, University of Zurich, Winterthurerstrasse 190, 8057 Zurich, Switzerland\label{aff157}
\and
INAF - Osservatorio Astronomico d'Abruzzo, Via Maggini, 64100, Teramo, Italy\label{aff158}
\and
Theoretical astrophysics, Department of Physics and Astronomy, Uppsala University, Box 524, 751 20 Uppsala, Sweden\label{aff159}
\and
Mathematical Institute, University of Leiden, Einsteinweg 55, 2333 CA Leiden, The Netherlands\label{aff160}
\and
School of Physical Sciences, The Open University, Milton Keynes, MK7 6AA, UK\label{aff161}
\and
School of Physics \& Astronomy, University of Southampton, Highfield Campus, Southampton SO17 1BJ, UK\label{aff162}
\and
ASTRON, the Netherlands Institute for Radio Astronomy, Postbus 2, 7990 AA, Dwingeloo, The Netherlands\label{aff163}
\and
Center for Advanced Interdisciplinary Research, Ss. Cyril and Methodius University in Skopje, Macedonia\label{aff164}
\and
Institute of Astronomy, University of Cambridge, Madingley Road, Cambridge CB3 0HA, UK\label{aff165}
\and
Center for Astrophysics and Cosmology, University of Nova Gorica, Nova Gorica, Slovenia\label{aff166}
\and
Kobayashi-Maskawa Institute for the Origin of Particles and the Universe, Nagoya University, Chikusa-ku, Nagoya, 464-8602, Japan\label{aff167}
\and
Institute for Particle Physics and Astrophysics, Dept. of Physics, ETH Zurich, Wolfgang-Pauli-Strasse 27, 8093 Zurich, Switzerland\label{aff168}
\and
Department of Astrophysical Sciences, Peyton Hall, Princeton University, Princeton, NJ 08544, USA\label{aff169}
\and
Fakult\"at f\"ur Physik, Universit\"at Bielefeld, Postfach 100131, 33501 Bielefeld, Germany\label{aff170}
\and
School of Mathematics, Statistics and Physics, Newcastle University, Herschel Building, Newcastle-upon-Tyne, NE1 7RU, UK\label{aff171}
\and
Space physics and astronomy research unit, University of Oulu, Pentti Kaiteran katu 1, FI-90014 Oulu, Finland\label{aff172}
\and
Mullard Space Science Laboratory, University College London, Holmbury St Mary, Dorking, Surrey RH5 6NT, UK\label{aff173}
\and
SRON Netherlands Institute for Space Research, Landleven 12, 9747 AD, Groningen, The Netherlands\label{aff174}
\and
Department of Physics and Astronomy, University of British Columbia, Vancouver, BC V6T 1Z1, Canada\label{aff175}}

\abstract{
The images collected in photometric mode by the Near-Infrared Spectrometer and Photometer (NISP) instrument of the \Euclid satellite are used to model the point-spread function (PSF) of a selection of sources detected in single and combined frames. We used Q1 and DR1 data to improve the development of the PSF module of the near-infrared processing function (NIR PF) and to refine the models currently provided. We explored a few methods normally used to evaluate the quality of the derived models and discuss possible variations related to the position of the sources in the focal plane and to the wavelength and time of the observations. The procedure of fitting the PSF profile, compared to other methods used for flux estimation, such as aperture photometry, indicates that updated models are better than those provided by Q1 images. The current results are in agreement with several key requirements defined for the mission, but there is still room for further improvements, which we plan to investigate and implement in future releases.
}


\titlerunning{NISP-P point-spread function modelling pipelines for DR1}

\maketitle
\nolinenumbers

\section{\label{sec:intro} Introduction}

The \Euclid mission \citep{EuclidSkyOverview} of the European Space Agency started the nominal observations on 14 February 2024. It has already provided us with data from the \citet{Q1cite}, which covers 63.1\,deg$^2$ of the sky \citep{Q1-TP001}. Data products for the \citet[DR1,][]{DR1cite} are now available and comprise about 1900~deg$^2$ \citep{DR1-TP001}. The observations include data acquired with the two \Euclid instruments: the visible imager \citep[VIS,][]{EuclidSkyVIS} and the Near-Infrared Spectrometer and Photometer \citep[NISP,][]{EuclidSkyNISP}. The NISP instrument was designed to perform slitless spectroscopy and to collect images in three photometric passbands in the near-infrared (NIR): \YE, \JE, and \HE \citep{Schirmer-EP18}, reaching a median depth of $m_{\rm AB}$\,=\,24.4~mag for 5$\sigma$ point-like sources \citep{Scaramella-EP1}.

To achieve scientific goals, \Euclid needs to measure the shapes and colours of galaxies with high precision \citep{Laureijs11}, which in turn requires precise models of the point-spread function (PSF) across the entire field of view (FoV). The primary purpose of the VIS instrument is to measure weak gravitational lensing, also known as cosmic shear. While VIS measures shapes, NISP measures the brightness of galaxies in the infrared bands (photometry) and spreads their light into spectra (slitless spectroscopy). To map the three-dimensional distribution of the Universe, we need to know how far away these galaxies are. This is done by measuring their flux across different wavelength bands to estimate their photometric redshift (\mbox{photo-$z$}). If the PSF is poorly modelled, the measurement of how much light falls into a specific pixel will be wrong, skewing the colour measurements and resulting in incorrect distance estimates.

The observation strategy of the mission consists of taking a sequence of pointings, each comprising multiple exposures following a predefined dithering pattern \citep{Scaramella-EP1,DR1-TP027}. In the Euclid Wide Survey (EWS), the pointings include a set of four dithered exposures in each spectroscopic and photometric channel, while deep fields and calibration fields require repeated visits and a much larger number of exposures. The single-dither exposures are processed by dedicated pipelines, which apply pre-processing steps, perform astrometric and photometric calibrations, resample and combine single exposures in stacked images, extract catalogues, and model the PSF profile\footnote{Here, the PSF profile or PSF shape is the two-dimensional brightness pattern (or one-dimensional cross-section) of a single point source after passing through an optical system, defining its sharpness, symmetry, and degree of blurring.} of point-like sources extracted from individual and stacked frames.

Instead of using traditional optical charge-coupled devices, \Euclid NIR data was acquired with HgCdTe H2RG complementary metal-oxide-semiconductor (CMOS) detectors, arranged in a matrix of 4$\times$4 in the focal plane array (FPA). In this architecture, an infrared-sensitive semiconductor layer is bonded directly to a readout integrated circuit, enabling independent, non-destructive sampling of each pixel \citep[see][for more details]{Kubiketal2016}. Given the diffraction limit of the 1.2~m primary mirror and a pixel scale of \mbox{$\sim$\,\ang{;;0.3}} across the three photometric passbands, the sampling falls below the Nyquist criterion, resulting in an undersampled PSF \citep[see also][for an overview of the PSF generated by the NISP instrument]{EuclidSkyNISP}. Consequently, a dedicated code is required to accurately model and process these undersampled profiles.

Here we describe the {\tt NIR\_PointSpreadFunction} module of the near-infrared processing function \citep[NIR PF,][]{Q1-TP003,DR1-TP003} developed to model the PSF profile of point-like sources observed by NISP under undersampled conditions. Using data collected during Q1 and DR1 observations, we discuss recent improvements in the code and demonstrate its performance through photometric measurements based on PSF models derived from single-dither exposures and stacked images.

This paper is structured as follows. In Sect.~\ref{sec:obs_data_requirem} we present the observation data used and some key requirements for optimal mission performance. In Sect.~\ref{sec:psf_derivation}, we describe the NIR PSF module and the procedure adopted to derive the PSF models. In Sect.~\ref{sec:results}, we present the models obtained on the basis of individual exposures and stacked frames, explore their application to fitting the PSF profile, discuss possible refinements, and evaluate the dependencies on position, wavelength, and time. A brief summary and some final remarks are provided in Sect.~\ref{sec:summary}. Finally, in the Appendices from \ref{sec:appendix_tables} to \ref{sec:appendix_dqc} we present additional tables and plots that illustrate complementary results.

\section{\label{sec:obs_data_requirem} Observation data and mission requirements}

\subsection{\label{sec:data} Data description and processing}

In the work presented in the current paper, we first used Q1 data processed with the initial versions of the code, which were previously developed based on simulated images and on the first data collected during the performance verification (PV) phase. The code has been improved since then, and the available PSF models for DR1 have been refined.

Beginning with level~1 (LE1) data, the raw images observed with the NISP instrument in photometric mode (\mbox{NISP-P}) are then handled by the NIR PF up to the production of level~2 (LE2) data. As described in \citet{Q1-TP003,DR1-TP003}, the NIR PF comprises several dedicated pipelines, designed to process data from LE1 to LE2 or to produce specific calibration data required to process observed images. Here, we focus on the NIR scientific and PSF calibration pipelines.

\paragraph{NIR scientific pipeline} The first modules of this pipeline, or pre-processing steps, account for instrument and detector effects. Next, single-dither exposures are astrometrically and photometrically calibrated, followed by catalogue extraction. This second step includes the first derivation of the PSF models performed on the basis of individual exposures. The four single exposures of wide-field observations are later resampled onto a common grid and combined into a single stacked image. Finally, the pipeline performs a second derivation of the PSF models and a second catalogue extraction, this time based on stacked frames.

\paragraph{NIR PSF calibration pipeline} In this configuration, only the pre-processing steps are executed before deriving the PSF models, which is performed on individual exposures using observations of a dedicated self-calibration field (SCF). The other tasks (astrometric and photometric calibration, catalogue extraction, resampling, stacking, and the derivation of PSF models using stacked frames) are not executed. \\

To produce the results discussed in the following sections, we used data from different calibration observation blocks, which consist of selected areas of the sky dedicated to performing specific calibrations \citep{DR1-TP026}. For images and catalogues from single-dither exposures, we used: SCF (\mbox{CALBLOCK-F-001}) to create image stamps, derive PSF models, perform PSF fitting and report some results from the data quality control (DQC) pipeline (Appendix~\ref{sec:appendix_dqc}); and photo-$z$ calibration fields (\mbox{CALBLOCK-F-005}) to report further results from the DQC pipeline. For images and catalogues from stacked frames, we used data from an observation of \mbox{CALBLOCK-F-005} to create image stamps, derive PSF models, and perform PSF fitting. Table~\ref{table:data} shows some information on the DR1 data products used, which are publicly available at the European Space Astronomy Centre (ESAC), in the \Euclid science archive at the ESAC Science Data Centre (ESDC)\footnote{\url{https://eas.esac.esa.int/sas/}}.

\subsection{\label{sec:requirem} Photometric requirements and goals}

Based on the context and data outlined in the previous sections, several key conditions must be met to achieve optimal technical and scientific performance. To define these baseline requirements, we draw directly from the guidelines established by the mission requirements document \citep{MRD:2015} and by the NIR PF requirements document \citep{NIR_Req:2020}, both internal to the \Euclid mission. While the former focuses primarily on a general set of mission objectives, the latter emphasises the flow-down of these upper-level requirements to the software modules of the NIR pipelines. Combining the insights from both, we have identified the mandatory requirements and desirable goals related to our work, which are detailed below.

\begin{description}

\item[(i)] The radii encircling 50\% (rEE50) and 80\% (rEE80) of energy shall be $<$\,\ang{;;0.4} and $<$\,\ang{;;0.7}, respectively, for all wavelengths below 1486~nm.

\item[(ii)] The PSF centroid of the sources shall be determined to better than 0.1 NISP pixels.

\item[(iii)] The relative and absolute photometric error of the calibrated images shall be $<$\,1.5\% and $<$\,5\% (goal), respectively.

\item[(iv)] The NIR PF shall provide PSF models as a function of position on the FoV, wavelength, and time.

\end{description}

\section{\label{sec:psf_derivation} Point-spread function derivation}

\subsection{\label{sec:psf_task} The NIR PSF module}

The {\tt NIR\_PointSpreadFunction} is one of the tasks of the NIR PF executed by the scientific and calibration pipelines. This module was designed to derive PSF models either inside the 16 NISP detectors of single-dither exposures or inside a stacked frame. For individual frames, it follows the pre-processing steps of the two pipelines, and, in the case of the scientific pipeline, it is followed by the astrometric calibration and other subsequent steps. For the stacked frame, the PSF task is executed after the combination of photometrically calibrated exposures, properly resampled on a common grid.

The NIR PSF software can be executed in two different modes, in Mission Database (MDB) mode or in Inflight mode. As explained below, the MDB mode is applicable only for single-dither exposures, whereas the Inflight mode is appropriate for all NIR frames, single or stacked.

\subsubsection{\label{sec:psf_task_mdb} PSF modelling in MDB mode}

In MDB mode, the software takes as input a library of PSF models stored in the MDB. These models, derived by the NISP instrument team through physical-optical calculations, account for dependencies on focal plane position and wavelength. Since the models correspond to the light reaching individual detectors, the MDB mode is not applicable to stacked frames, which combine the properties of individual images. A total of 2160 PSF shapes were derived on a coarse grid of focal plane positions (five per each of the 16 detectors) and wavelengths (optical monochromatic shapes at nine wavelengths over each of the three \mbox{NISP-P} passbands). Although the MDB mode is not used to produce and deliver the PSF models, this option can be activated if needed, for instance, to perform tests or to help with the implementation of new functionalities in the code.

Considering that an idealised point source of light is changed when passing through the instrument optics and reaching the detector, some broadening effects must be taken into account when predicting the shape of point-like sources in the images based on the library of PSF profiles available in the MDB. The most significant effects typically seen in HgCdTe and CMOS devices are (i) pixelisation, related to the discretisation of the signal over the detector pixels, (ii) intrapixel sensitivity (IPS), which modifies the incoming light depending on where the photons reach the pixels in a sub-pixel scale \citep[e.g.][]{KavaldjievNinkov2001}, and (iii) interpixel capacitance (IPC), which allows the charge of a pixel to be partially transferred to neighbouring pixels \citep[e.g.][]{LeGraetetal2022,EuclidSkyNISP}. IPS and IPC are actually second-order effects compared to pixelisation. There might be other second-order broadening effects, such as the brighter-fatter effect (BFE), but they required further investigation.

To reduce the effects of pixelisation, the MDB PSF models are oversampled by a factor of 6 (pixel size of 3\,$\mu{\rm m}$) compared to the NISP detectors (pixel size of 18\,$\mu{\rm m}$)\footnote{Hereafter, we refer to the pixels of the NISP detectors as ``detector pixels'' and to the size S$\times$S of the oversampled PSF models as ``S/6$\times$S/6 detector pixels''.}. To account for broadening effects, the software then performs a convolution of the original MDB models with two different functions. The first function simulates the IPS as a non-uniform pixel response. For this, we adopted a parabolic sensitivity function. The second function is a 3$\times$3 pixel matrix, which is intended to model the IPC coupling. Appendix~\ref{sec:appendix_ips_ipc} shows more details on the adopted IPS and IPC kernels.

Once the execution of the software finishes, a FITS file is created containing 16 extensions, one for each detector, for a single position within the detectors, and for a single monochromatic wavelength. The position and wavelength of the output models are set in a configuration file, passed as input to the code.

\subsubsection{\label{sec:psf_task_inflight} PSF modelling in Inflight mode}

In Inflight mode, the software derives the PSF models directly from single-dither exposures or from stacked frames. This is the standard mode used for survey data processing. In the case of the NIR scientific pipeline, the models are used mainly for validation purposes, as discussed in Sect.~\ref{sec:psf_variab_valid}. Instead, the dedicated NIR PSF calibration pipeline is used to derive PSF models from SCF data, allowing us to draw on a considerably larger sample of sources, as outlined in the Appendix~\ref{appendix:psfex_tuning}. In this case, the models are passed as input to the NIR astrometric calibration task to compute the centroids of the extracted sources and to the photometric calibration task to calculate the aperture corrections, which in turn are used to compute photometric zero points. We refer to \citet{Q1-TP003,DR1-TP003} for more details on the NIR data-processing steps.

The NIR PSF task requires the utilisation of two external packages that are part of the {\tt AstrOmatic} software\footnote{\url{https://www.astromatic.net/software/}}. First, the {\tt SourceExtractor} package \citep{BertinArnouts1996} is used to extract source catalogues from the observed images. Then, the PSF profile modelling is performed using the {\tt PSFEx} package \citep{Bertin2011}. The extracted catalogues are passed as input to {\tt PSFEx}, which performs source selection by limiting the sample to unflagged detections likely to be point sources.

The extraction of sources by {\tt SourceExtractor} is dictated by some parameters set in a configuration file. Among the parameters, it is worth mentioning those related to the size limit of the sources to be extracted, {\tt DETECT\_MINAREA} and {\tt DETECT\_MAXAREA}, and those related to the detection threshold, {\tt DETECT\_THRESH} and {\tt ANALYSIS\_THRESH}. Moreover, it is important to ensure that the parameters {\tt GAIN}, {\tt SATUR\_LEVEL}, and {\tt PHOT\_APERTURES} are set to the correct values, since they will affect the number of detections and the profile-fitting procedure. The values set in the NIR PSF task for these and other parameters of {\tt SourceExtractor} are listed in Table~\ref{table:sextractor_conf}. A description of all the parameters discussed in the present work is available in Table~\ref{table:sextractor_psfex_params}.

In the NIR pipelines, the flagging of detected sources is based on two independent approaches. First, pixels associated with a given detection might be flagged by the initial steps as saturated, of poor quality, contaminated, or hit by spurious sources such as cosmic rays. This information is propagated through the pipelines and stored in the data quality layer of the multi-extension FITS files created by each task. Then, when {\tt SourceExtractor} is executed, the code takes this flagging information into account and, in addition, might flag some of the detections for other different reasons (truncated objects, blends, etc.). In the current work, sources flagged by the NIR pipelines as {\tt INVALID} or flagged by {\tt SourceExtractor} for some reason are all excluded from the catalogues used in our analysis.

With {\tt PSFEx}, we can (i) define the size and sampling of the derived models; (ii) account for the variation of the PSF profile across the FPA; and (iii) set the level of detail of the models. Similarly to the MDB models, we adopt an oversampling factor of 6 with respect to the pixel size of the NISP detectors, which is higher than the critical (Nyquist) sampling. When running the scientific pipeline (for both single dithers and stacked frames), we set the size of the PSF models to 61/6 detector pixels ($\sim$\,\ang{;;3}). This is 5 times the full width at half maximum (FWHM) measured in the \HE filter, which is typically smaller than 2 detector pixels ($\sim$\,\ang{;;0.6}). When using the PSF calibration pipeline (only for single dithers) to produce more detailed models, the size of the models is set to 141/6 detector pixels ($\sim$\,\ang{;;7}), which is more than 10 times the FWHM in the \HE filter, and large enough to include the most prominent diffraction patterns (rings and spikes)\footnote{Hereafter, we refer to the models derived with the NIR scientific pipeline as ``simplified PSF models'', and to the models derived with the NIR PSF calibration pipeline as ``extended PSF models''.}. Regarding the variation of the PSF across the FPA, the models are set to be constant within each detector and within the stacked frame ({\tt PSFVAR\_DEGREES}\,=\,0), but variable among different detectors. We refer to Sect.~\ref{sec:psf_variab_valid} for more details on the variation of the derived models.

Some important parameters of {\tt PSFEx}, also defined by means of a configuration file, are the following: with {\tt PSF\_SIZE} and {\tt PSF\_SAMPLING} we define the size and sampling of the models; {\tt SAMPLE\_FWHMRANGE}, {\tt SAMPLE\_MINSN}, and {\tt SAMPLE\_MAXELLIP} are used to select bright enough point-like sources; {\tt BASIS\_TYPE} refers to the basis vectors and is currently set to use the Gauss--Laguerre basis functions; {\tt BASIS\_SCALE} affects the width of the PSF profile, and {\tt BASIS\_NUMBER} defines the level of detail in the reconstruction of the core and wings. Another relevant parameter is named {\tt VIGNET}, provided by {\tt SourceExtractor}. It defines the size of the image stamps, centred on each detected source, used by {\tt PSFEx} to derive the PSF models. {\tt BASIS\_NUMBER} and {\tt BASIS\_SCALE} are particularly important since they represent key parameters in the PSF modelling process. They depend on wavelength and require some tuning to set the optimal values (see the Appendix~\ref{appendix:psfex_tuning} for a detailed description of the tuning procedure that we adopted).

\begin{figure*}[p]
\centering
\begin{minipage}[t]{0.33\textwidth}
\centering
\resizebox{\hsize}{!}{\includegraphics{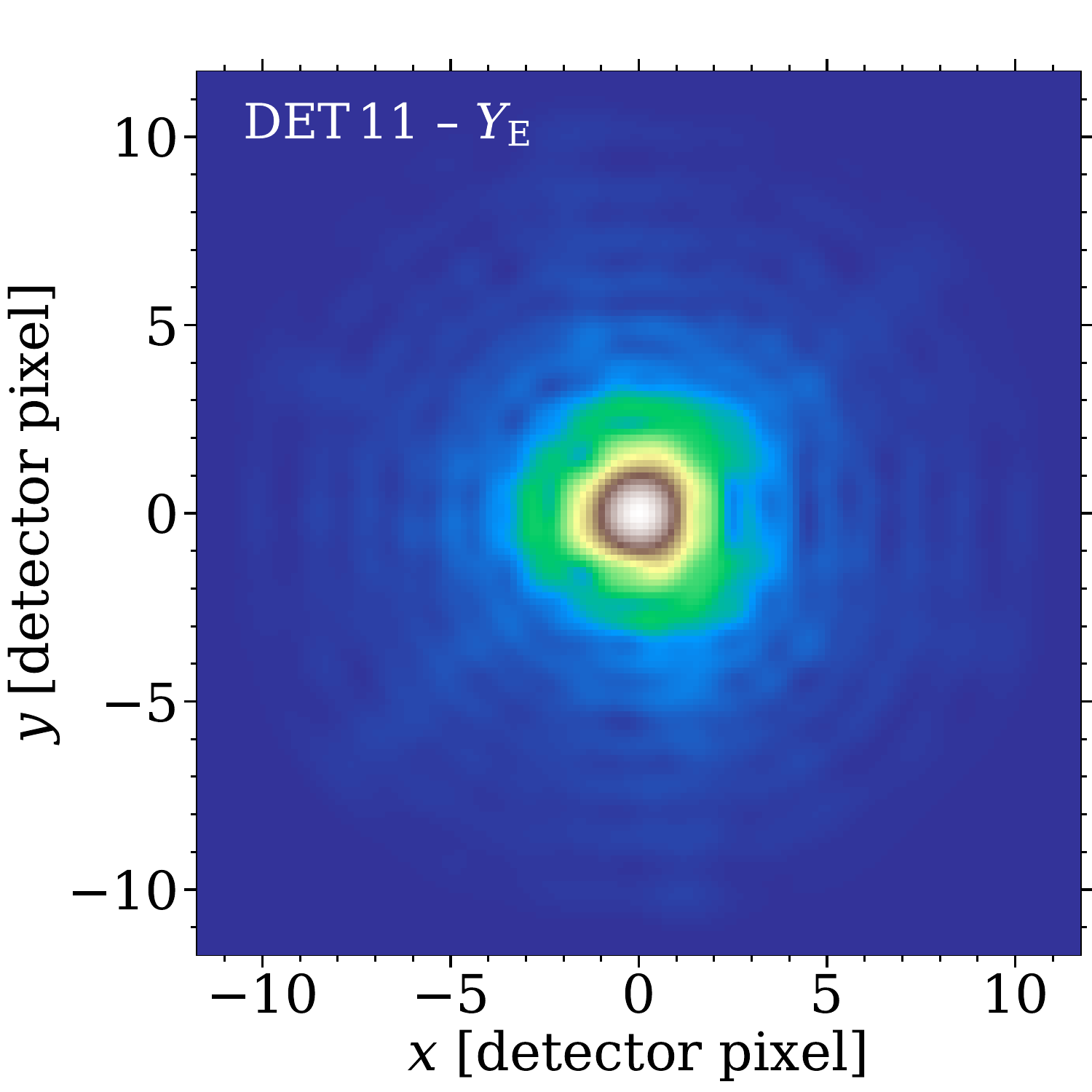}}
\end{minipage}
\begin{minipage}[t]{0.33\textwidth}
\centering
\resizebox{\hsize}{!}{\includegraphics{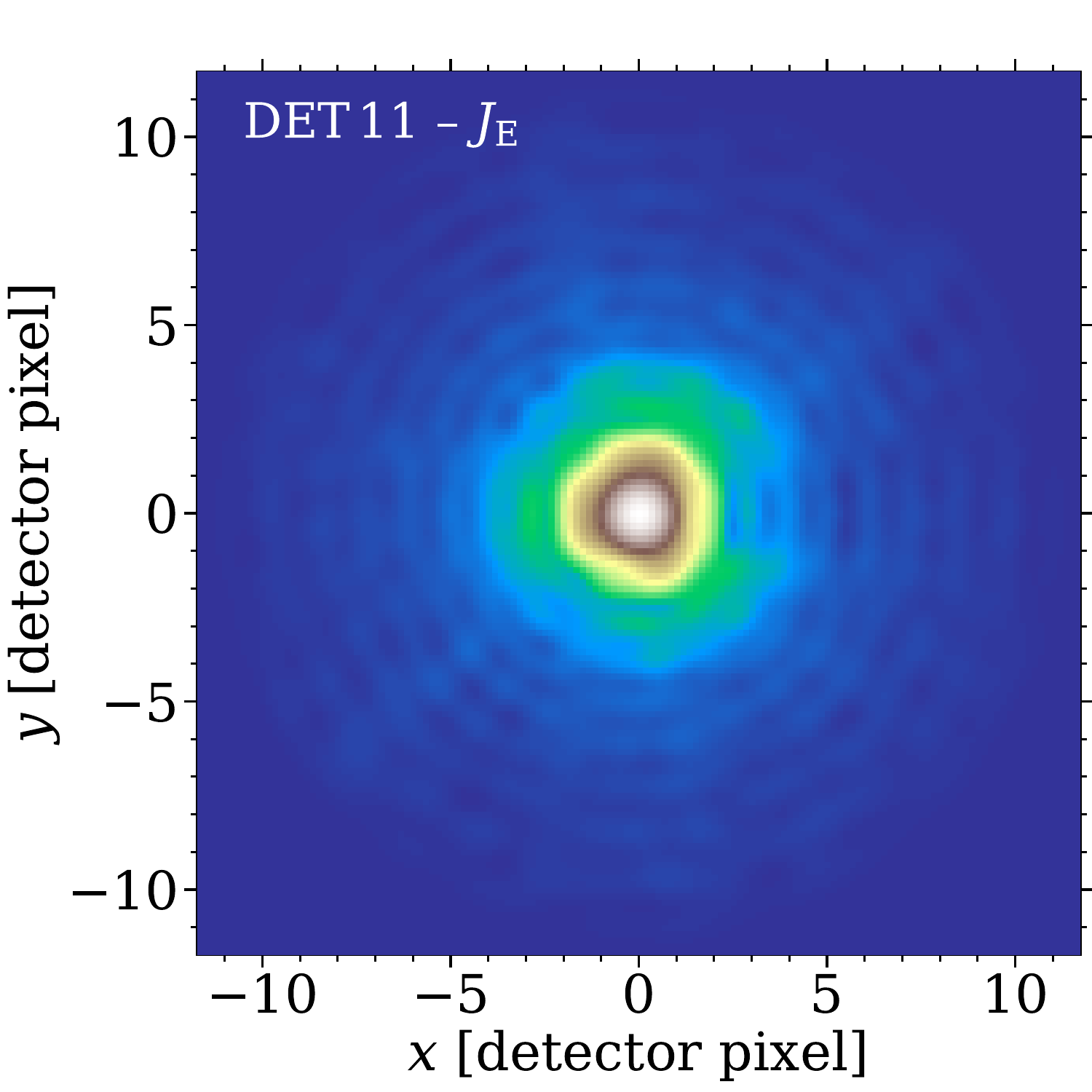}}
\end{minipage}
\begin{minipage}[t]{0.33\textwidth}
\centering
\resizebox{\hsize}{!}{\includegraphics{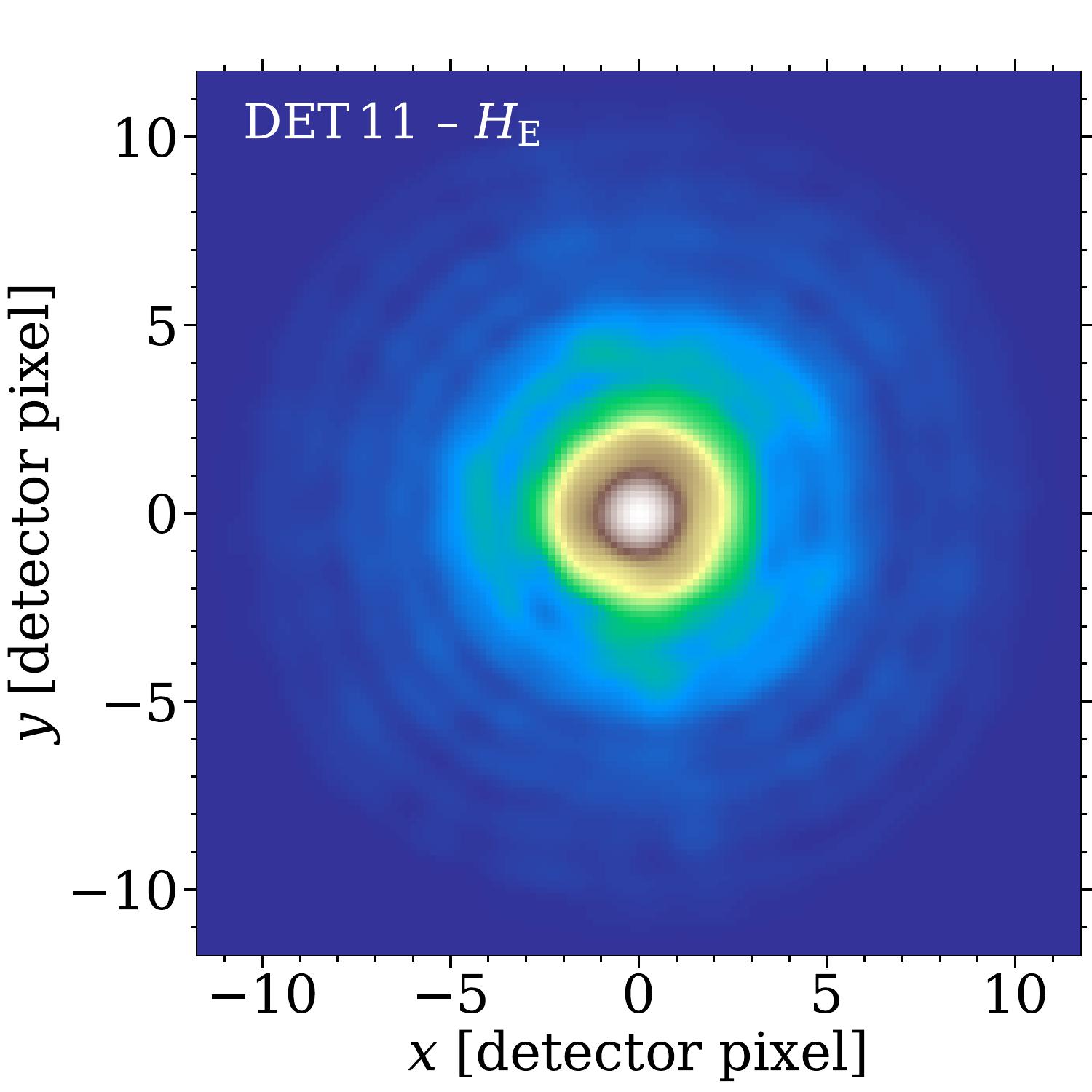}}
\end{minipage} \\
\begin{minipage}[t]{0.33\textwidth}
\centering
\resizebox{\hsize}{!}{\includegraphics{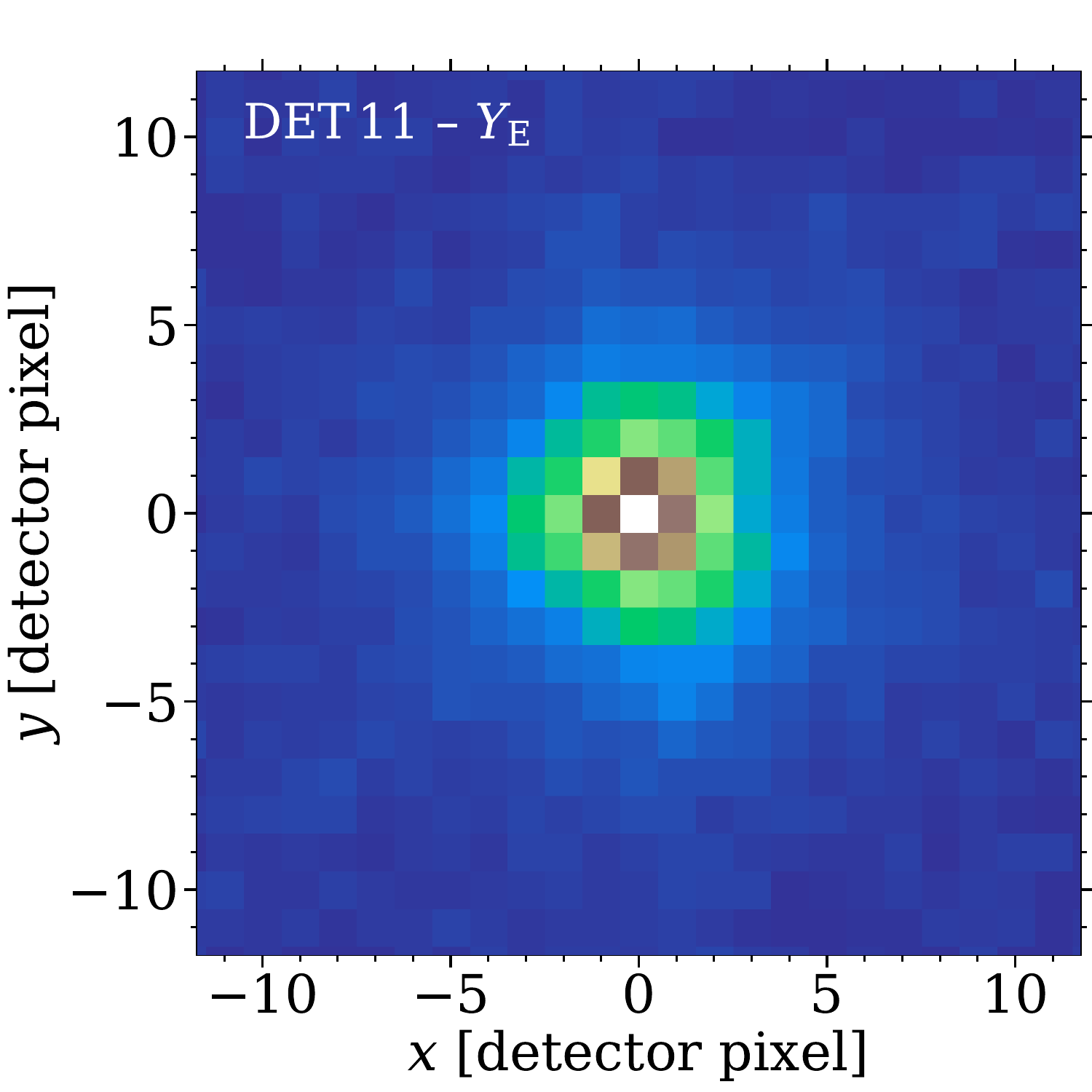}}
\end{minipage}
\begin{minipage}[t]{0.33\textwidth}
\centering
\resizebox{\hsize}{!}{\includegraphics{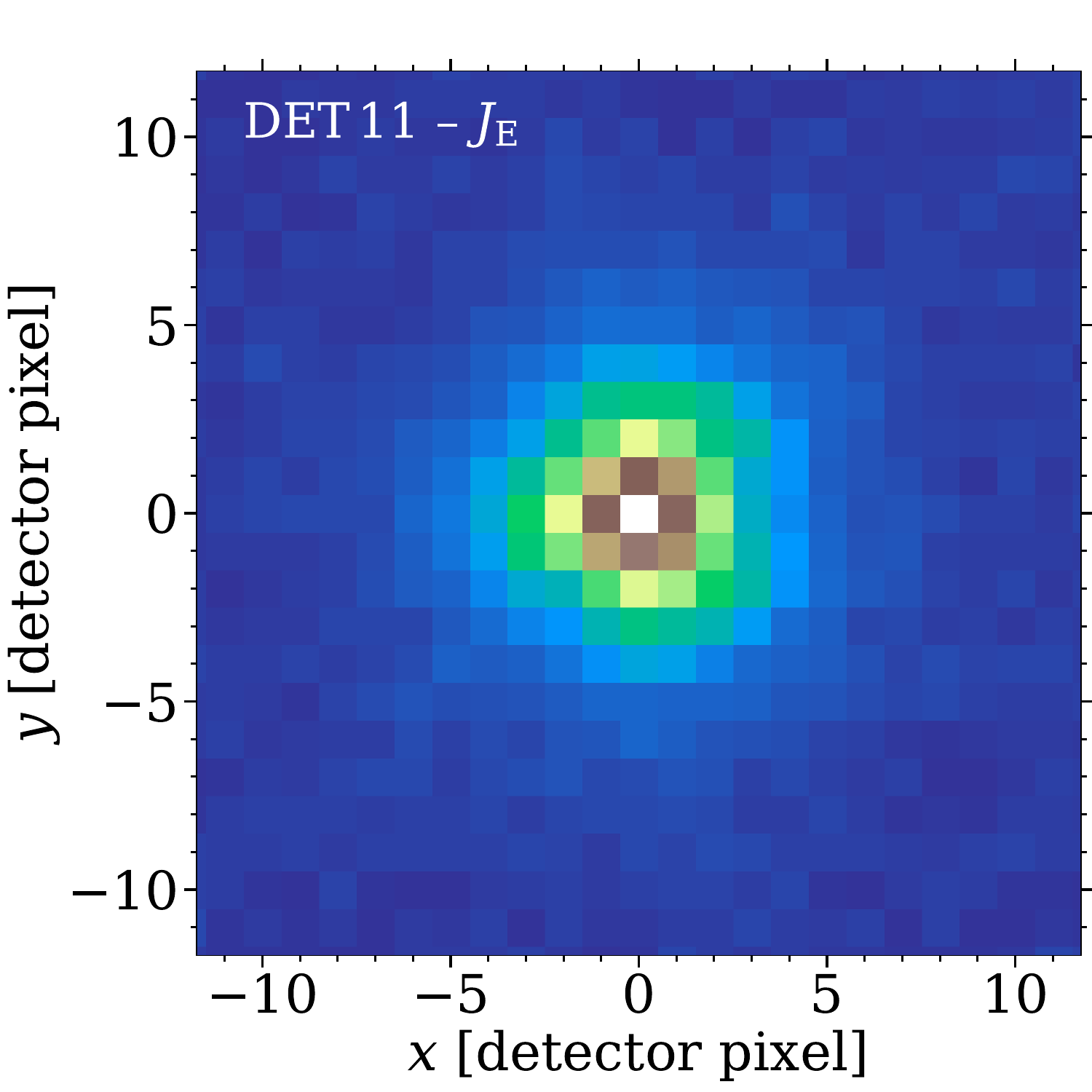}}
\end{minipage}
\begin{minipage}[t]{0.33\textwidth}
\centering
\resizebox{\hsize}{!}{\includegraphics{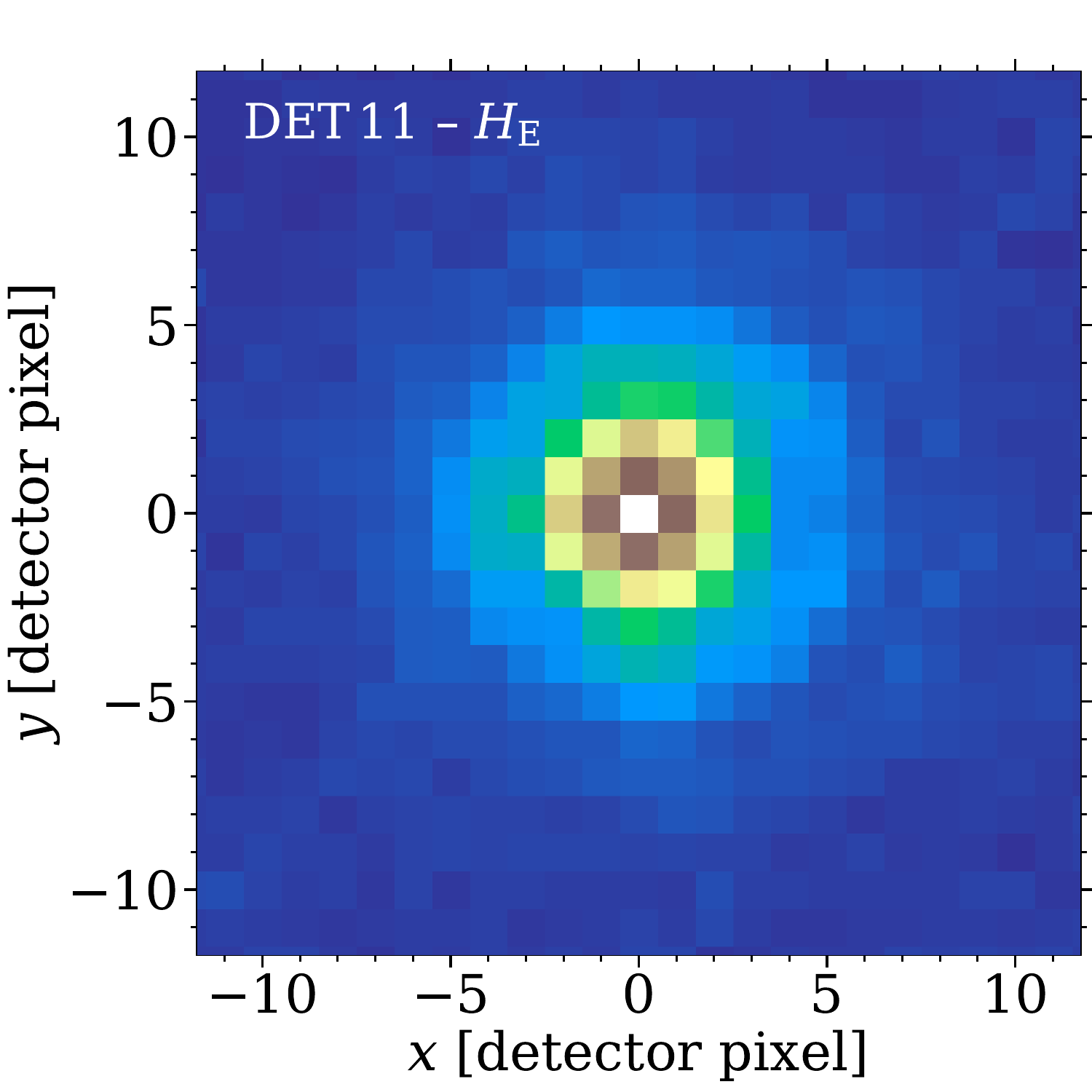}}
\end{minipage} \\
\begin{minipage}[t]{\textwidth}
\hspace*{28pt}
\resizebox{0.94\hsize}{!}
{\includegraphics{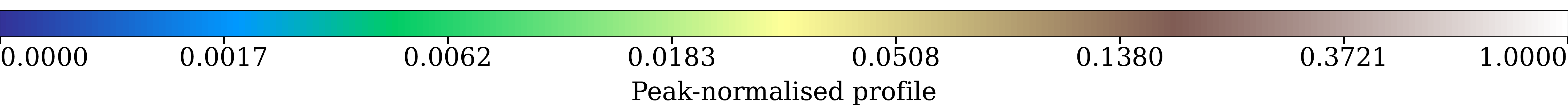}}
\end{minipage}
\caption{Examples of \mbox{NISP-P} extended PSF models (\emph{top panels}) and image stamps (\emph{bottom panels}) for single-dither exposures of the SCF (Observation IDs 519 and 6185). The number of sources used in the reconstruction of the PSF profile is 3131 for \YE, 2936 for \JE, and 2991 for \HE, and the number of basis vectors, $N_{\rm max}$, are listed in Table~\ref{table:psfex_tuning}. The image stamps are a median of 47, 68, and 182 point-like sources for \YE, \JE, and \HE passbands, respectively. The sampling of the PSF models and image stamps are \ang{;;0.05} and \ang{;;0.3}, respectively.}
\label{figure:psf_models_vignets}
\vspace{10pt}
\centering
\begin{minipage}[t]{0.33\textwidth}
\centering
\resizebox{\hsize}{!}{\includegraphics{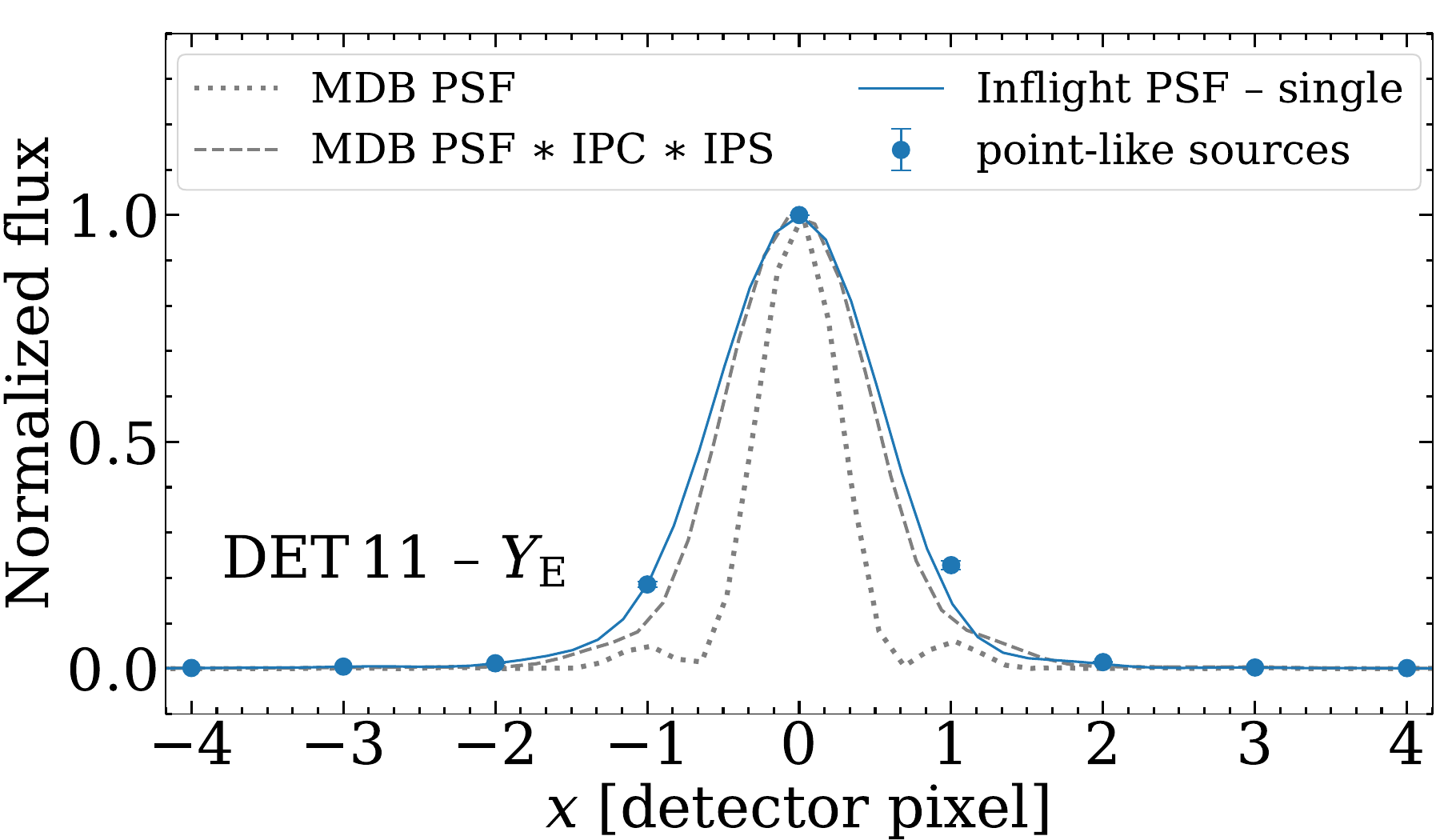}}
\end{minipage}
\begin{minipage}[t]{0.33\textwidth}
\centering
\resizebox{\hsize}{!}{\includegraphics{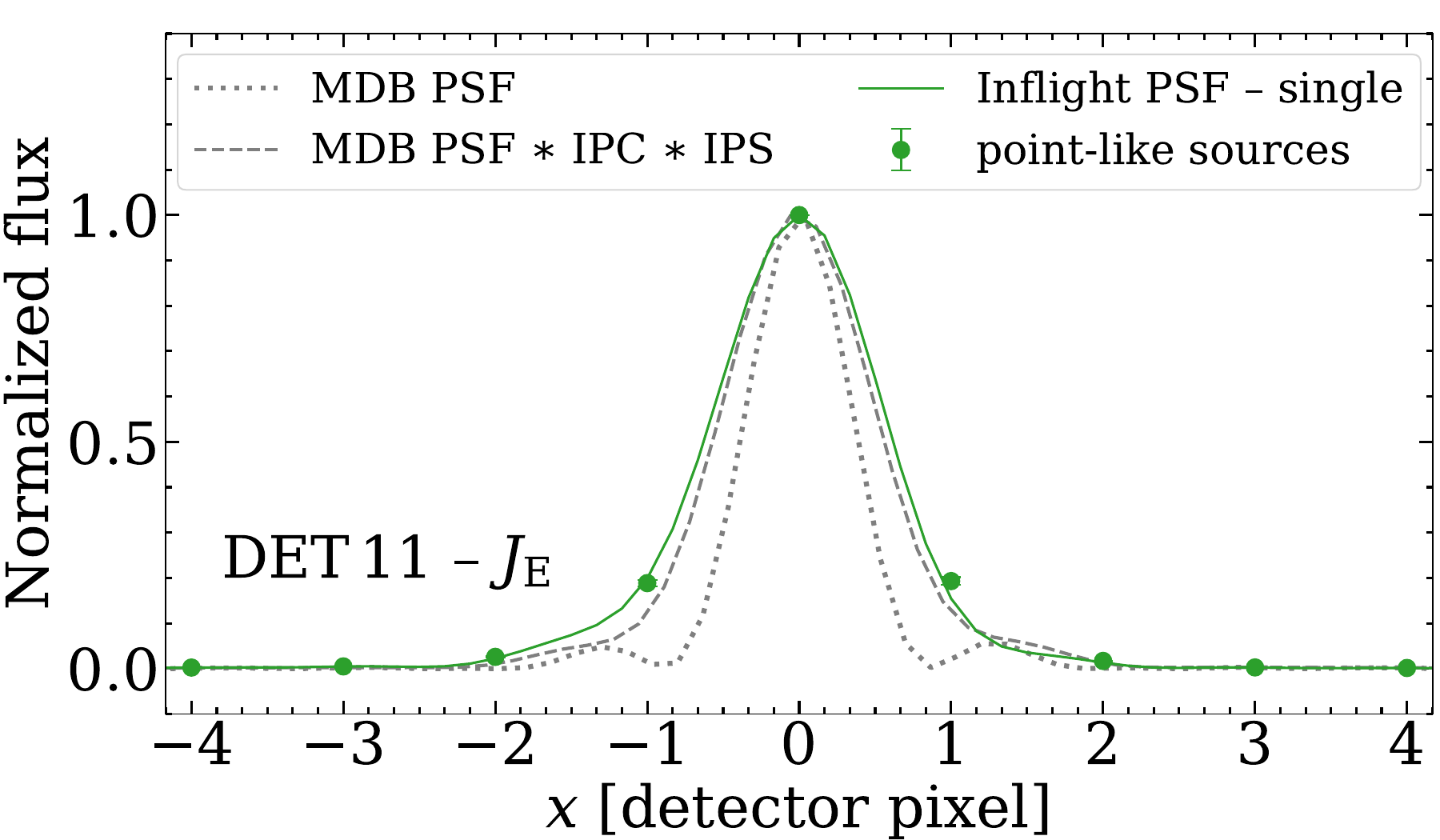}}
\end{minipage}
\begin{minipage}[t]{0.33\textwidth}
\centering
\resizebox{\hsize}{!}{\includegraphics{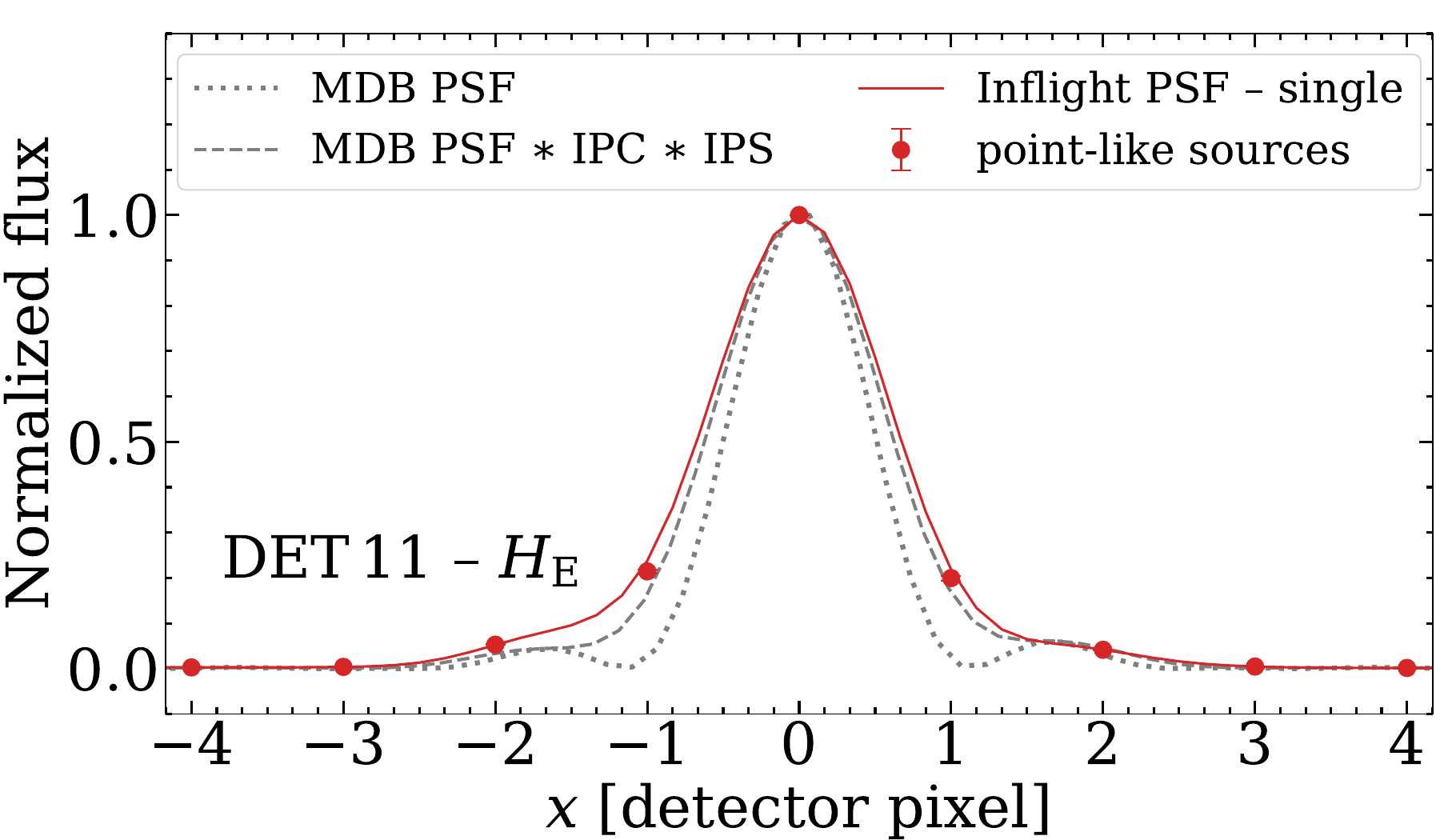}}
\end{minipage} \\
\begin{minipage}[t]{0.33\textwidth}
\centering
\resizebox{\hsize}{!}{\includegraphics{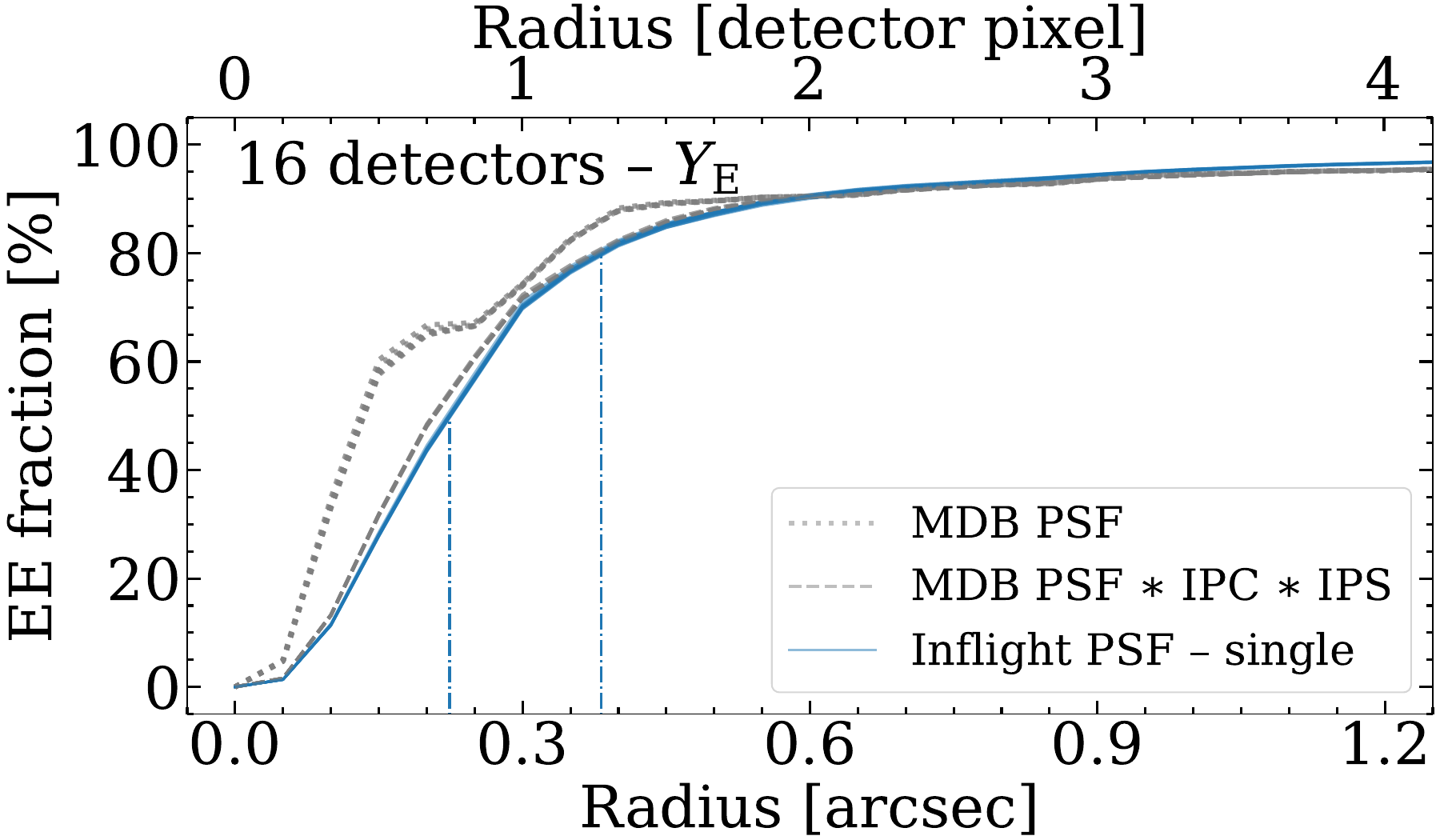}}
\end{minipage}
\begin{minipage}[t]{0.33\textwidth}
\centering
\resizebox{\hsize}{!}{\includegraphics{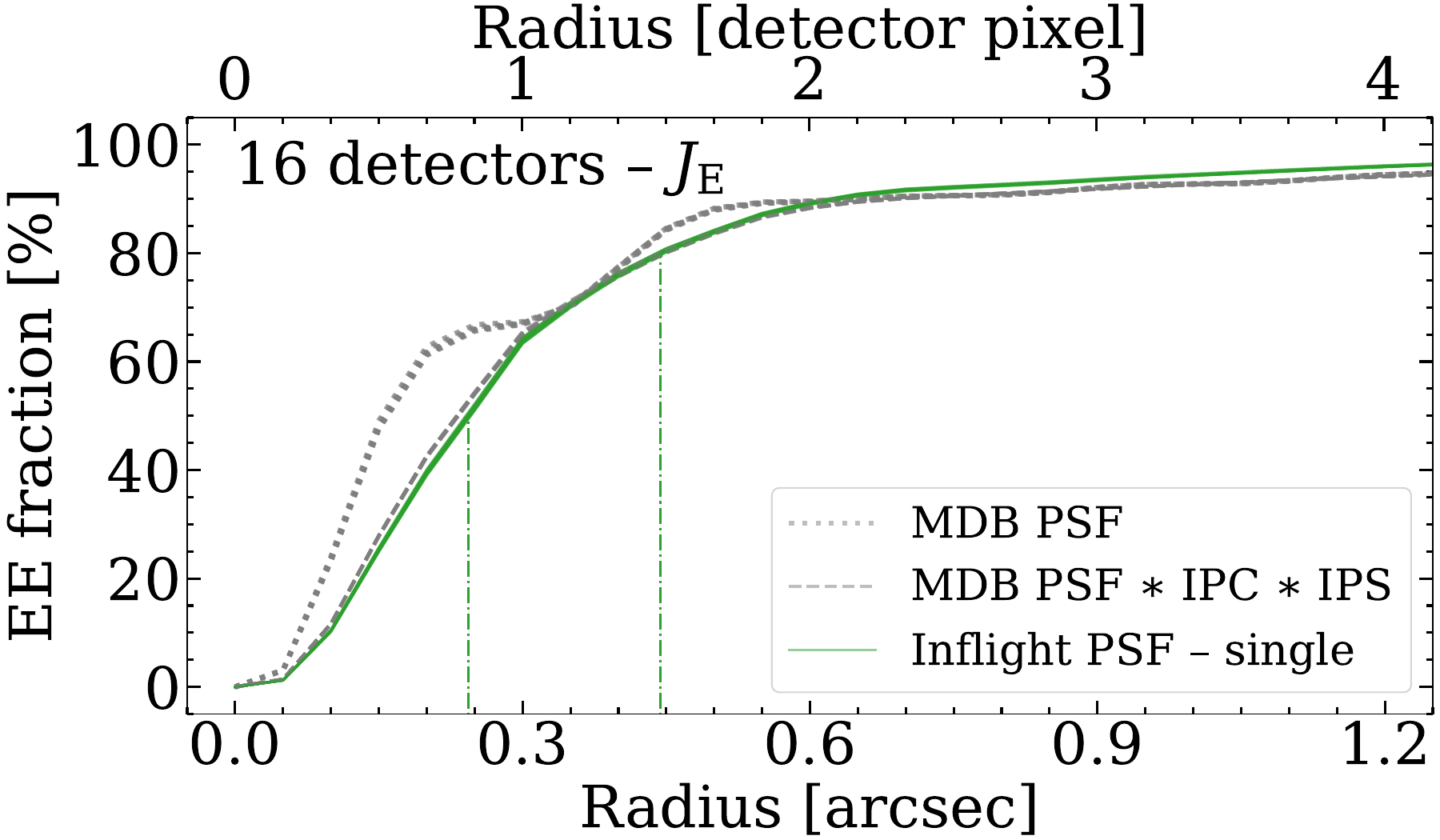}}
\end{minipage}
\begin{minipage}[t]{0.33\textwidth}
\centering
\resizebox{\hsize}{!}{\includegraphics{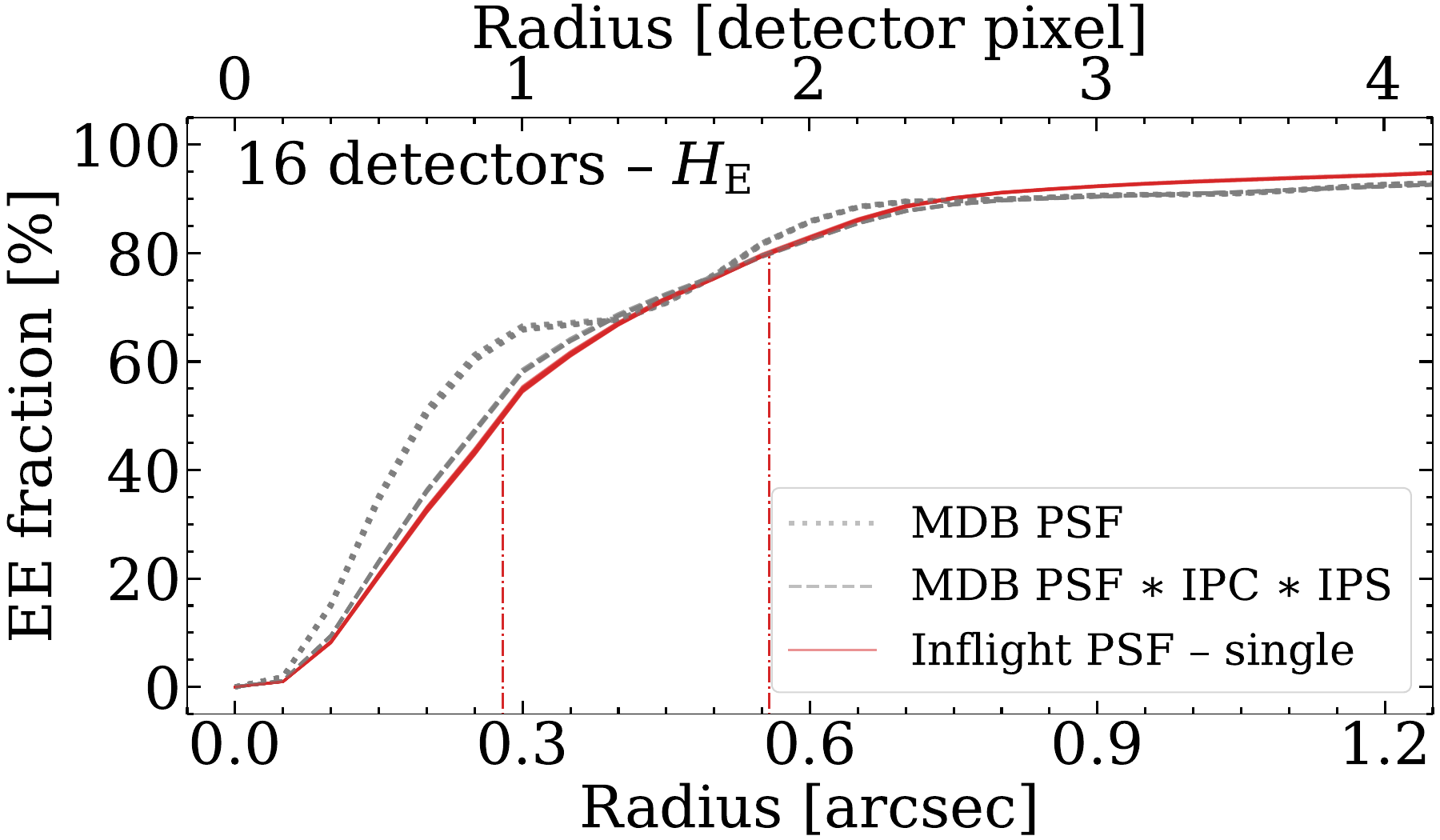}}
\end{minipage}
\caption{One-dimensional representations of the PSF models and image stamps shown in Fig.~\ref{figure:psf_models_vignets}. {\it Top panels}: $x$-axis cross sections comparing the PSF profiles for Detector ID {\tt 11} derived in MDB and Inflight modes with the median profile of point-like sources. {\it Bottom panels}: fraction of encircled energy as a function of the distance to the peak of the PSF profiles for the 16 NISP detectors (represented by independent curves). The vertical dashed-dotted lines indicate the mean radii encircling 50\% (rEE50) and 80\% (rEE80) of the total flux contained in each model. Both the original library of MDB PSF models (dotted lines) and those convolved with the IPC and IPS kernels (dashed lines) are shown for comparison.}
\label{figure:psf_profile_ee}
\end{figure*}

\begin{figure*}[p]
\centering
\begin{minipage}[t]{0.33\textwidth}
\centering
\resizebox{\hsize}{!}{\includegraphics{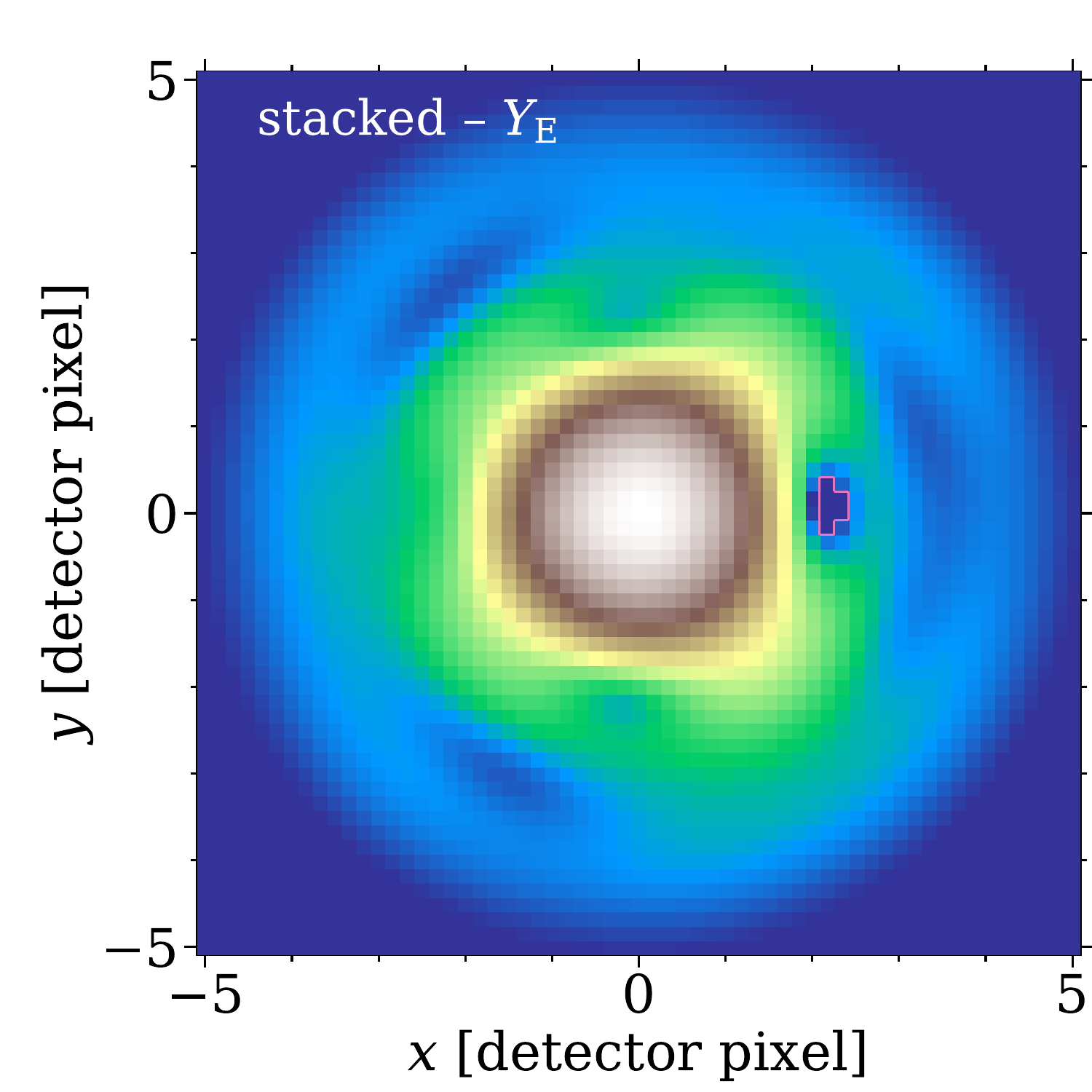}}
\end{minipage}
\begin{minipage}[t]{0.33\textwidth}
\centering
\resizebox{\hsize}{!}{\includegraphics{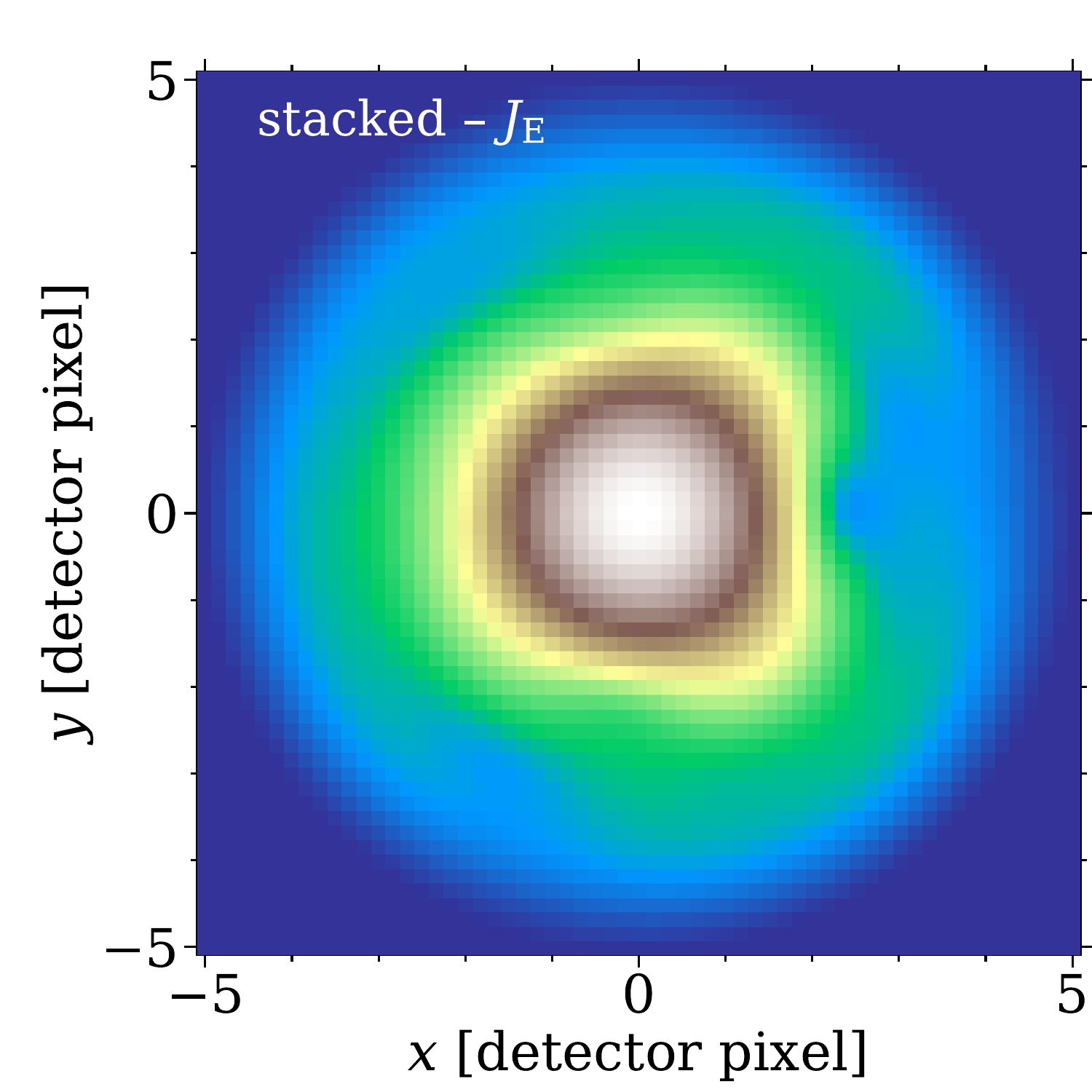}}
\end{minipage}
\begin{minipage}[t]{0.33\textwidth}
\centering
\resizebox{\hsize}{!}{\includegraphics{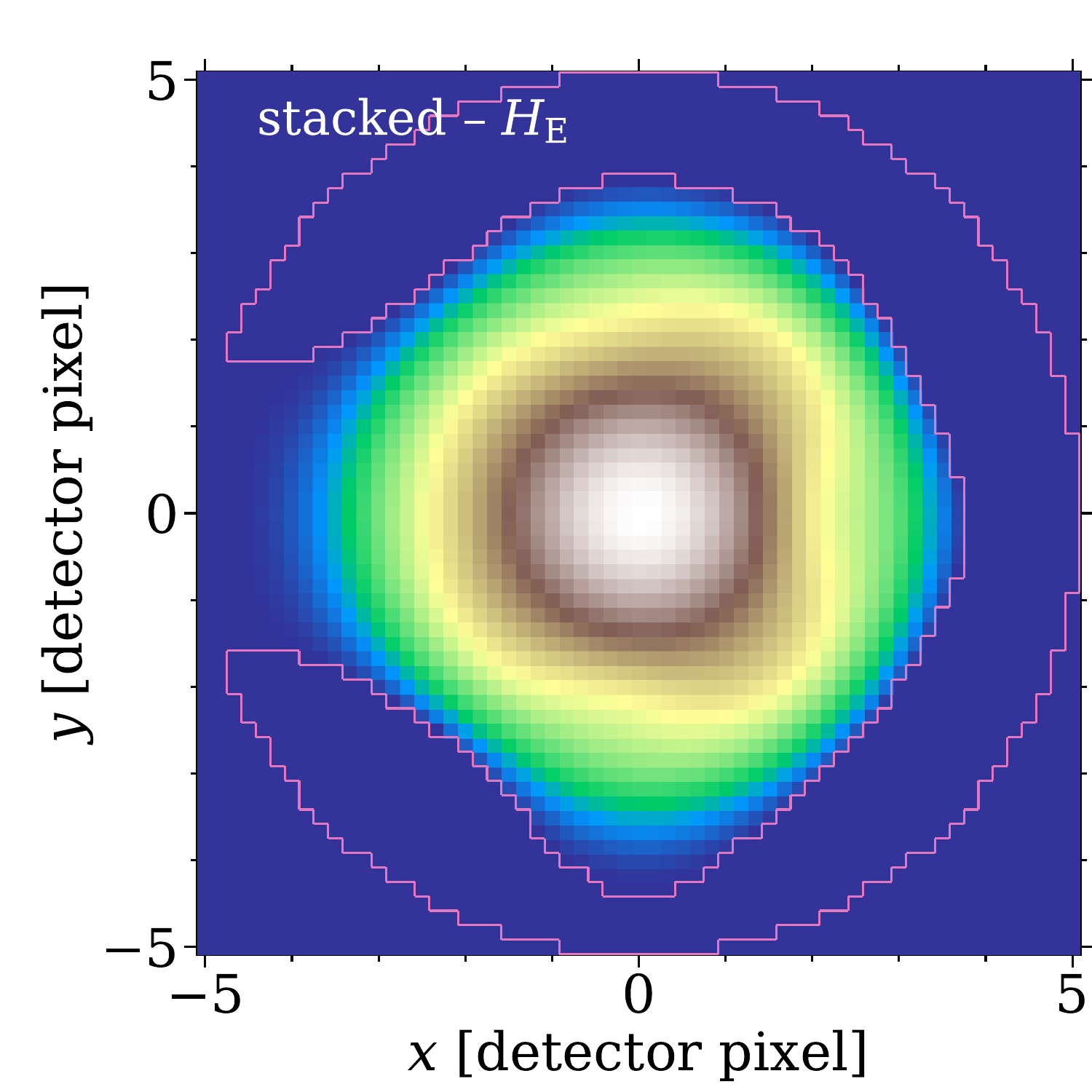}}
\end{minipage} \\
\begin{minipage}[t]{0.33\textwidth}
\centering
\resizebox{\hsize}{!}{\includegraphics{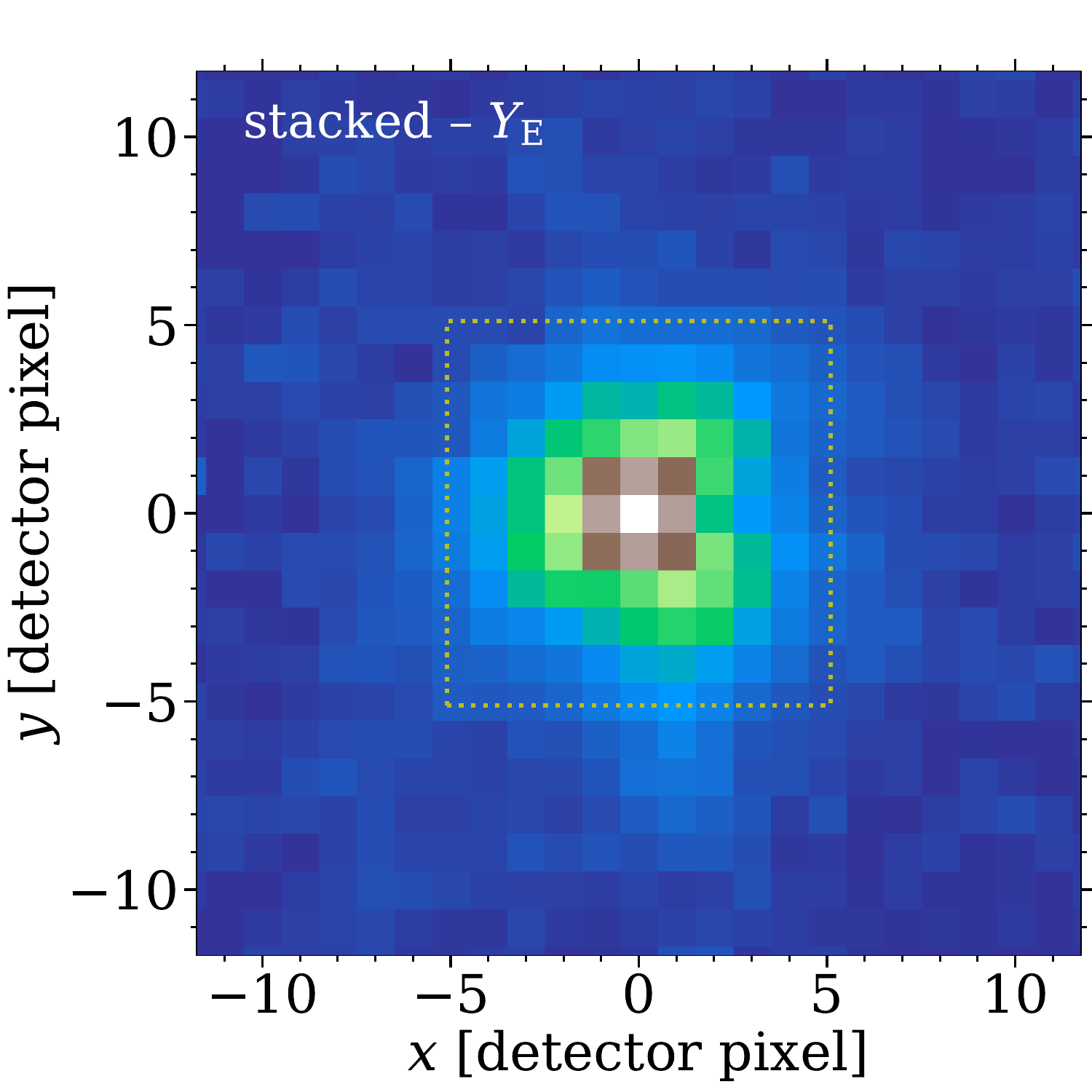}}
\end{minipage}
\begin{minipage}[t]{0.33\textwidth}
\centering
\resizebox{\hsize}{!}{\includegraphics{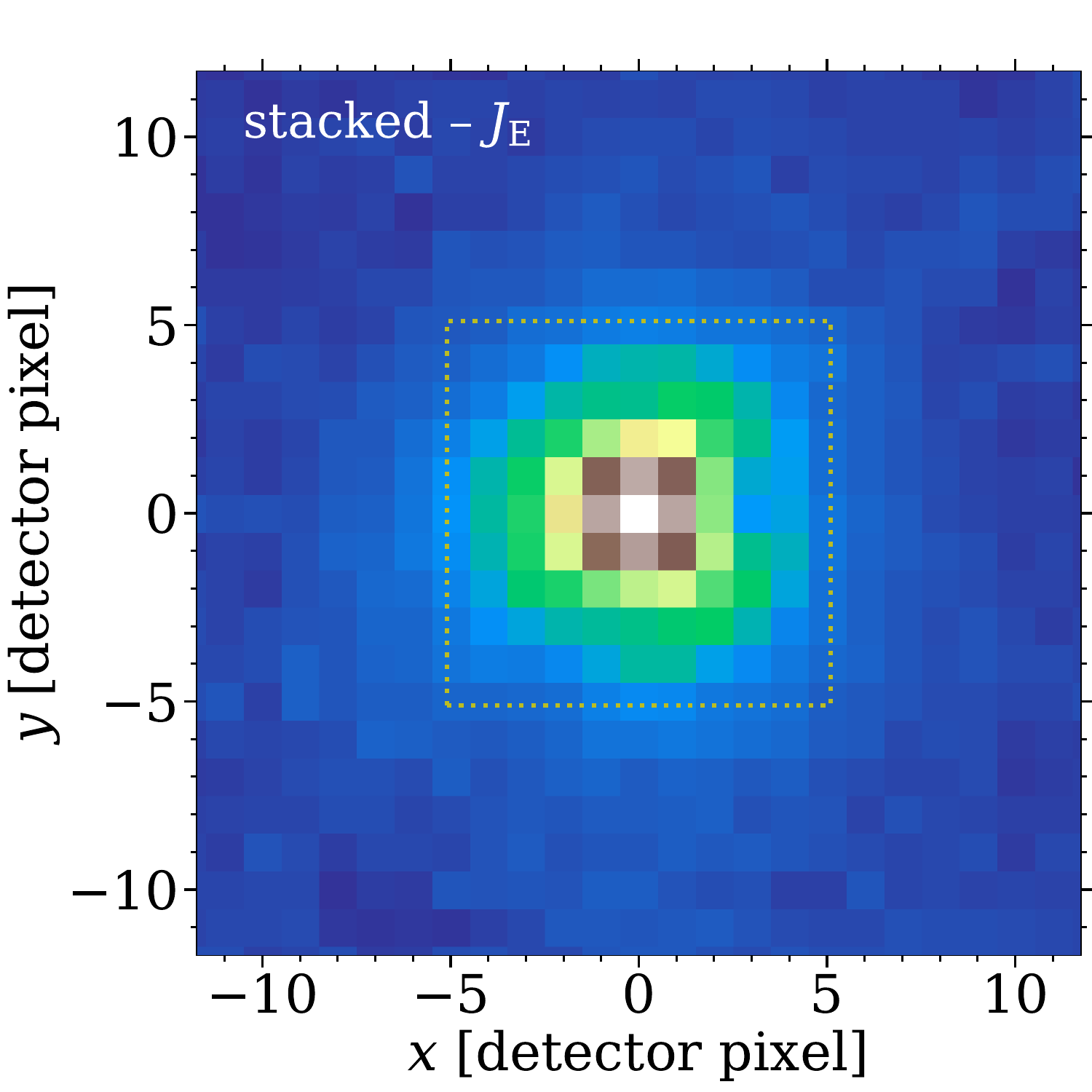}}
\end{minipage}
\begin{minipage}[t]{0.33\textwidth}
\centering
\resizebox{\hsize}{!}{\includegraphics{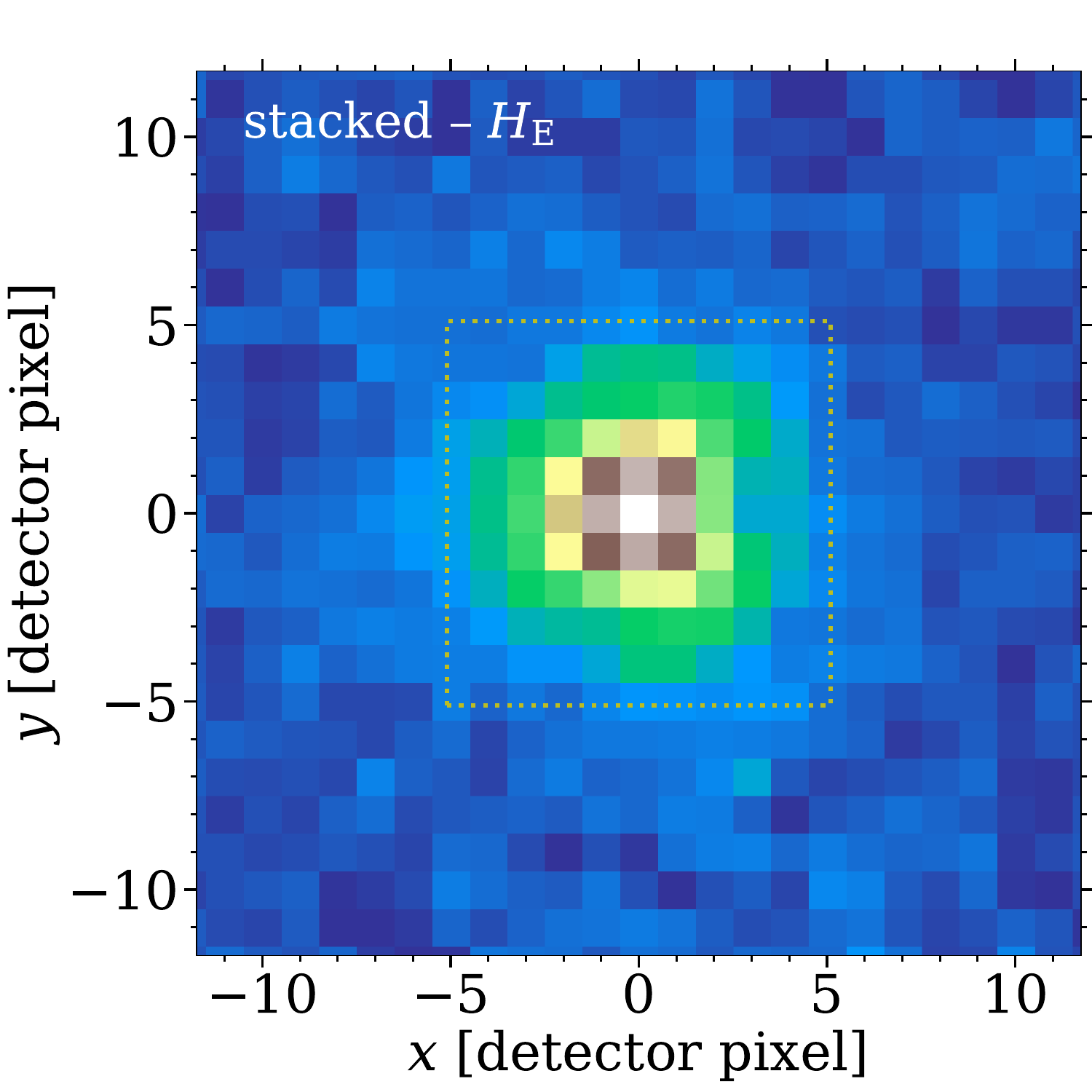}}
\end{minipage} \\
\begin{minipage}[t]{\textwidth}
\hspace*{28pt}
\resizebox{0.94\hsize}{!}
{\includegraphics{figures/psf_model_image_vignet_colorbar.pdf}}
\end{minipage}
\caption{Same as in Fig.~\ref{figure:psf_models_vignets}, but showing examples of \mbox{NISP-P} PSF models (\emph{top panels}) and image stamps (\emph{bottom panels}) for stacked frames of the EWS (Observation ID 2955). The number of sources used in the reconstruction of the PSF profile is 810, 863, and 614 while the image stamps are a median of 151, 136, and 116 point-like sources for \YE, \JE, and \HE passbands, respectively. The yellow dotted lines indicate the size of stacked PSF models shown the in top panels (61/6$\times$61/6 detector pixels) compared with those derived by the PSF calibration pipeline from single exposures (141/6$\times$141/6 detector pixels). The magenta solid lines encircle pixels for which the PSF models have slightly negative values ($>$\,$-$0.0015).}
\label{figure:psf_models_vignets_stacked}
\vspace{10pt}
\centering
\begin{minipage}[t]{0.33\textwidth}
\centering
\resizebox{\hsize}{!}{\includegraphics{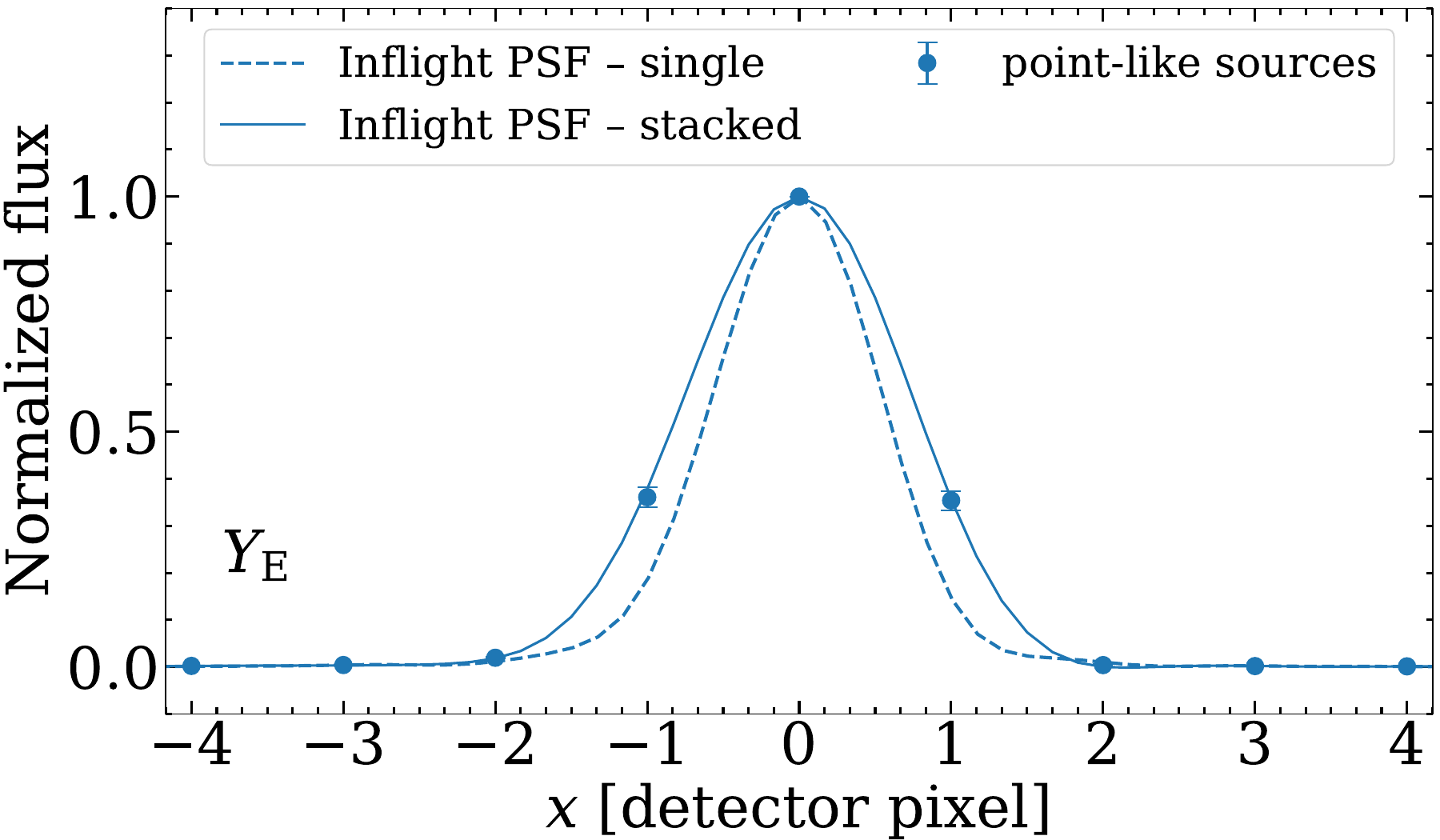}}
\end{minipage}
\begin{minipage}[t]{0.33\textwidth}
\centering
\resizebox{\hsize}{!}{\includegraphics{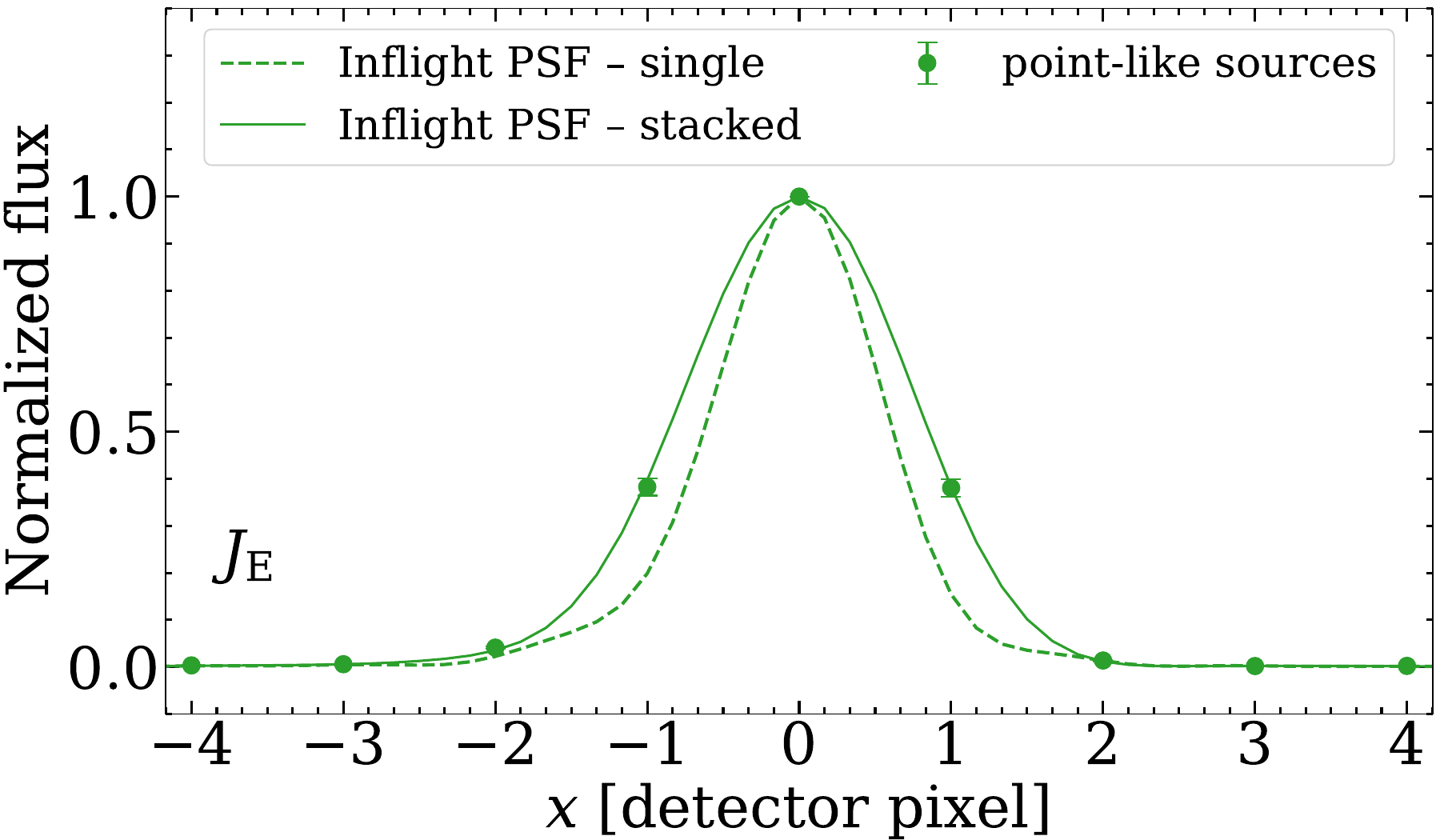}}
\end{minipage}
\begin{minipage}[t]{0.33\textwidth}
\centering
\resizebox{\hsize}{!}{\includegraphics{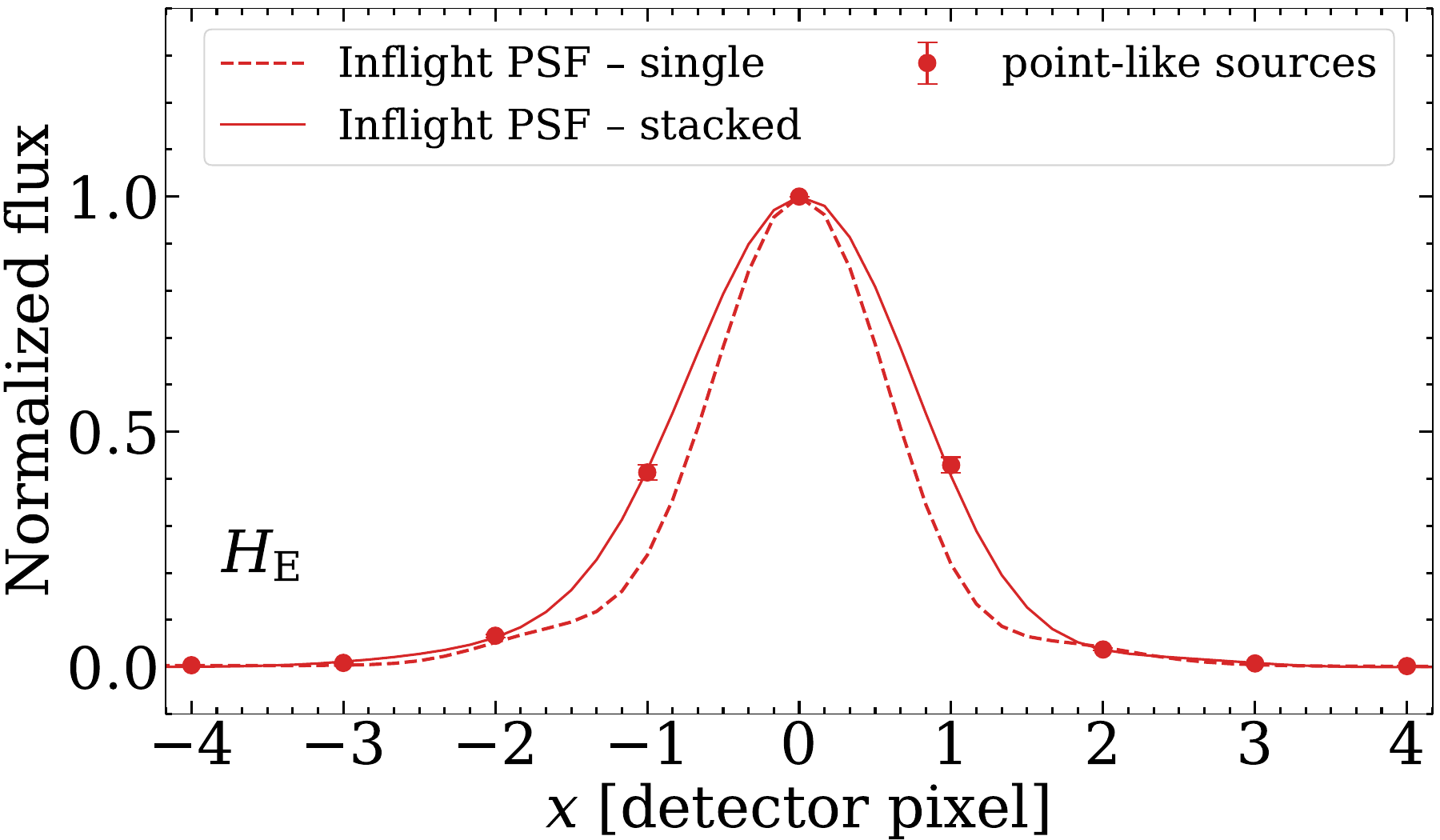}}
\end{minipage} \\
\begin{minipage}[t]{0.33\textwidth}
\centering
\resizebox{\hsize}{!}{\includegraphics{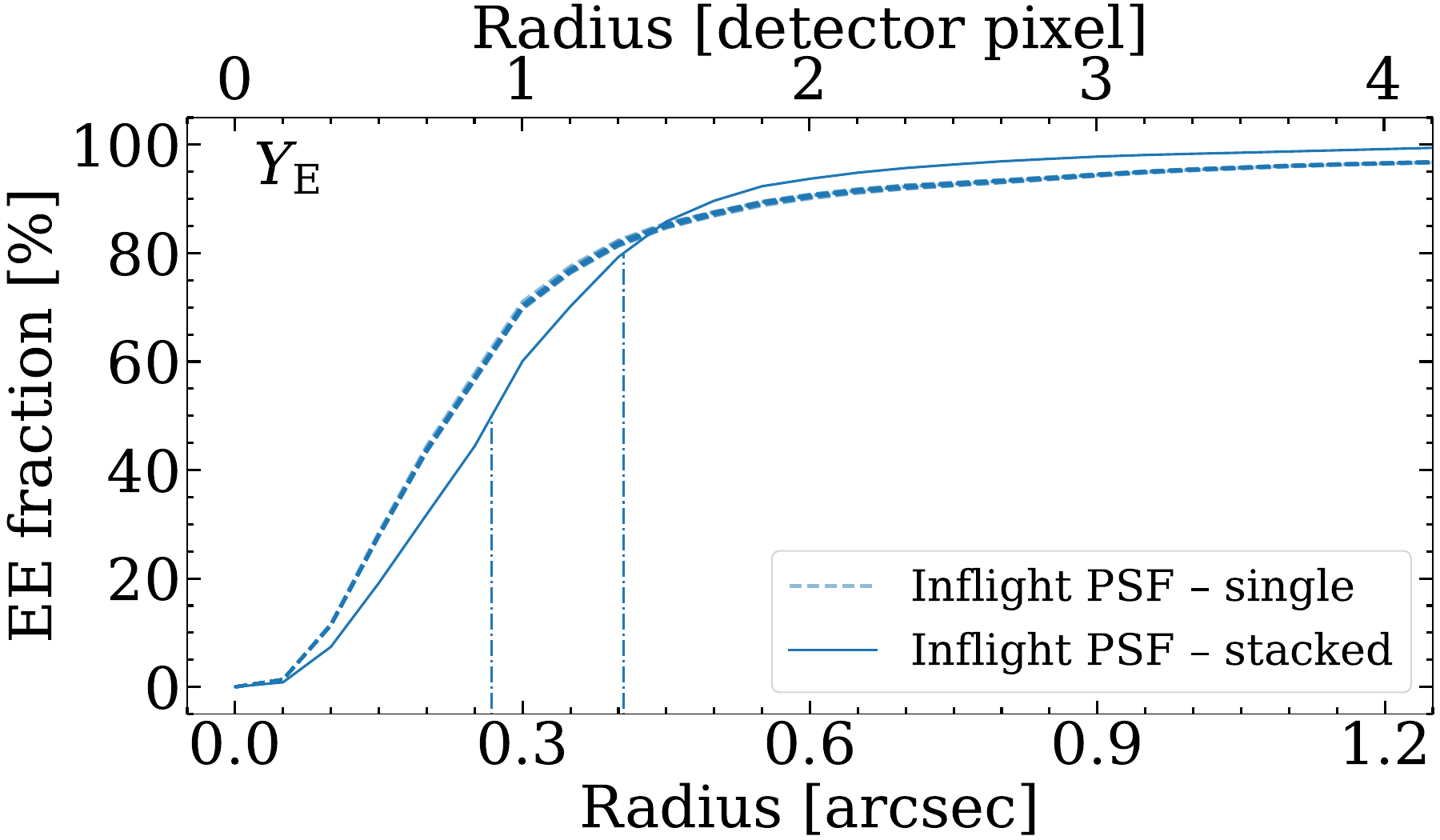}}
\end{minipage}
\begin{minipage}[t]{0.33\textwidth}
\centering
\resizebox{\hsize}{!}{\includegraphics{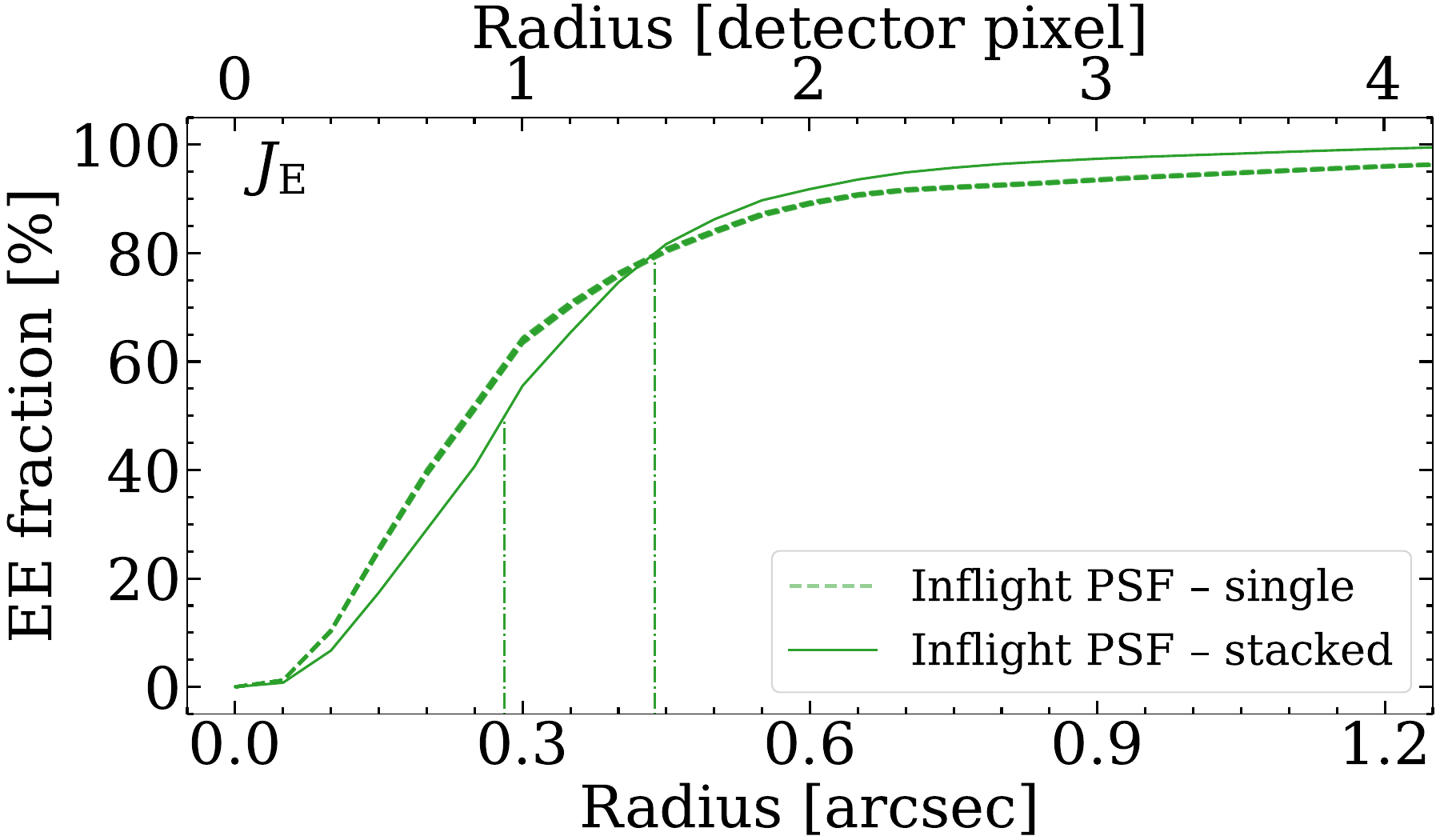}}
\end{minipage}
\begin{minipage}[t]{0.33\textwidth}
\centering
\resizebox{\hsize}{!}{\includegraphics{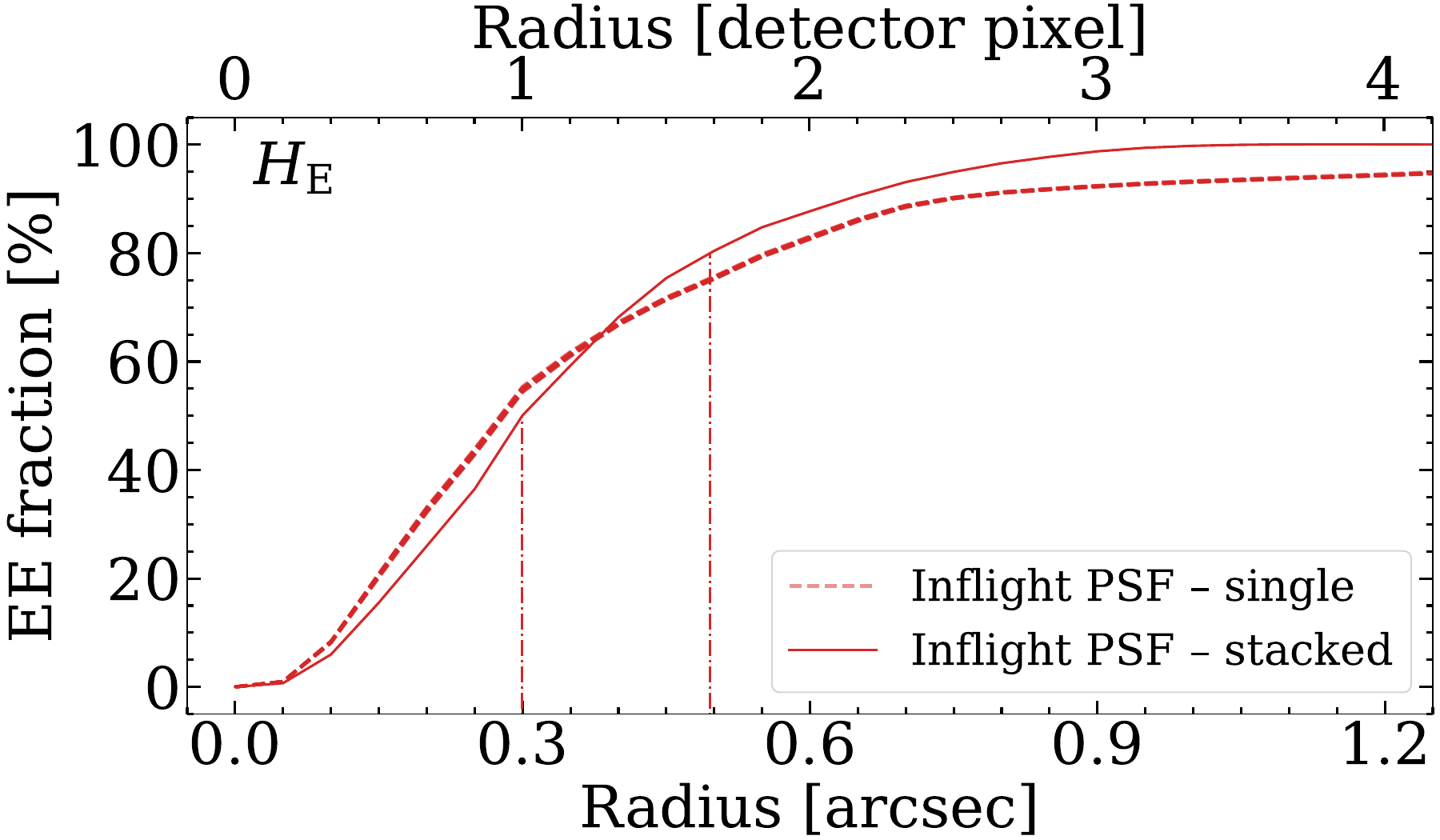}}
\end{minipage}
\caption{Same as in Fig.~\ref{figure:psf_profile_ee}, but plotting one-dimensional representations of the PSF models and image stamps shown in Fig.~\ref{figure:psf_models_vignets_stacked} for the stacked-frame case. The Inflight PSF models shown in Fig.~\ref{figure:psf_profile_ee} and derived from single-dither exposures are overplotted for comparison. The top panels refer to the PSF models for Detector ID {\tt 11} while the bottom panels illustrate the encircled energy curves for the 16 NISP detectors.}
\label{figure:psf_profile_ee_stacked}
\end{figure*}

\section{\label{sec:results} Results and discussion}

\subsection{\label{sec:psf_models} Derived PSF models}

The top panels of Fig.~\ref{figure:psf_models_vignets} show examples of extended PSF models derived on the basis of single-dither exposures from SCF data observed in the \mbox{NISP-P} \YE, \JE, and \HE passbands. As discussed in the Appendix~\ref{appendix:psfex_tuning}, these models are data products of the PSF calibration pipeline and are calculated using a large number of sources. The PSF models clearly reproduce typical diffraction patterns (rings and spikes) commonly seen in astronomical images, especially in the profile of bright stars. Moreover, some of the primary optical aberrations, such as the coma (asymmetric tail) and the trefoil (three-lobed footprint), are also observed. For non-saturated stars, these patterns are so weak that they are almost indistinguishable from background noise. However, by combining the stamp of several stars, weak features become clearly discernible, revealing structures that are undetectable in one single source. The bottom panels of the same figure show, for each passband, an image stamp calculated as the median profile of point-like sources. As expected, the data exhibit the identical structural signatures seen in the PSF models. Point-like sources were selected according to the FWHM ranges and the minimum signal-to-noise ratio (S/N) listed in Table~\ref{table:psfex_tuning}, and also considering sources with ellipticity $<$\,0.05. Moreover, only sources with centroid position close to the centre of the pixel (fraction of the $x$- and $y$-axes positions $<$\,0.1~pixel) were used to minimise sub-pixel spatial variations.

The top panels of Fig.~\ref{figure:psf_profile_ee} show the $x$-axis cross section of the same PSF models and image stamps shown in Fig.~\ref{figure:psf_models_vignets}. For comparison, the cross section of the original MDB models (from the library of PSF models) and the models derived in MDB mode (original models convolved with the IPC and IPS kernels) are also plotted. Although the profiles of the convolved MDB models are affected by pixelisation, IPC, and IPS, these effects alone (as currently defined) are not enough to accurately reproduce the profile of the PSF models or the image stamps. In the NIR pipelines, the IPC and IPS are represented by simplified kernels, kept constant across the FoV. However, both exhibit distinct spatial variations and dynamically scale with underlying physical parameters (temperature, signal strength, wavelength, or illumination angle), particularly in near-infrared HgCdTe and CMOS detectors \citep{Donlonetal2017,LeGraetetal2022,Zhongetal2021}. Further investigation is surely needed, also to quantify additional sources of broadening, such as the BFE \citep{Plazasetal2018}, that might affect the shape of the derived PSF models.

For a complete overview, Fig.~\ref{figure:psf_models} shows two-dimensional representations of the PSF models derived for each detector in each \mbox{NISP-P} passband. For clarity of visualisation, these are cropped portions (51/6$\times$51/6 detector pixels) of the full-frame PSF models, which are shown in Fig.~\ref{figure:psf_models_vignets} for Detector ID {\tt 11}.

Examples of PSF models derived on the basis of stacked frames are shown in the top panels of Fig.~\ref{figure:psf_models_vignets_stacked}. These models are data products of the NIR scientific pipeline, executed using four single-dither exposures of EWS observations. The panels show the peak-normalised profiles from zero to one. However, some of the pixels have negative values, which are artefacts introduced by the resampling functions used by {\tt PSFEx} to create models with finer sampling. The negative values are $>$\,$-$0.0015 and $>$\,$-$0.0008 for the \YE and \HE bands, respectively, and their presence does not affect the use of these models in applications such as PSF-fitting photometry. The bottom panels of this figure show the image stamps calculated as the median profile of point-like sources detected in the stacked frame. These stamps are the same size (25$\times$25 detector pixels) as the PSF models and the image stamps shown in Fig.~\ref{figure:psf_models_vignets}. By contrast, stacked PSF models are currently created with a smaller size (61/6$\times$61/6 detector pixels), which is indicated by the square plotted over the image stamps. Based on the results obtained when using models produced by the PSF calibration pipeline, we conclude that it is better to use larger PSF sizes to allow more accurate modelling of the real profile.

The top panels of Fig.~\ref{figure:psf_profile_ee_stacked} compare the $x$-axis cross section of the stacked PSF models and image stamps with the models derived from single-dither exposures. The agreement between the model and the stamp profiles is excellent for the three \mbox{NISP-P} passbands. As expected, the PSF profiles derived from stacked frames are clearly wider than those from individual exposures.

\subsection{\label{sec:ee} Encircled energy curves}

Two of the requirements stated in Sect.~\ref{sec:requirem} are strictly related to the energy encircled within a given circle centred on the profile of point sources. Once the PSF models are derived, the fraction of encircled energy (EE) as a function of radius is obtained by an azimuthally averaged curve constructed from the peak to the borders of the PSF profile. The values of rEE50 and rEE80 are then calculated in the three \mbox{NISP-P} passbands for single and stacked frames, as illustrated in the bottom panels of Fig.~\ref{figure:psf_profile_ee} (single-dither case) and Fig.~\ref{figure:psf_profile_ee_stacked} (stacked case). Stacked frames, since they are a combination of multiple dithered exposures, are expected to have broader profiles compared to single exposures. This demonstrated is evidenced by both the PSF profiles and the encircled energy curves shown in these figures.

\begin{table*}[t!]
\caption{PSF parameters for the 16 NISP detectors estimated using the PSF models derived in the three \mbox{NISP-P} passbands from the single-dither exposure and stacked-frame cases.}
\centering
\smallskip
\label{table:psf_params}
\smallskip
\begin{tabular}{l@{\hskip 2.5em} ccc@{\hskip 2.5em} ccc@{\hskip 2.5em} ccc@{\hskip 2.5em} ccc}
\hline\hline\noalign{\vskip 2pt}
 & \multicolumn{3}{c@{\hskip 2.5em}}{rEE50 [arcsec]} & \multicolumn{3}{c@{\hskip 2.5em}}{rEE80 [arcsec]} & \multicolumn{3}{c@{\hskip 2.5em}}{FWHM [arcsec]} & \multicolumn{3}{c@{\hskip 2.5em}}{Aperture correction} \\
\noalign{\vskip 2pt}\hline\noalign{\vskip 2pt}
Det. ID & \YE & \JE & \HE & \YE & \JE & \HE & \YE & \JE & \HE & \YE & \JE & \HE \\
\noalign{\vskip 2pt}\hline\noalign{\vskip 2pt}
{\tt 11} & 0.224 & 0.246 & 0.281 & 0.378 & 0.444 & 0.557 & 0.355 & 0.358 & 0.372 & 1.059 & 1.069 & 1.081 \\
{\tt 21} & 0.225 & 0.245 & 0.281 & 0.382 & 0.443 & 0.560 & 0.355 & 0.356 & 0.373 & 1.057 & 1.070 & 1.082 \\
{\tt 31} & 0.224 & 0.242 & 0.278 & 0.382 & 0.441 & 0.558 & 0.353 & 0.352 & 0.367 & 1.057 & 1.070 & 1.084 \\
{\tt 41} & 0.224 & 0.245 & 0.281 & 0.382 & 0.445 & 0.561 & 0.354 & 0.356 & 0.374 & 1.058 & 1.070 & 1.084 \\
{\tt 12} & 0.225 & 0.246 & 0.280 & 0.383 & 0.448 & 0.556 & 0.355 & 0.354 & 0.370 & 1.059 & 1.070 & 1.083 \\
{\tt 22} & 0.225 & 0.243 & 0.280 & 0.386 & 0.441 & 0.559 & 0.355 & 0.355 & 0.368 & 1.060 & 1.070 & 1.082 \\
{\tt 32} & 0.224 & 0.241 & 0.278 & 0.386 & 0.440 & 0.560 & 0.351 & 0.352 & 0.365 & 1.061 & 1.070 & 1.084 \\
{\tt 42} & 0.220 & 0.241 & 0.276 & 0.375 & 0.441 & 0.555 & 0.349 & 0.350 & 0.364 & 1.058 & 1.069 & 1.083 \\
{\tt 13} & 0.224 & 0.244 & 0.277 & 0.384 & 0.445 & 0.552 & 0.353 & 0.353 & 0.366 & 1.061 & 1.067 & 1.083 \\
{\tt 23} & 0.225 & 0.245 & 0.279 & 0.383 & 0.445 & 0.554 & 0.355 & 0.356 & 0.369 & 1.059 & 1.070 & 1.083 \\
{\tt 33} & 0.225 & 0.243 & 0.278 & 0.387 & 0.446 & 0.559 & 0.352 & 0.352 & 0.366 & 1.060 & 1.071 & 1.085 \\
{\tt 43} & 0.222 & 0.242 & 0.277 & 0.381 & 0.444 & 0.555 & 0.351 & 0.351 & 0.365 & 1.058 & 1.070 & 1.083 \\
{\tt 14} & 0.224 & 0.242 & 0.277 & 0.385 & 0.440 & 0.554 & 0.351 & 0.353 & 0.365 & 1.062 & 1.068 & 1.081 \\
{\tt 24} & 0.224 & 0.245 & 0.281 & 0.382 & 0.447 & 0.558 & 0.354 & 0.355 & 0.373 & 1.058 & 1.072 & 1.084 \\
{\tt 34} & 0.223 & 0.246 & 0.280 & 0.379 & 0.448 & 0.557 & 0.352 & 0.355 & 0.373 & 1.059 & 1.070 & 1.084 \\
{\tt 44} & 0.223 & 0.243 & 0.279 & 0.383 & 0.445 & 0.560 & 0.352 & 0.352 & 0.368 & 1.059 & 1.069 & 1.084 \\
\noalign{\vskip 2pt}\hline\noalign{\vskip 2pt}
Stacked  & 0.268 & 0.281 & 0.300 & 0.405 & 0.438 & 0.496 & 0.465 & 0.488 & 0.511 & 1.023 & 1.027 & 1.014 \\
\hline
\end{tabular}
\tablefoot{The table lists the radii encircling 50\% and 80\% of the energy (cols. 2--7), the FWHM (cols. 8--10) calculated as an average of the values measured in the $x$ and $y$ directions, and the aperture corrections (cols. 11--13) estimated for a circular aperture with diameter of six detector pixels. The measurements are based on the PSF models shown in Figs.~\ref{figure:psf_models_vignets}, \ref{figure:psf_profile_ee}, and \ref{figure:psf_models} for the single exposure, and in Figs.~\ref{figure:psf_models_vignets_stacked} and \ref{figure:psf_profile_ee_stacked} for the stacked frame.}
\end{table*}

Table~\ref{table:psf_params} lists our estimate of rEE50 and rEE80 for the 16 NISP detectors and for the stacked frame shown as examples in Figs.~\ref{figure:psf_profile_ee} and \ref{figure:psf_profile_ee_stacked}. For different detectors and passbands in the single-dither exposure case, rEE50 ranges from \ang{;;0.22} to \ang{;;0.28} and rEE80 ranges from \ang{;;0.27} to \ang{;;0.30}. The encircled energy curves for different detectors are nearly coincident, clearly indicating that these variations are driven primarily by the passband, while remaining relatively insensitive to the choice of detector. For the stacked frame case, rEE50 and rEE80 range from \ang{;;0.38} to \ang{;;0.56} and from \ang{;;0.41} to \ang{;;0.50}, respectively, depending on the passband. All these values are well below the thresholds mentioned in Sect.~\ref{sec:requirem}, undoubtedly satisfying the first requirement.

As discussed in \citet{EuclidSkyNISP}, the encircled energy curves calculated from PSF profiles, as those derived in this work, are slightly wider than the PSF derived from the wavefront reconstruction. Again, this is an expected behaviour, considering the assumption of overly simplified optical models and the several broadening effects that affect the incoming light (pixelisation, IPC, IPS, pointing jitter, etc.) from just before reaching the telescope up to readout (we note that some simulations we performed revealed that the IPS also changes the PSF profile but makes it slightly narrow instead). The bottom panels of Fig.~\ref{figure:psf_profile_ee} compare the original library of MDB PSF models with the original profiles convolved with the IPC and IPS kernels, derived in MDB mode, and with models derived for single-dither exposures in Inflight mode.

The PSF models derived from stacked frames are still not optimised and require some improvements. In the bottom panel of Fig.~\ref{figure:psf_profile_ee_stacked}, the fraction of encircled energy reaches the level 100\% at a radius of about 1\arcsec\ for the \HE filter, but at a much larger radius for the \YE and \JE filters. This behaviour is unexpected, as seen in the real image stamps shown in the bottom panels of Fig.~\ref{figure:psf_models_vignets_stacked}. The first diffraction rings of the PSF profile are observed to be brighter for \HE than for \JE, and even fainter for \YE, an effect not reproduced by the models illustrated in the top panels of the same figure. A new tuning of the input parameters may help to improve the reproduction of the PSF diffraction patterns and to avoid unwanted artefacts such as the negative values discussed in the previous section.

\subsection{\label{sec:phot} PSF-fitting and aperture photometry}

Although the derived PSF models perform well within the mission requirements on the fractions of encircled energy, relying solely on these top-level metrics may hide fine details regarding the profile's exact shape. A more robust quality assessment involves evaluating how these models impact the photometric accuracy itself. This is often accomplished by fitting the derived PSF models to the profile of observed sources. The fitting of the PSF profile can be performed by {\tt SourceExtractor}, which receives as input a file containing the PSF models (one for each detector) and provides as output additional columns, namely the {\tt MAG\_PSF} parameter and the corresponding errors {\tt MAGERR\_PSF}.

During the development phase of the mission, when only simulated data were available, we compared the {\tt MAG\_PSF} photometry with the magnitude of the sources in a simulated catalogue used as reference. During true observations, this PSF-derived photometry must be validated against an independent photometric indicator. While cross-referencing with external {\it Gaia} photometry was considered, it introduced a significantly larger scatter. Consequently, we chose fixed-aperture photometry as our reference baseline because it provides a tighter, more reliable comparison for point-like sources. For this, we adopted the {\tt MAG\_APER} photometry also provided by {\tt SourceExtractor}.

\begin{figure*}
\centering
\begin{minipage}[t]{0.49\textwidth}
\centering
\resizebox{\hsize}{!}{\includegraphics{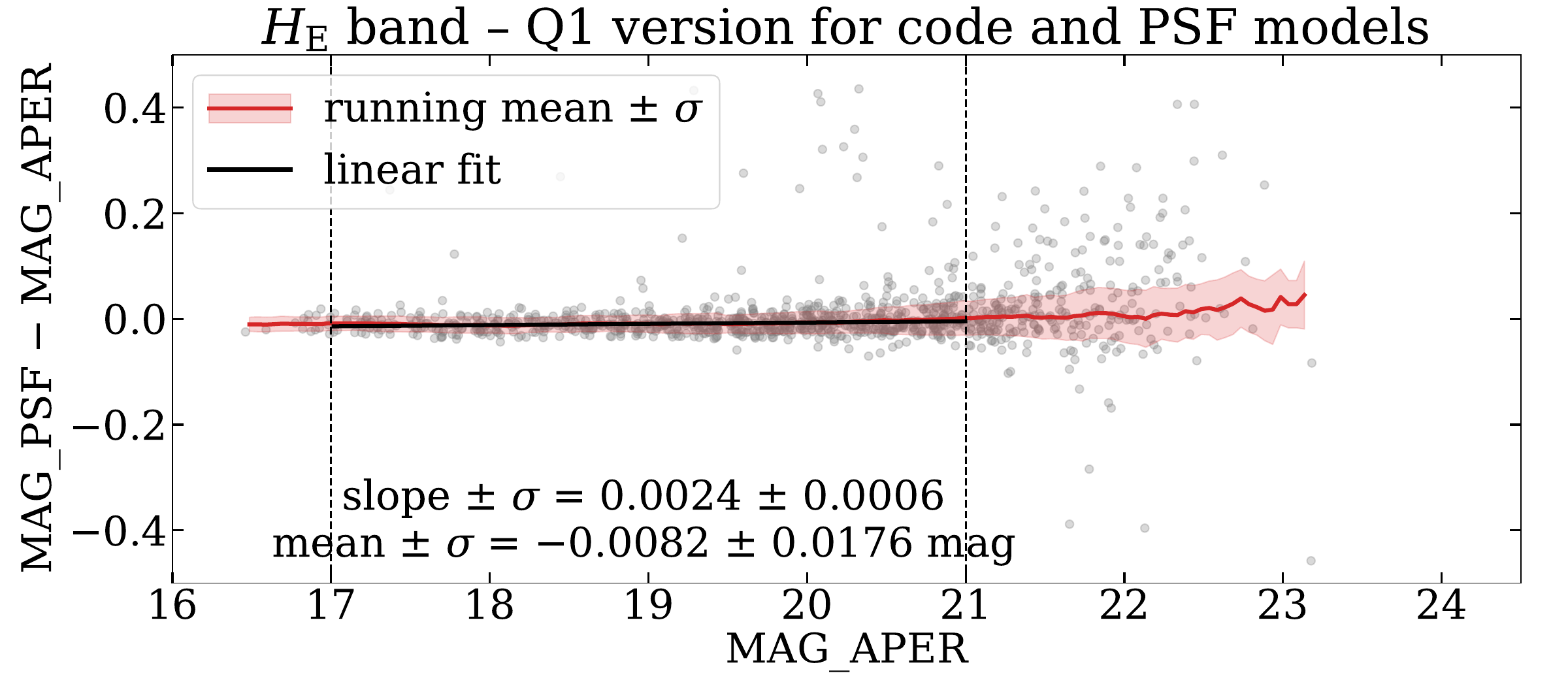}}
\end{minipage}
\begin{minipage}[t]{0.49\textwidth}
\centering
\resizebox{\hsize}{!}{\includegraphics{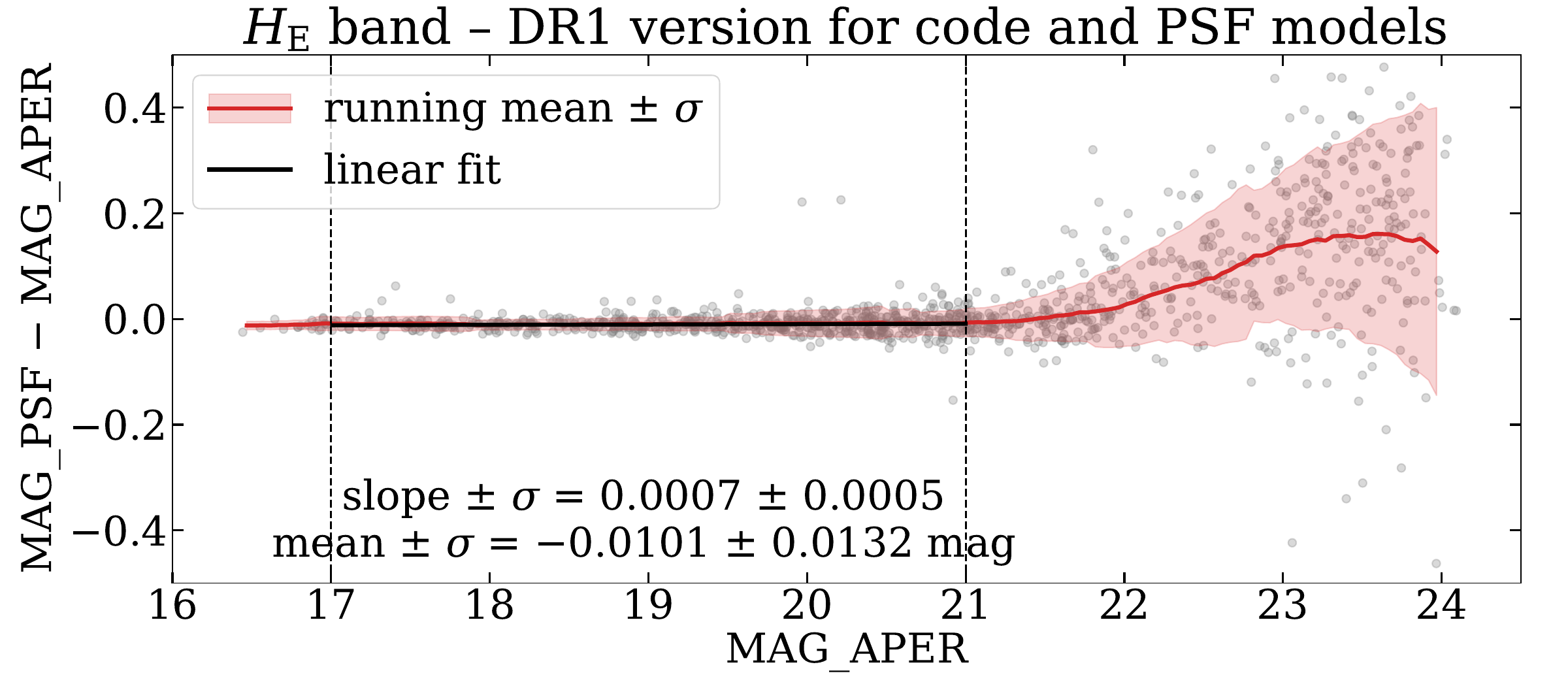}}
\end{minipage} \\
\begin{minipage}[t]{0.49\textwidth}
\centering
\resizebox{\hsize}{!}{\includegraphics{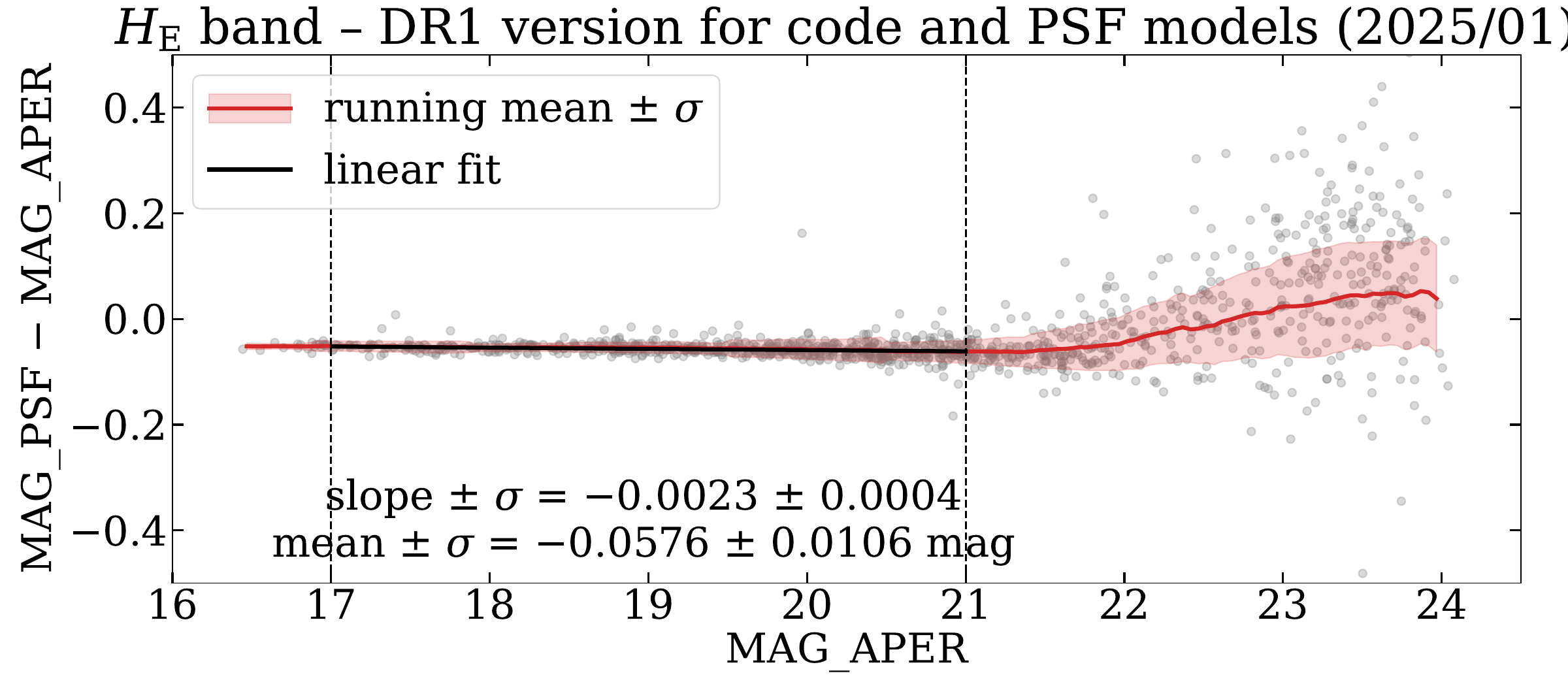}}
\end{minipage}
\begin{minipage}[t]{0.49\textwidth}
\centering
\resizebox{\hsize}{!}{\includegraphics{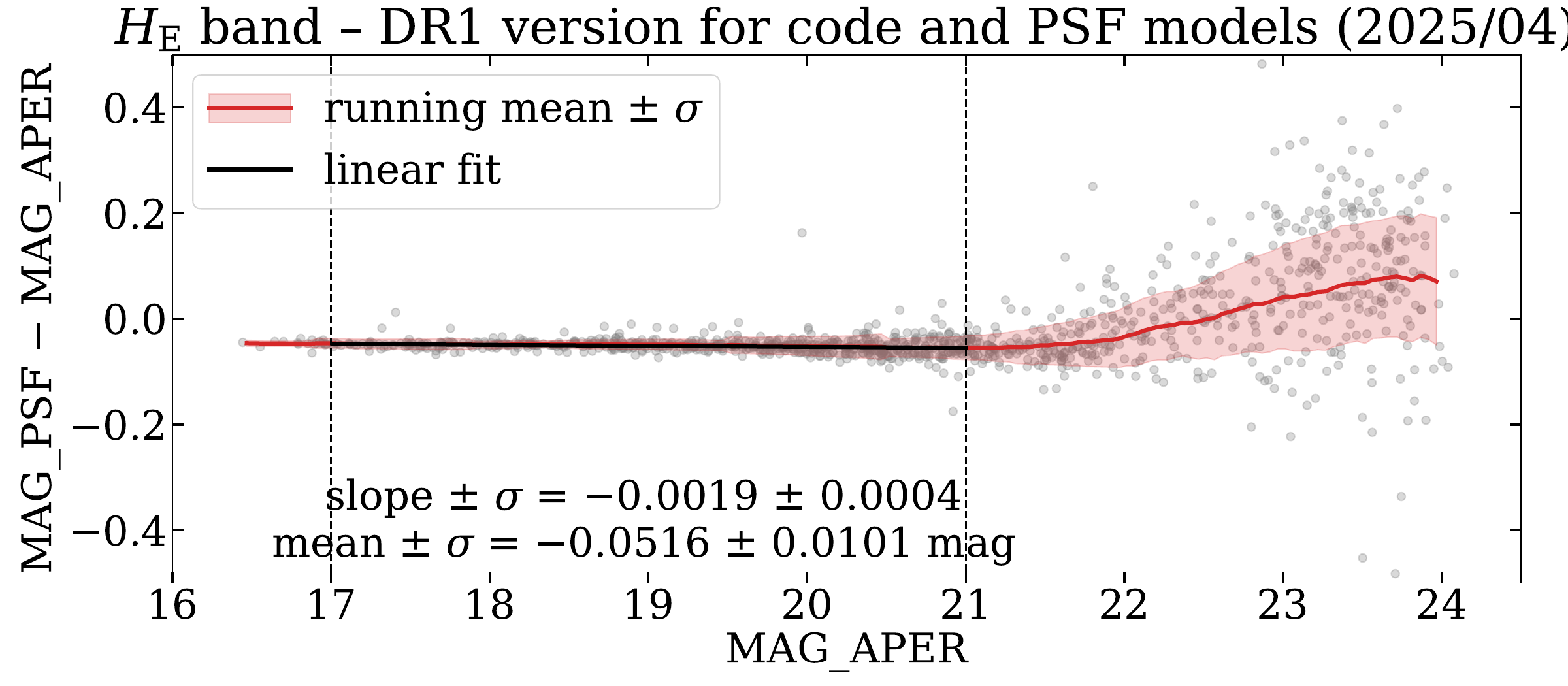}}
\end{minipage}
\caption{Comparison of PSF-fitting and aperture photometry for Q1 and DR1 versions of code and PSF models (using \HE band data from Observation ID 2955). {\it Top left panel}: results from Q1 release. {\it Top right panel}: same model as in the left panel, but with updated code. {\it Bottom panels}: updated code with extended PSF models derived from SCF data (Observation IDs 2174 and 519 processed in January and April 2025, respectively).}
\label{figure:mag_psf_mag_aper}
\vspace{10pt}
\centering
\begin{minipage}[t]{0.49\textwidth}
\centering
\resizebox{\hsize}{!}{\includegraphics{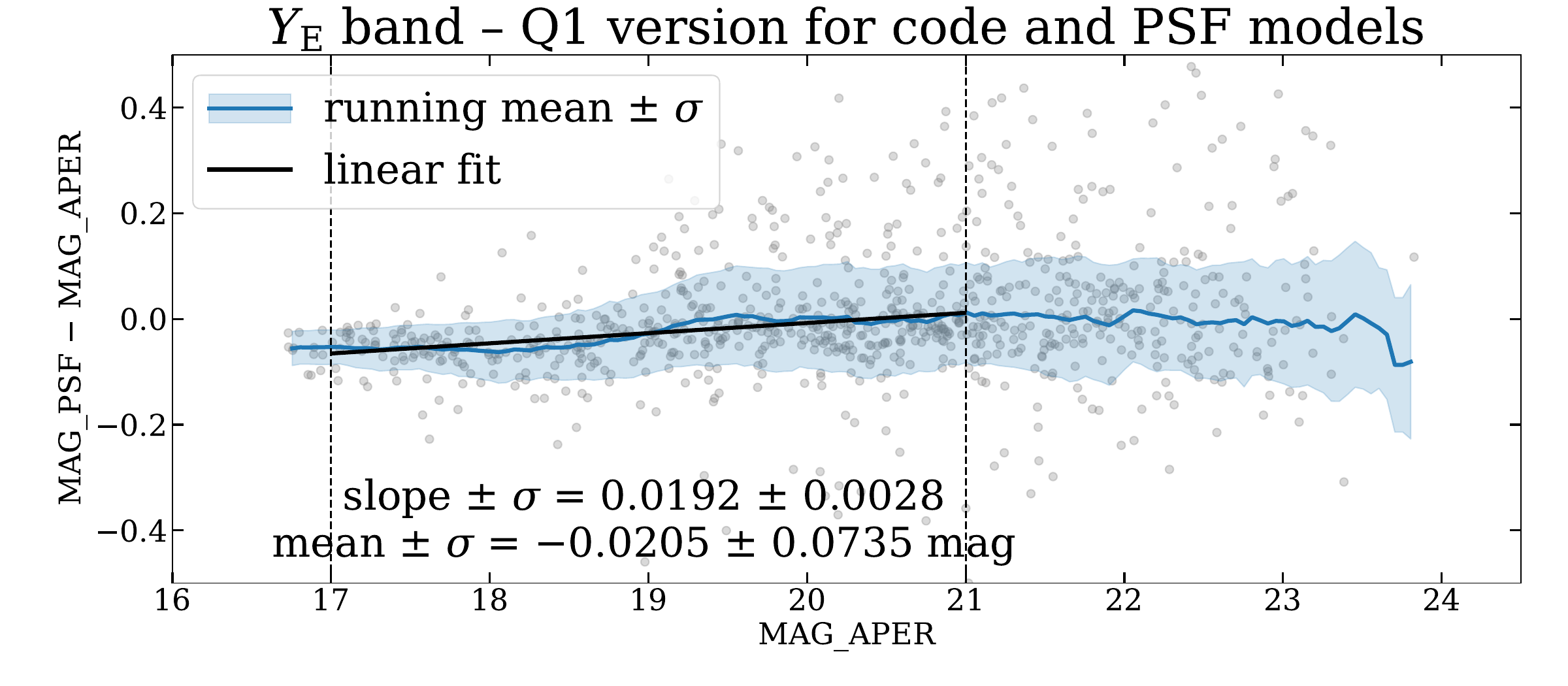}}
\end{minipage}
\begin{minipage}[t]{0.49\textwidth}
\centering
\resizebox{\hsize}{!}{\includegraphics{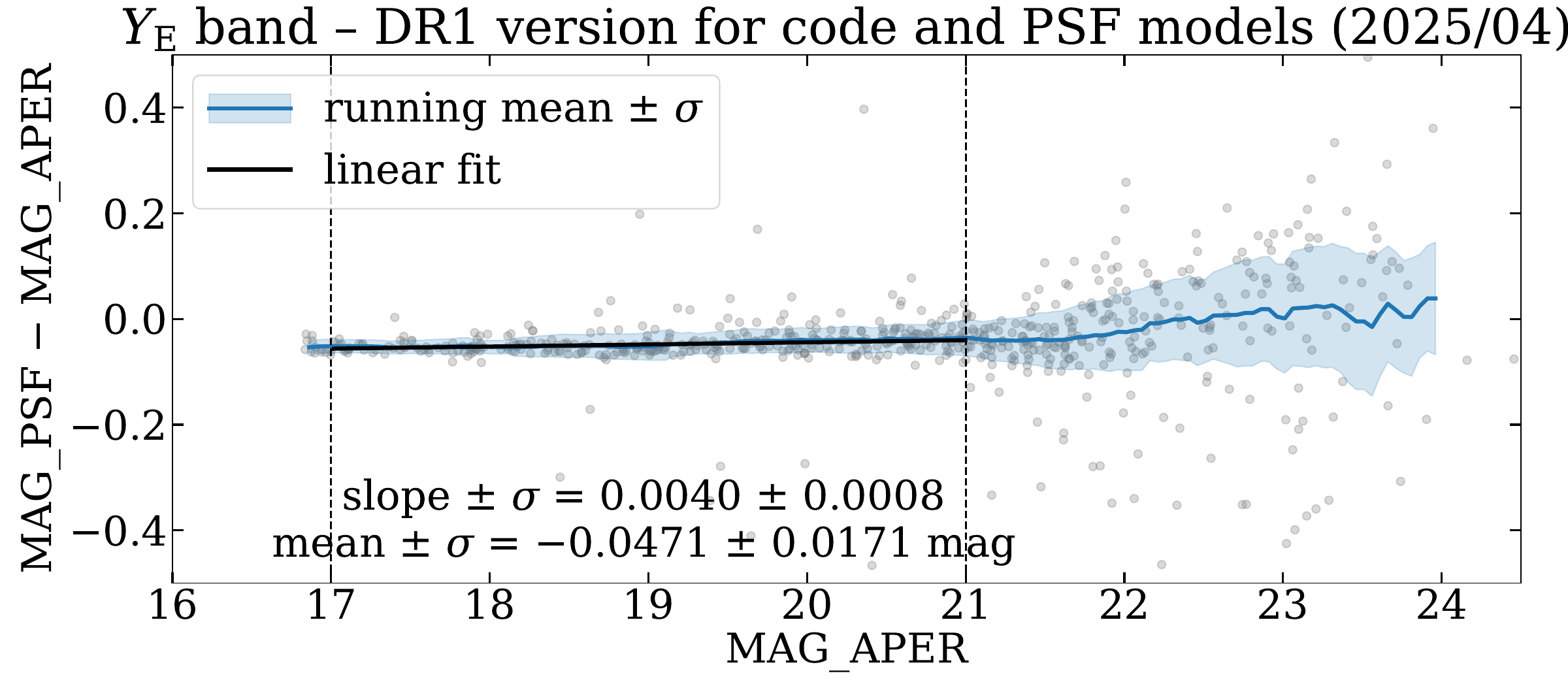}}
\end{minipage} \\
\begin{minipage}[t]{0.49\textwidth}
\centering
\resizebox{\hsize}{!}{\includegraphics{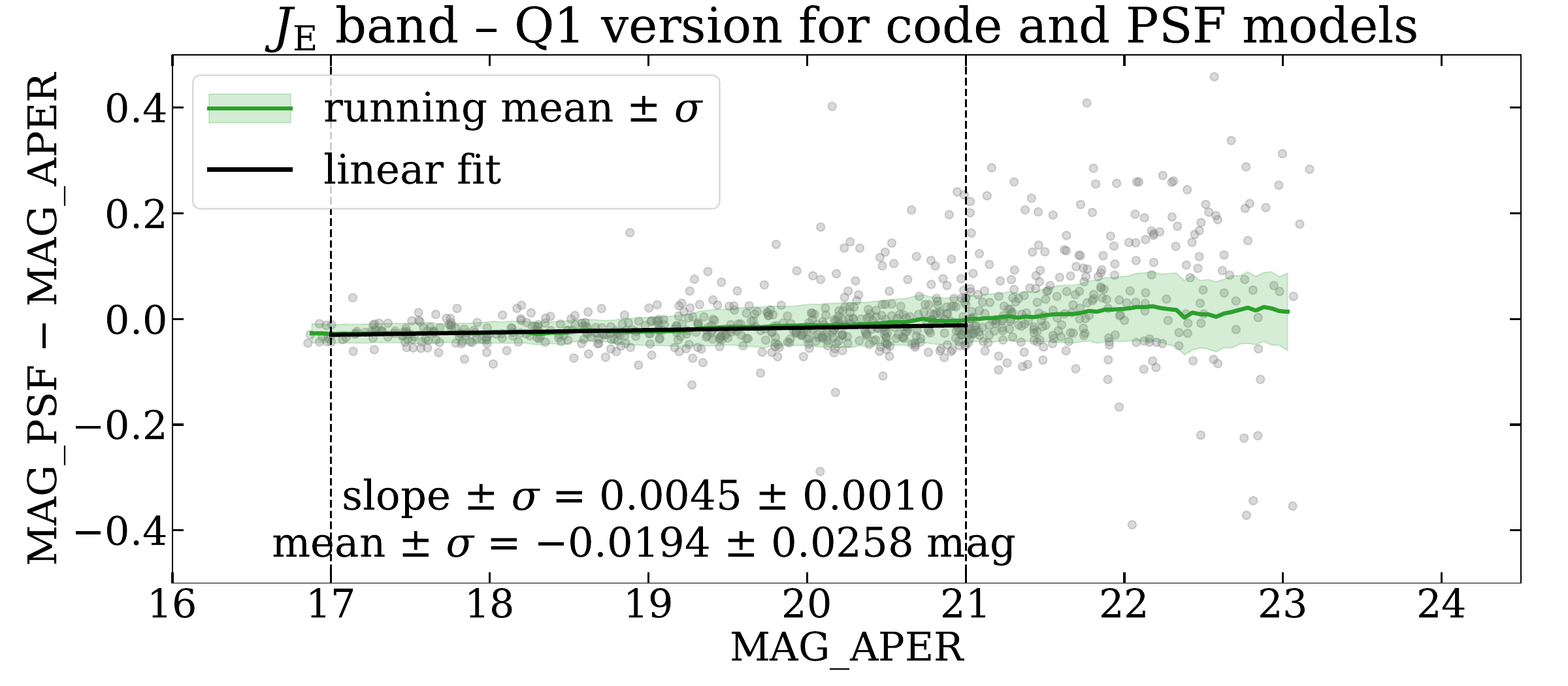}}
\end{minipage}
\begin{minipage}[t]{0.49\textwidth}
\centering
\resizebox{\hsize}{!}{\includegraphics{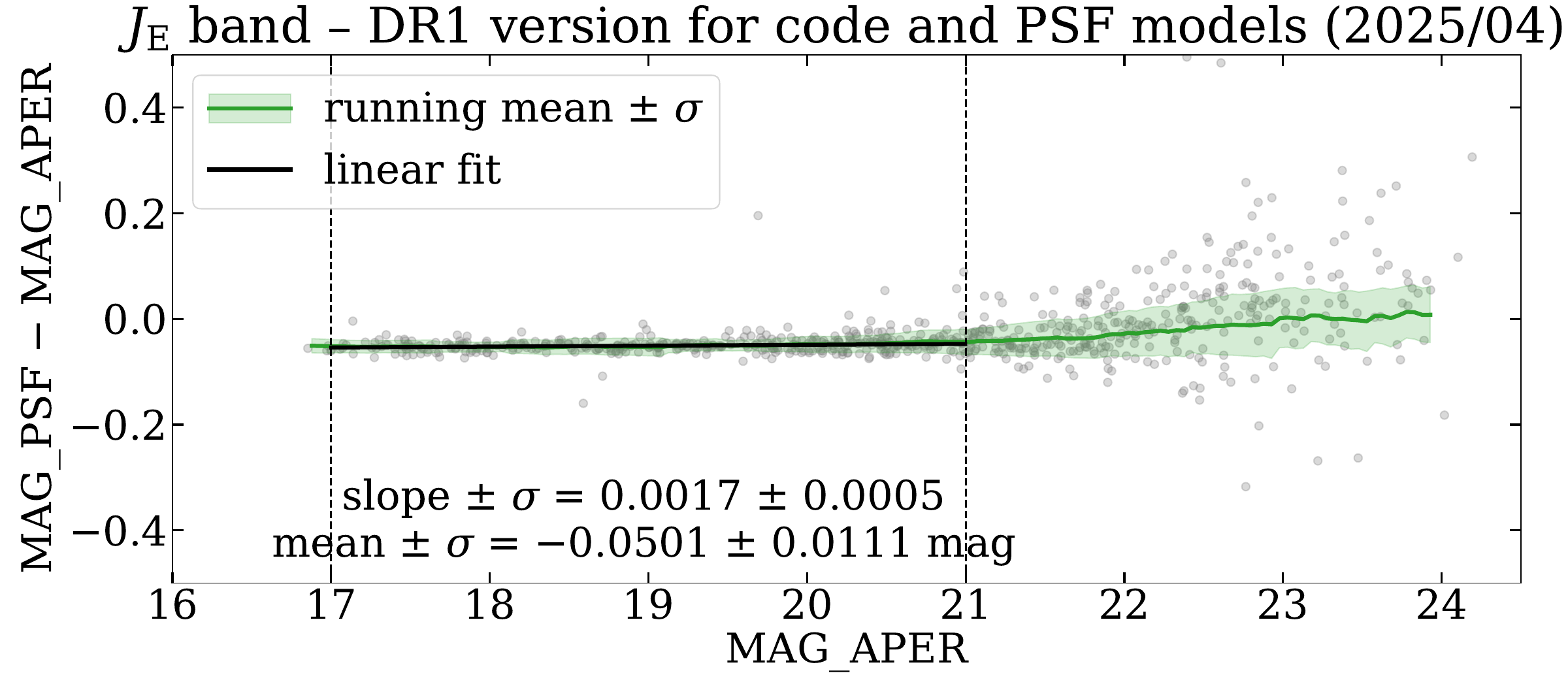}}
\end{minipage}
\caption{Same as in Fig.~\ref{figure:mag_psf_mag_aper}, but comparing the Q1 (\emph{left panels}) and DR1 (\emph{right panels}) versions of code and PSF models for the \YE (\emph{top panels}) and \JE (\emph{bottom panels}) passbands.}
\label{figure:mag_psf_mag_aper_yj}
\end{figure*}

The {\tt PHOT\_APERTURES} parameter of {\tt SourceExtractor} sets the diameter of the circular area used for aperture photometry measurements. In the input configuration file, we initially set it at six detector pixels (\ang{;;1.8}), which is almost 5 times larger than the FWHM in the \HE filter listed in Table~\ref{table:psf_params}. Even if this aperture comprises the central peak and the main diffraction patterns of the PSF profile, it encloses about 94\%, 93\%, and 92\% of the total flux measured for the \YE, \JE, and \HE passbands, respectively, inside the stamps of 141/6$\times$141/6 detector pixels. Therefore, to account for these flux losses, we need to apply a correction to the fluxes measured using this method. Table~\ref{table:psf_params} also lists the aperture corrections estimated from the PSF models derived for the three NIR filters, which are applied to each NISP detector during absolute photometric calibration \citep{Q1-TP003}.

It is worth noting that the 141/6$\times$141/6 detector pixel stamps do not contain the true flux expected to be measured for point-like sources. A small fraction of the signal, as part of the long diffraction spikes typical of the \mbox{NISP-P} observations \citep{EuclidSkyNISP}, is still lost. From the PSF models obtained in MDB mode, which have much larger stamps (512/6$\times$512/6 detector pixels), we estimate that about 1\% (for \YE), 1.8\% (for \JE), and 2.1\% (for \HE) of the true flux remain outside the stamps of 141/6$\times$141/6 detector pixels. However, we prefer not to use larger stamps to minimise contamination from neighbouring sources.

\begin{figure*}
\centering
\begin{minipage}[t]{0.49\textwidth}
\centering
\resizebox{\hsize}{!}{\includegraphics{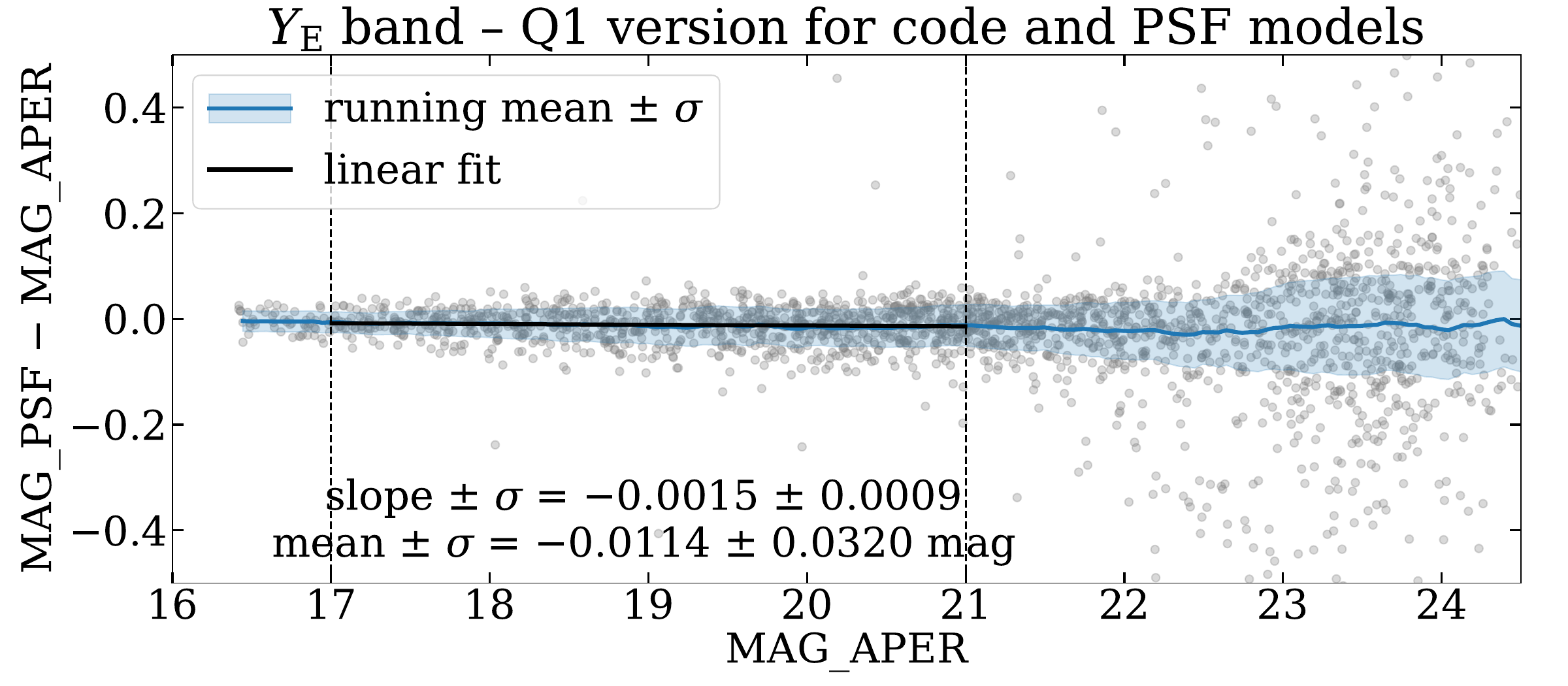}}
\end{minipage}
\begin{minipage}[t]{0.49\textwidth}
\centering
\resizebox{\hsize}{!}{\includegraphics{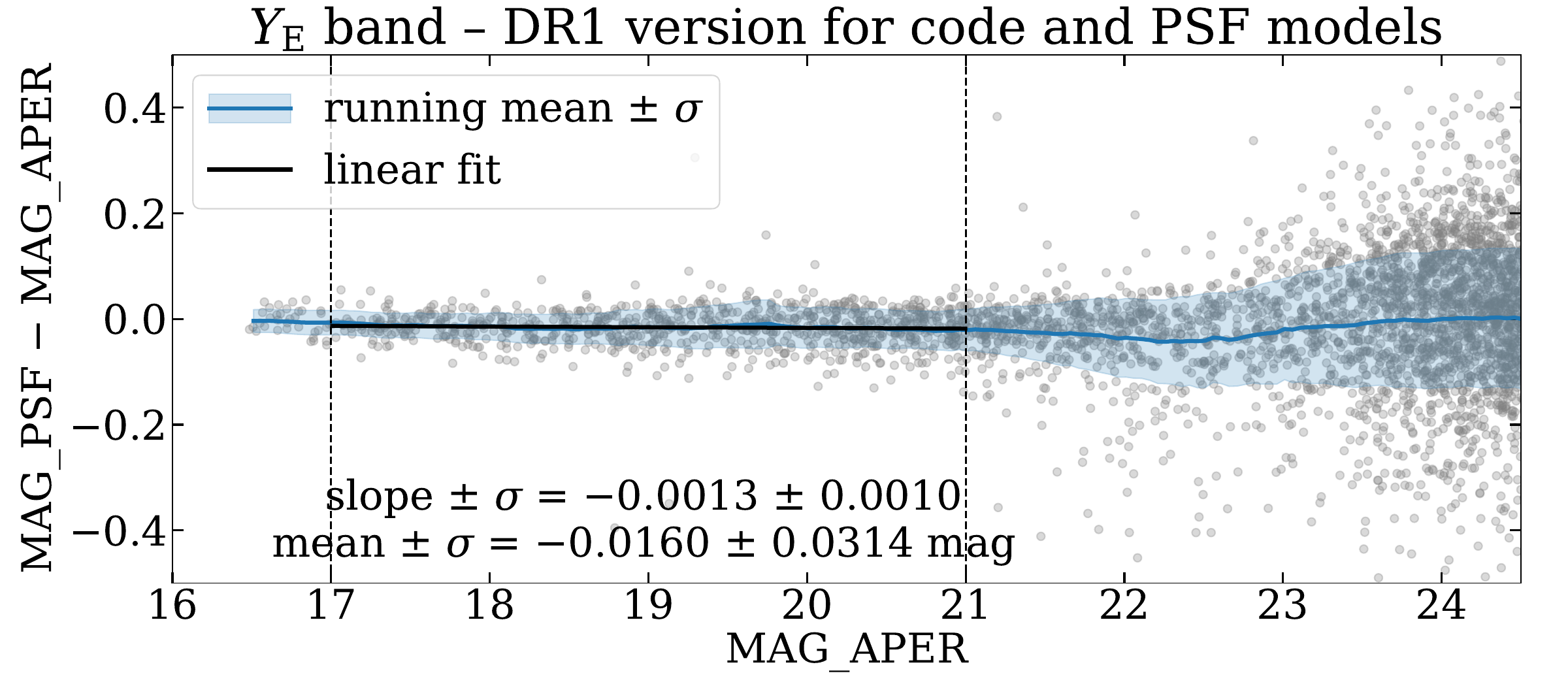}}
\end{minipage} \\
\begin{minipage}[t]{0.49\textwidth}
\centering
\resizebox{\hsize}{!}{\includegraphics{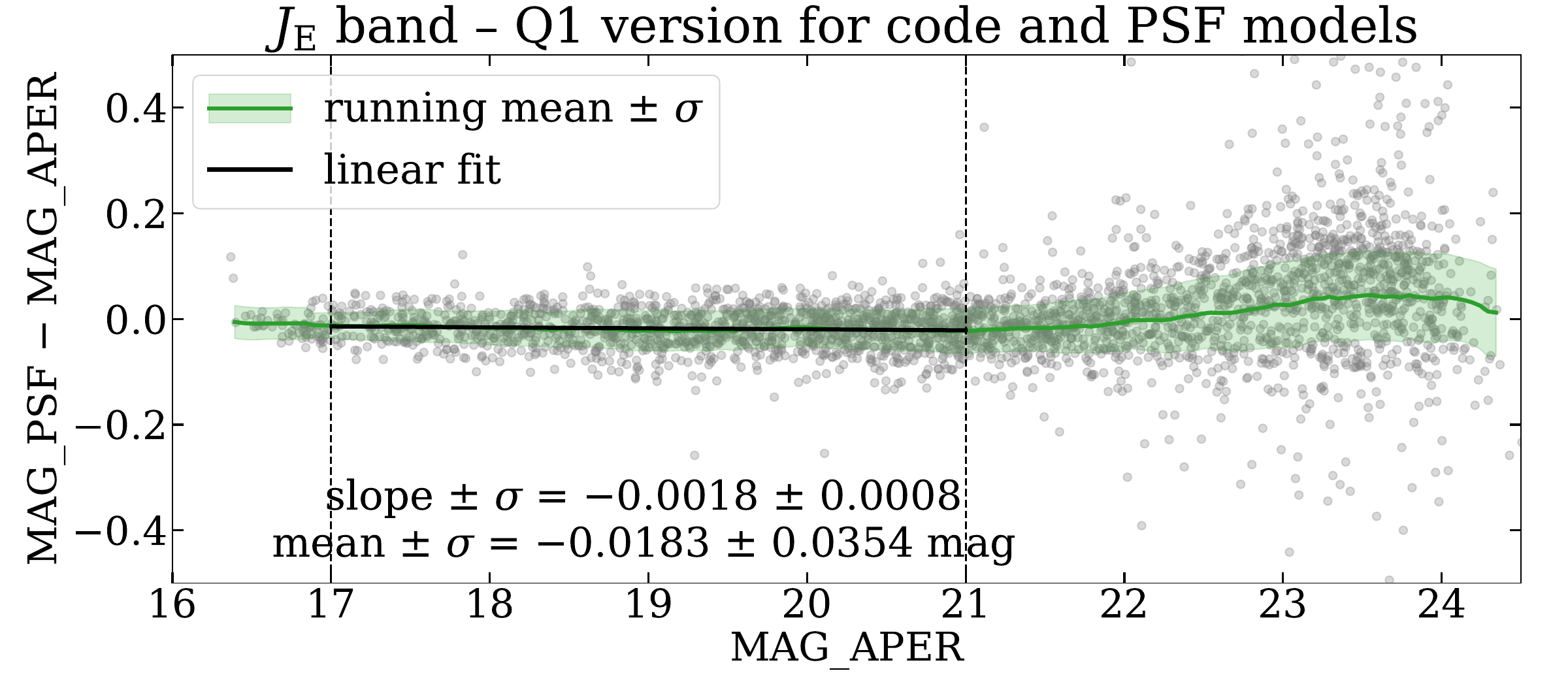}}
\end{minipage}
\begin{minipage}[t]{0.49\textwidth}
\centering
\resizebox{\hsize}{!}{\includegraphics{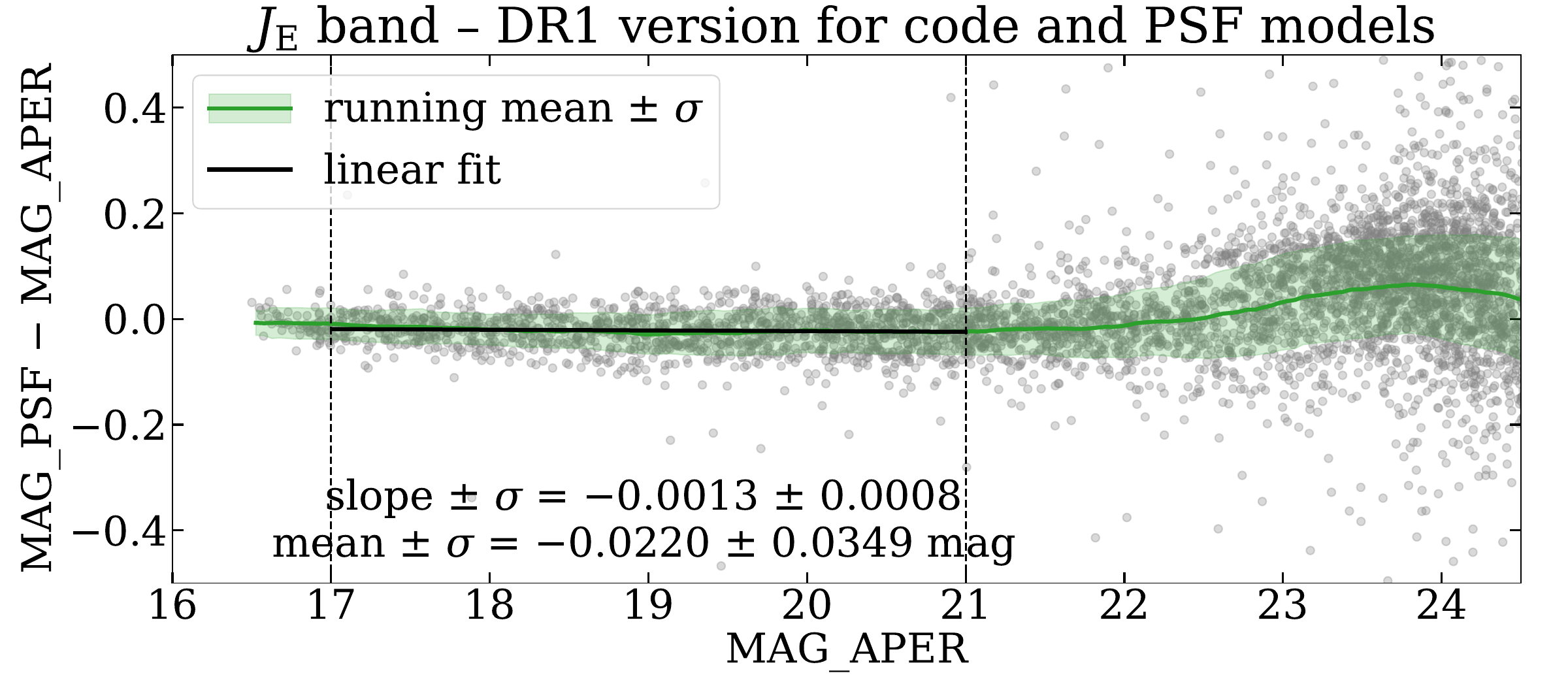}}
\end{minipage} \\
\begin{minipage}[t]{0.49\textwidth}
\centering
\resizebox{\hsize}{!}{\includegraphics{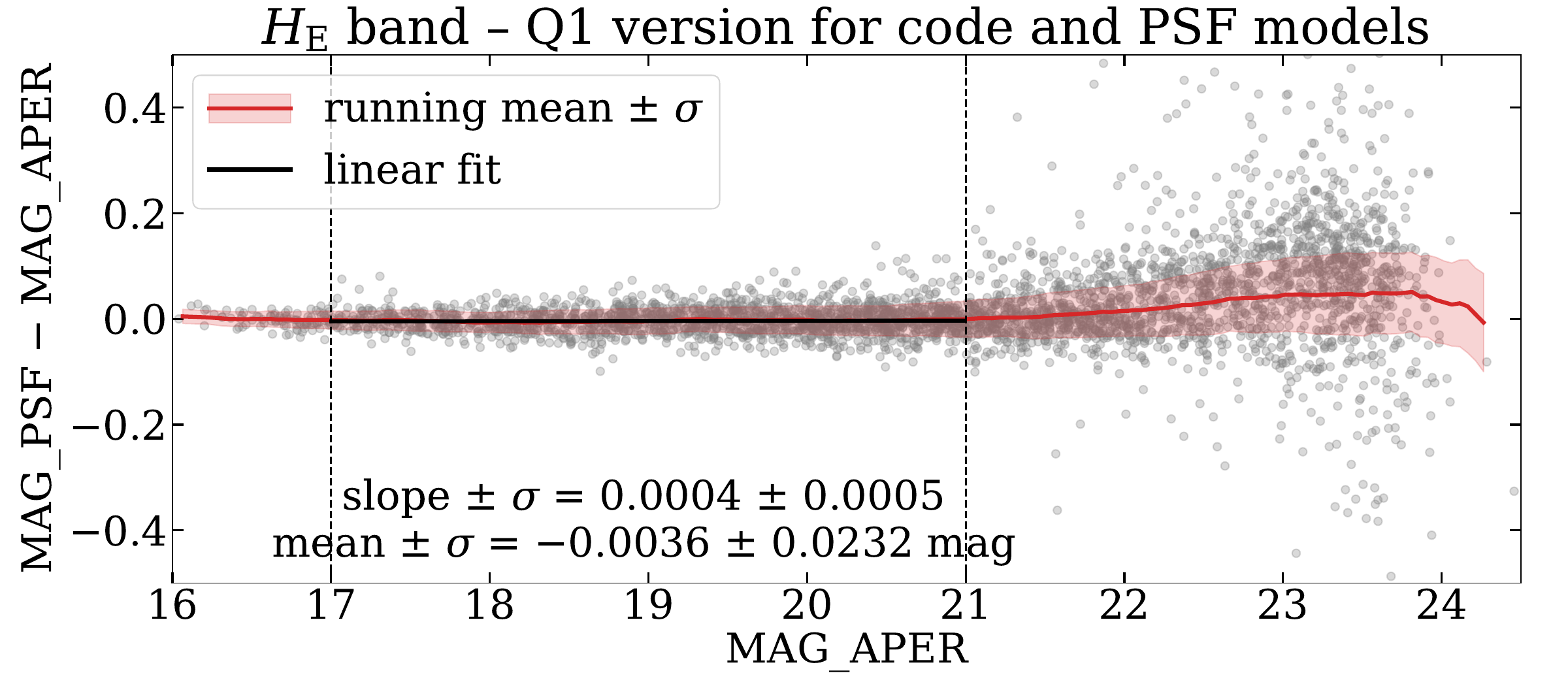}}
\end{minipage}
\begin{minipage}[t]{0.49\textwidth}
\centering
\resizebox{\hsize}{!}{\includegraphics{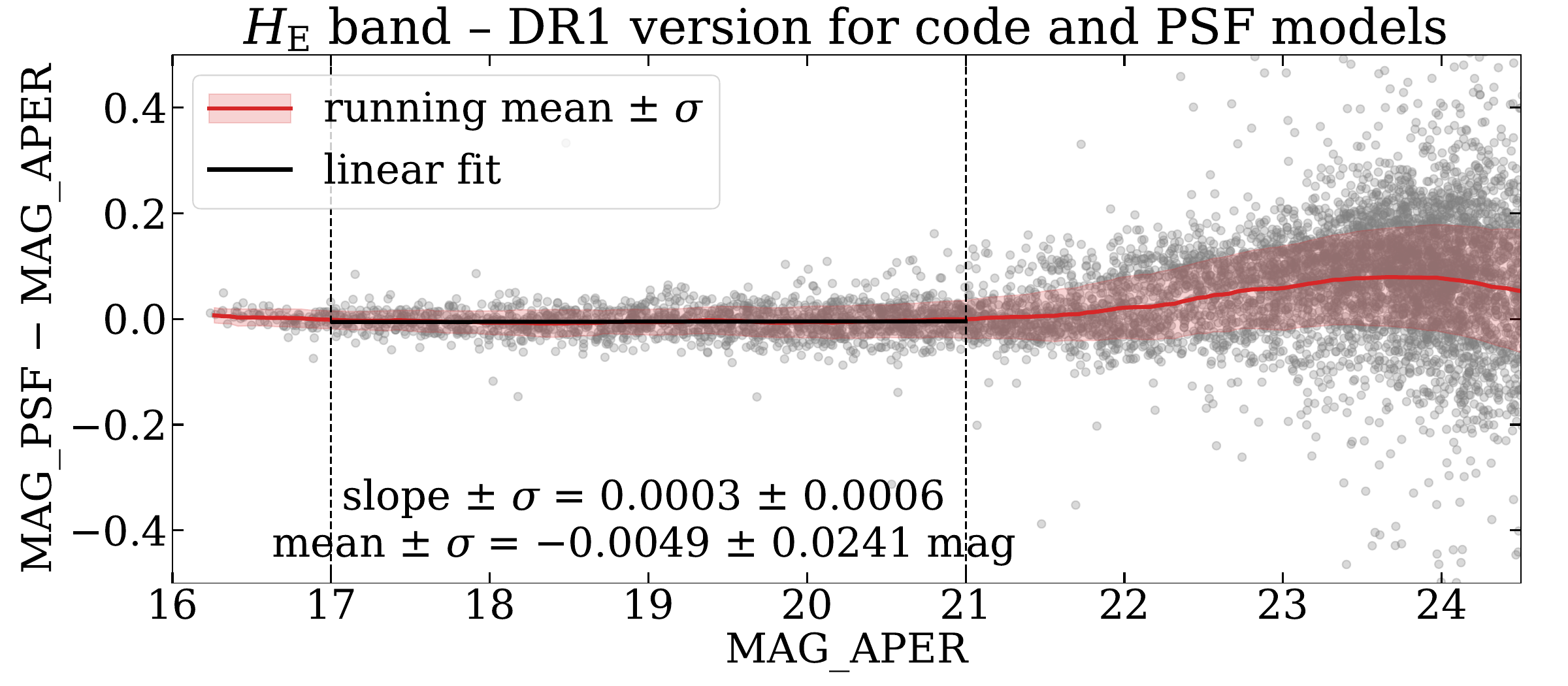}}
\end{minipage}
\caption{Same as in Fig.~\ref{figure:mag_psf_mag_aper}, but comparing the Q1 (\emph{left panels}) and DR1 (\emph{right panels}) versions of code and PSF models based on the stacked catalogues obtained for the three \mbox{NISP-P} bands.}
\label{figure:mag_psf_mag_aper_stack}
\end{figure*}

A by-product of the PSF-fitting procedure is the $x$ and $y$ coordinates of the sources available in the {\tt SourceExtractor} output catalogues. As part of the NIR validation plan and software test specifications, we developed a method to examine the performance of the PSF modelling task of the NIR pipelines. In this approach, we use the derived PSF models to perform PSF fitting on simulated images, for which we know the exact position of the sources. The results show that the differences between the PSF-fitting and simulated positions remain far below 0.1~detector pixels (the mean and standard deviation are smaller than 0.003 and 0.03~detector pixels, respectively, for both coordinates), therefore satisfying the second requirement specified in Sect.~\ref{sec:requirem}.

\subsubsection{\label{sec:phot_fixed} Fixed-aperture photometry}

As described in \citet{Q1-TP003,DR1-TP003}, one of the products of the NIR scientific pipeline are the catalogues created by the {\tt NIR\_CatalogExtraction} task for each astrometrically and photometrically calibrated image. In this task, the creation of the catalogue currently includes two steps. First, a source detection is performed using the {\tt SourceExtractor} code based on a source map produced from the reconstruction of a second derivative image. Then, in a second execution of the code, this source map is passed as a constraint to {\tt SourceExtractor} along with the scientific image to perform PSF-fitting and aperture photometry. In the upcoming paragraphs, we discuss some improvements achieved with the implementation of a new version of the catalogue extraction code together with the utilisation of the extended models derived by the PSF calibration pipeline.

Figure~\ref{figure:mag_psf_mag_aper} illustrates, in four comparative panels, the photometric improvements resulting from the refinements in the modelling and fitting of the PSF. The top left panel corresponds to the baseline photometry of Q1 data and reflects the state of the catalogue and the PSF models prior to any enhancements. The top right panel shows the same data after incorporating updates to the catalogue extraction code, which improved both source detection and initial flux estimation. We note that, in this and the other figures discussed in this section, only point-source objects are plotted, which were selected according to the FWHM values listed in Table~\ref{table:psfex_tuning}.

The Q1 code version was originally tested and optimised using data from simulations and from the PV phase. Although fixed and automatic aperture photometry methods performed adequately, PSF fitting showed systematic issues. A key limitation was the use of an excessively high initial detection threshold, which led to an underestimate of the size of the sources on the source map. This had a limited impact on fixed and automatic aperture photometry (where extraction sizes are explicitly defined in the configuration file), but it significantly affected the PSF-fitting procedure. {\tt SourceExtractor}, which uses the input source map to determine the region over which the PSF model is fitted, underestimated source sizes, resulting in biased or truncated fits.

In the new version of the NIR pipelines, the initial detection threshold has been reduced, allowing the algorithm to adaptively converge on a more accurate threshold. Additional parameters were introduced to filter out spurious detections, whose number could increase due to the lowering of the threshold. These changes led to a significant improvement in the PSF-fitting results, as is evident when comparing the two top panels of Fig.~\ref{figure:mag_psf_mag_aper}, which present the results obtained in the \HE band. Once a stable version of the code was achieved, it became possible to better isolate the effects of refinement of the PSF models. The lower panels of the same figure highlight the advantages of using extended PSF models derived from DR1 SCF data (processed in January and April 2025) compared to the simplified models used in Q1 data.

Similar enhancements found in the \HE band were also obtained for \YE and \JE, as seen in Fig.~\ref{figure:mag_psf_mag_aper_yj}. The left panels show a comparison between PSF-fitting and aperture photometry using the Q1 version of both catalogues and PSF models, whereas the right panels show the results for a combination of new catalogues and new PSF models available for DR1. These comparisons demonstrate a progressive reduction of the photometric residuals and improved consistency across the magnitude range, confirming the benefits of both code-level enhancements and a more robust PSF modelling strategy.

Regarding the stacked catalogues, the results currently obtained from the PSF fitting are quite different in comparison with the single-dither case. Although the changes implemented in the {\tt NIR\_CatalogExtraction} code also affect the construction of such catalogues, the PSF models derived from the stacked images, as previously discussed, do not have particular changes or improvements for DR1. Starting from the premise that the NIR stacked catalogues and images were not released as part of Q1, they were nonetheless processed using the same version of the pipelines employed for that release. Since these stacked data products are delivered as part of DR1 and therefore processed using the new version of the code, we are able to perform some preliminary comparisons.

As we had already noticed regarding the catalogues extracted from single-dither exposures, the changes made to the code allow us to obtain a PSF fit for many more sources and much fainter ones, as confirmed by Fig.~\ref{figure:mag_psf_mag_aper_stack} for catalogues extracted from stacked images. For the three \mbox{NISP-P} passbands, when going from Q1 to DR1 results, we cannot see any significant improvement in the distribution of residuals between the PSF-fitting and aperture photometry, which were instead present in the case of the single-dither catalogues. This is because the PSF fitting performed for the stacked image was already better than those for single exposures in Q1, since it was performed on the basis of a flatter background and a lower noise, allowing for a more ``clean'' fitting operation. We expect that the use of optimised PSF models, as the extended models derived for single dithers, will allow some improvement in the distribution of residuals and, therefore, provide even more accurate photometry from PSF fitting also in the case of the stacked catalogues.

The introduction of a lower initial detection threshold has significantly improved the PSF-fitting results. However, this makes us prone to the presence of spurious sources. Although partially removed, spurious detections are still a problem for the NIR PF, for both single-exposure and stacked catalogues. From the first tests of the NIR pipelines for DR1, we have been performing a detailed analysis of these spurious detections, investigating their cause (persistence residues, noise, saturated sources or cosmic rays not completely flagged, or detections on the diffraction spikes of bright sources) and their characteristics, so that they can be properly identified and flagged in future releases of the NIR catalogues. A better characterisation and removal of spurious sources will also allow us to obtain PSF models that are even more accurate, since they will be derived on the basis of only the good sources present in the images.

\subsubsection{\label{sec:phot_offset} PSF-fitting versus fixed-aperture photometry offset}

In the plots shown in Figs.~\ref{figure:mag_psf_mag_aper} and \ref{figure:mag_psf_mag_aper_yj}, although the scatter is significantly reduced for the three \mbox{NISP-P} filters, the offset between the PSF-fitting and aperture photometry actually increases when the new versions of code and PSF models are used. This is in part related to the PSF models themselves, particularly to the normalisation of their profiles. The {\tt PSFEx} code applies a normalisation factor to the derived models, which depends on a reference flux passed as input. By default, the code adopts the {\tt FLUX\_APER} photometry as a reference (or equivalently, the {\tt MAG\_APER} photometry). Therefore, although we expect no offset between {\tt MAG\_PSF} and {\tt MAG\_APER} under ideal conditions, an offset was actually observed.

\begin{figure*}
\centering
\begin{minipage}[t]{0.33\textwidth}
\centering
\resizebox{\hsize}{!}{\includegraphics{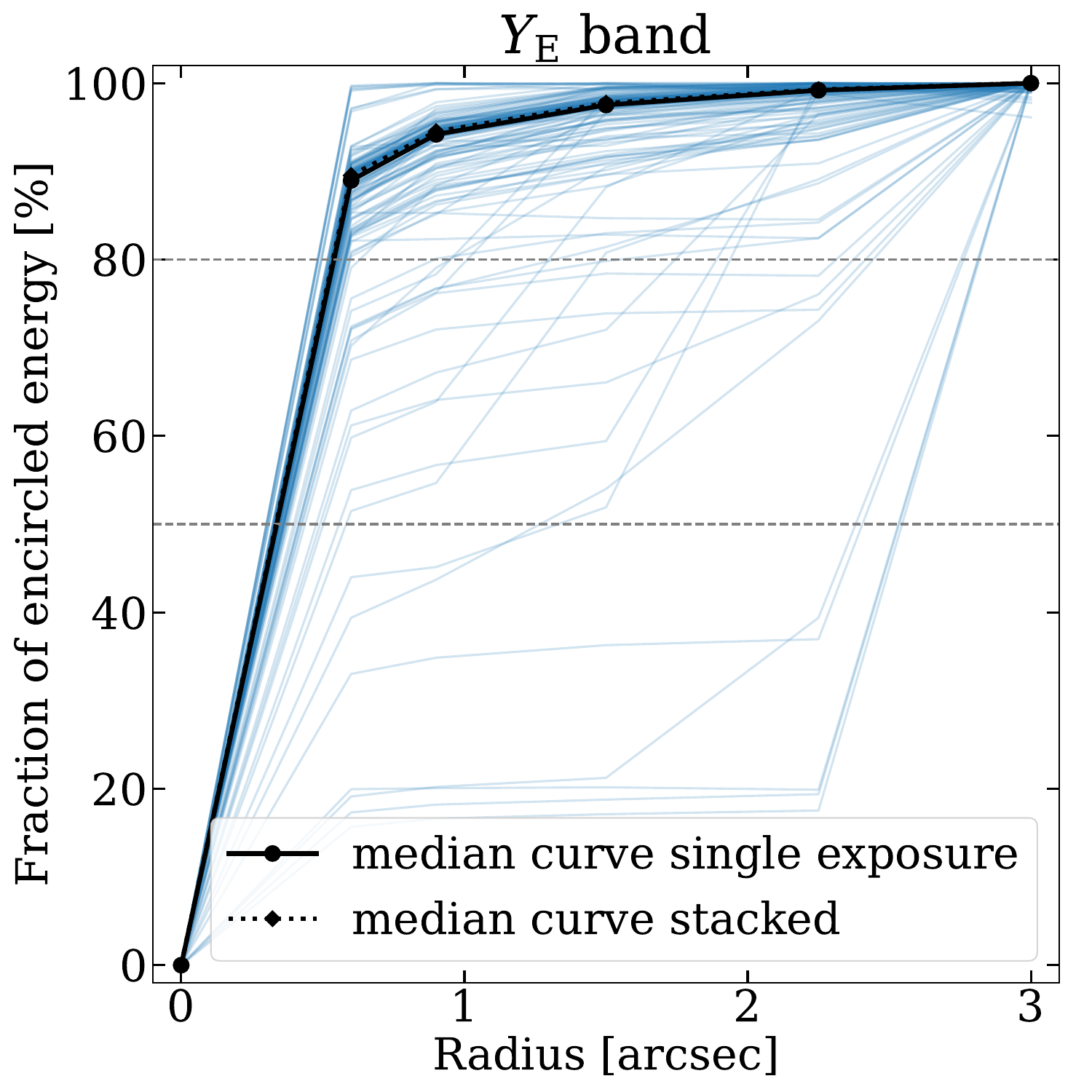}}
\end{minipage}
\begin{minipage}[t]{0.33\textwidth}
\centering
\resizebox{\hsize}{!}{\includegraphics{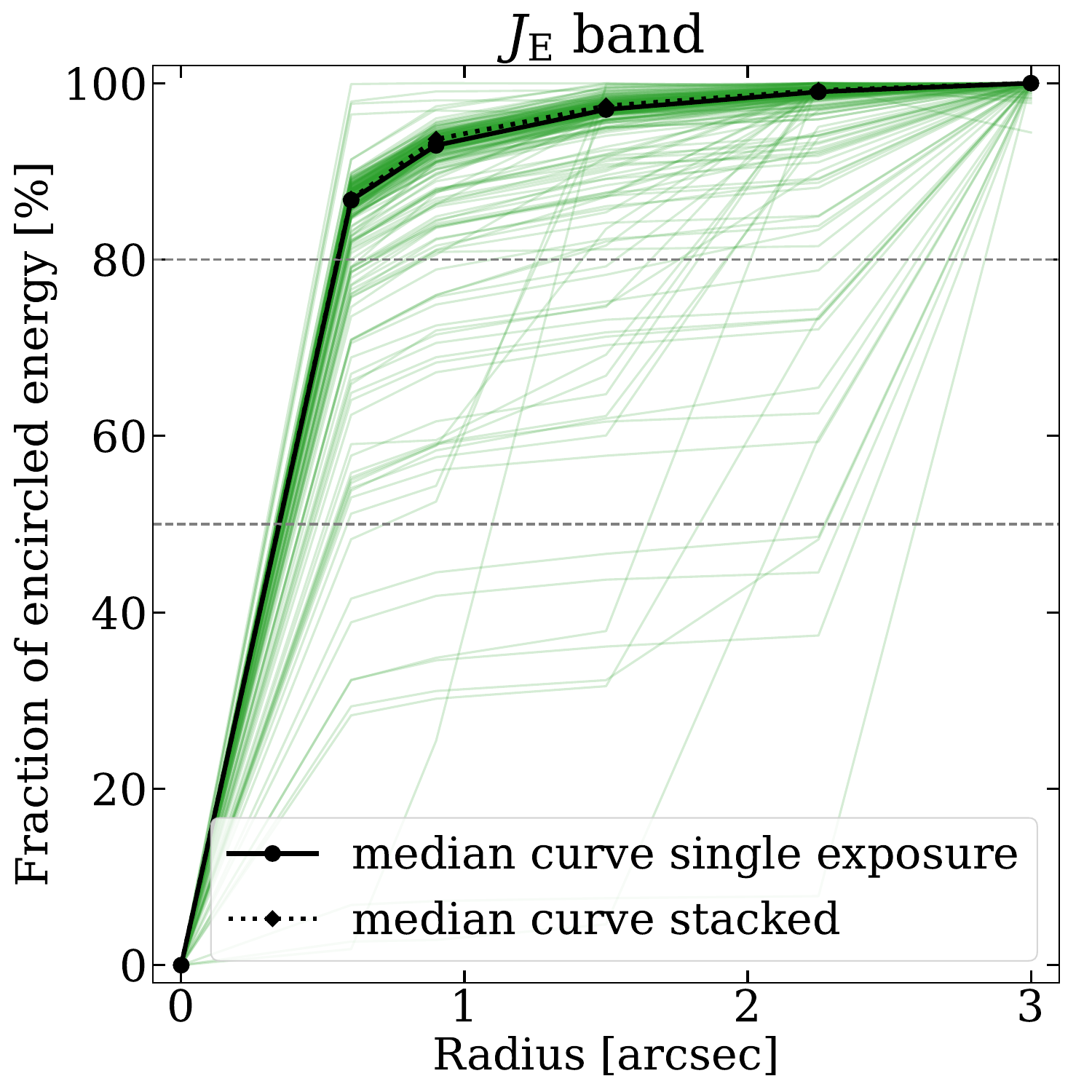}}
\end{minipage}
\begin{minipage}[t]{0.33\textwidth}
\centering
\resizebox{\hsize}{!}{\includegraphics{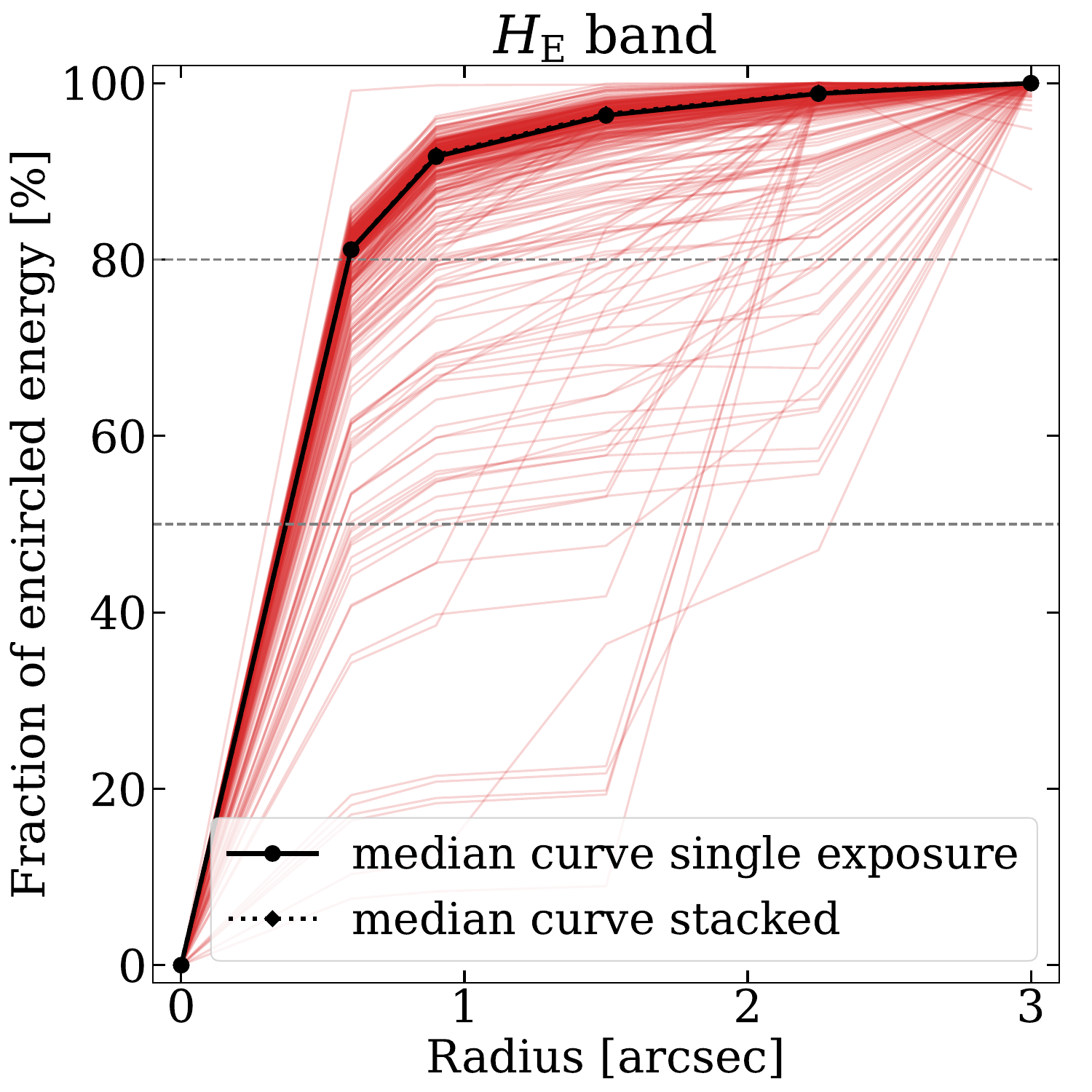}}
\end{minipage}
\caption{Encircled energy curves for a sample of point-like sources (from Detector ID {\tt 11} and stacked frames, Observation ID 2955), derived from multi-aperture photometry using diameters of \ang{;;0.6}, \ang{;;0.9}, \ang{;;1.5}, \ang{;;2.25}, and \ang{;;3}. The coloured lines represent individual sources, while the black ones indicate the median curves.}
\label{figure:EEphot}
\end{figure*}

When investigating additional causes of this offset, we noticed that the observed differences have a dependence on the centroid position of the source profiles on sub-pixel scales. For sources with a centroid close to the centre of a pixel, the differences between PSF-fitting and aperture photometry are close to zero. However, when going from the centre to the edges of a pixel, the differences increase in the sense that {\tt MAG\_PSF} becomes underestimated (i.e. the sources are brighter than expected) compared to {\tt MAG\_APER}. This effect is clearly seen in Fig.~\ref{figure:mag_psf_aper_offset}, which shows how the magnitude changes for sources observed in different dither exposures (i.e. in different sub-pixel positions). For all the sources observed in a minimal of ten dither exposures, we plot the differences between the magnitude of each source and the median value calculated for the same source. The left panel shows the differences for {\tt MAG\_APER}, which are on average close to zero and do not have significant variations. The right panel shows the differences for {\tt MAG\_PSF}, clearly indicating that sources with a centroid close to the edges of a pixel are systematically brighter.

The origin of the effect illustrated in the right panel of Fig.~\ref{figure:mag_psf_aper_offset} is likely related to the PSF-fitting procedure, which is sensitive to how the PSF profile is sampled on the detector under undersampled conditions. Moreover, this behaviour is amplified by the IPC and the IPS. Together, these effects modify the shape of point-like sources, thus impacting the fitting of the PSF profile. In the left panel, the results from fixed-aperture photometry demonstrate that the relative photometric error for {\tt MAG\_APER} is far below the third requirement described in Sect.~\ref{sec:requirem} (the change in magnitude of the same source remains within 0.005~mag, which corresponds to about 0.5\% in flux). A comparison of {\tt MAG\_APER} with {\it Gaia} photometry indicates that the goal of achieving an absolute photometric error of 5\% is also satisfied (currently, the differences remain within 2\%, 3\%, 5\% for \YE, \JE, and \HE, respectively).

Eliminating the differences observed between {\tt MAG\_PSF} and {\tt MAG\_APER} would require the application of a post-processing correction to the measured magnitudes or a change in the PSF-fitting procedure. Both of these solutions lie outside the scope of the present paper, which focusses on presenting results obtained from DR1 data. An appropriate solution is being investigated and is expected to be implemented in a future release of the \Euclid data. Before a suitable solution is available, we recommend applying a correction to the {\tt MAG\_PSF} values derived for point-like sources to account for the observed offset and therefore to be in accordance with the mission requirements and goals.

\subsubsection{\label{sec:phot_multi_aper} Multi-aperture photometry}

In the Q1 version of the {\tt NIR\_CatalogExtraction} task, we provided a single-aperture photometry estimate with radius \mbox{\ang{;;0.9}} (aperture with a diameter of 6 detector pixels). In the new version adopted for DR1, we implemented a multi-aperture photometry approach that allows flux measurements using a set of concentric apertures with radii \ang{;;0.6}, \ang{;;0.9}, \ang{;;1.5}, \ang{;;2.25}, and 3\arcsec\ (apertures with diameters of 4, 6, 10, 15, 20 detector pixels). This enhancement enables the reconstruction of the flux growth curve and, consequently, the encircled energy diagram. By analysing the flux as a function of aperture radii (instead of diameter, to be consistent with the encircled energy plots), we gain more detailed insight into the spatial distribution of the source light profile, which can be compared with the distribution obtained from the PSF model itself.

Figure~\ref{figure:EEphot} shows the encircled energy curves derived for a sample of sources extracted from calibrated images for the three \mbox{NISP-P} passbands. Each lighter line represents the normalised cumulative flux profile for an individual source, while the darker line represents the median curve with the five points at which it is calculated superimposed. Horizontal grey dashed lines indicate the encircled energy levels at 50\% and 80\%. The median curves intersect the level of 50\% and 80\% of the energy at the radii listed in Table~\ref{table:ee50_80}. This plot illustrates the effectiveness of the multi-aperture strategy in capturing the range of light concentration profiles across the dataset.

The same analysis was carried out on stacked catalogues, and the results are very similar to the case of single exposures. Figure~\ref{figure:EEphot} also shows the median curves in the three \mbox{NISP-P} filters for individual and stacked catalogues. These curves behave as expected among different bands, remaining lower for \HE and reaching larger radii at 50\% and 80\% than for \JE and \YE. However, for the same filter, the curves show only minimal differences between single and stacked cases, in contrast to those shown in Fig.~\ref{figure:psf_profile_ee_stacked} obtained directly from the PSF profiles. We defined the multi-aperture radii to allow us to sample the growing energy until its completeness and also to distinguish compact objects from extended galaxies. Therefore, we avoided the use of too small apertures of the size of a single pixel \mbox{($\sim$\,\ang{;;0.3})} or smaller, which means that the sampling does not accurately trace the encircled energy profile for small radii. This also explains the larger radii listed in Table~\ref{table:ee50_80} compared to the values in Table~\ref{table:psf_params}, considering that both rEE50 and rEE80 are smaller than the minimum aperture radius of \ang{;;0.6}. For radii larger than \ang{;;0.6}, the difference between single and stacked cases in the encircled energy curves plotted in Fig.~\ref{figure:psf_profile_ee_stacked} might be related to the fact that the PSF profiles of the sources in the stacked frame are not yet optimised. They are based on a limited number of sources and a small number of basis vectors, which makes it difficult to model the diffraction rings and spikes of the PSF profile.

\subsection{\label{sec:spread_model} Spread model}

Another parameter provided by {\tt SourceExtractor} and used as a morphological classifier is named {\tt SPREAD\_MODEL}. To estimate its value at the position of a given source, the code compares the best fit of the PSF model passed as input with the best fit of a slightly broader PSF model, simulating a large galaxy, explicitly
\be
{\tt SPREAD\_MODEL} = \frac{L_1 - L_2}{\sqrt{V_1 + V_2}}\,,
\ee
where $L_1$ and $L_2$ are the likelihood of the input and broader PSF models, respectively, and $V_1$ and $V_2$ are the variances of these models, used as a normalisation factor. Sources having {\tt SPREAD\_MODEL} close to zero are probably stars or compact galaxies. Extended galaxies would have positive values, whereas detections with narrow profiles, such as cosmic rays or detector artefacts, would have negative values.

\begin{table}[t!]
\caption{Encircled energy radii from multi-aperture photometry for a single-dither exposure and for a stacked frame (Observation ID 2955).}
\centering
\smallskip
\label{table:ee50_80}
\smallskip
{\small
\begin{tabular}{l@{\hskip 2em} ccc@{\hskip 2em} ccc}
\hline\hline\noalign{\vskip 2pt}
 & \multicolumn{3}{c@{\hskip 2em}}{rEE50 [arcsec]} & \multicolumn{3}{c}{rEE80 [arcsec]} \\
\noalign{\vskip 2pt}\hline\noalign{\vskip 2pt}
 & \YE & \JE & \HE & \YE & \JE & \HE \\
\noalign{\vskip 2pt}\hline\noalign{\vskip 2pt}
Det. ID {\tt 11}  & 0.336 & 0.346 & 0.370 & 0.538 & 0.554 & 0.591 \\
Stacked           & 0.335 & 0.345 & 0.370 & 0.536 & 0.552 & 0.591 \\
\hline
\end{tabular}}
\tablefoot{These measurements are based on the median encircled energy curves shown in Fig.~\ref{figure:EEphot}.}
\end{table}

\begin{figure*}
\centering
\begin{minipage}[t]{0.33\textwidth}
\centering
\resizebox{\hsize}{!}{\includegraphics{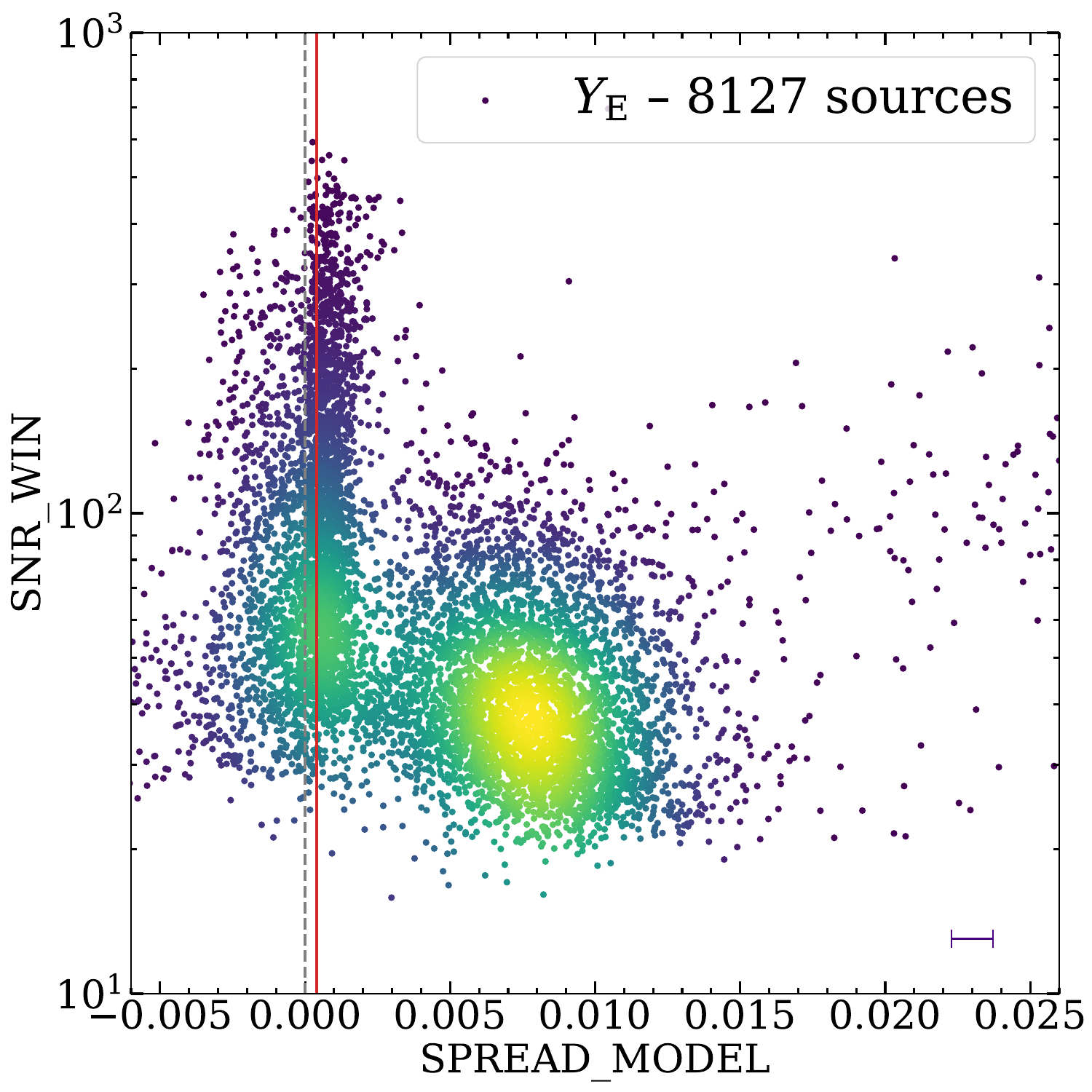}}
\end{minipage}
\begin{minipage}[t]{0.33\textwidth}
\centering
\resizebox{\hsize}{!}{\includegraphics{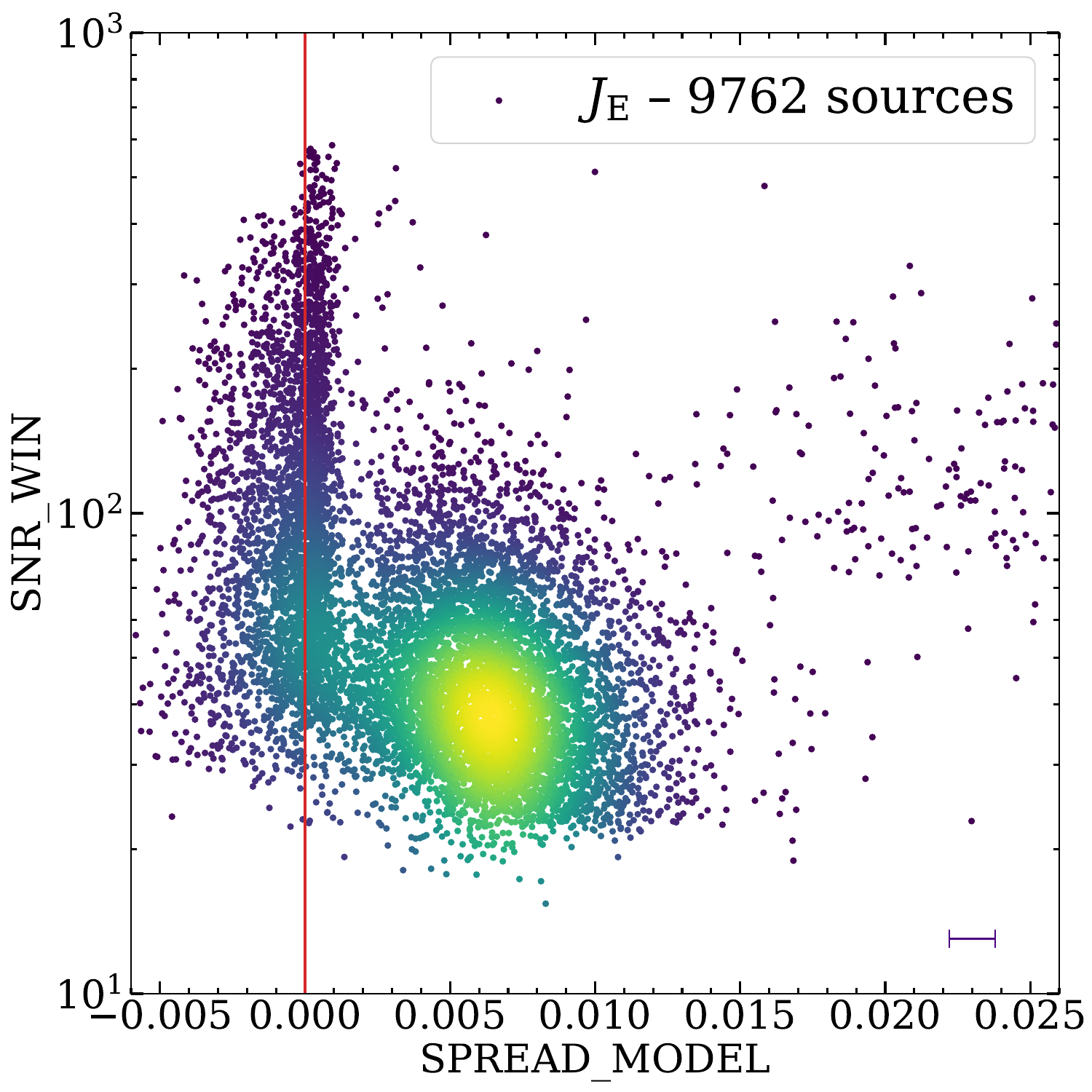}}
\end{minipage}
\begin{minipage}[t]{0.33\textwidth}
\centering
\resizebox{\hsize}{!}{\includegraphics{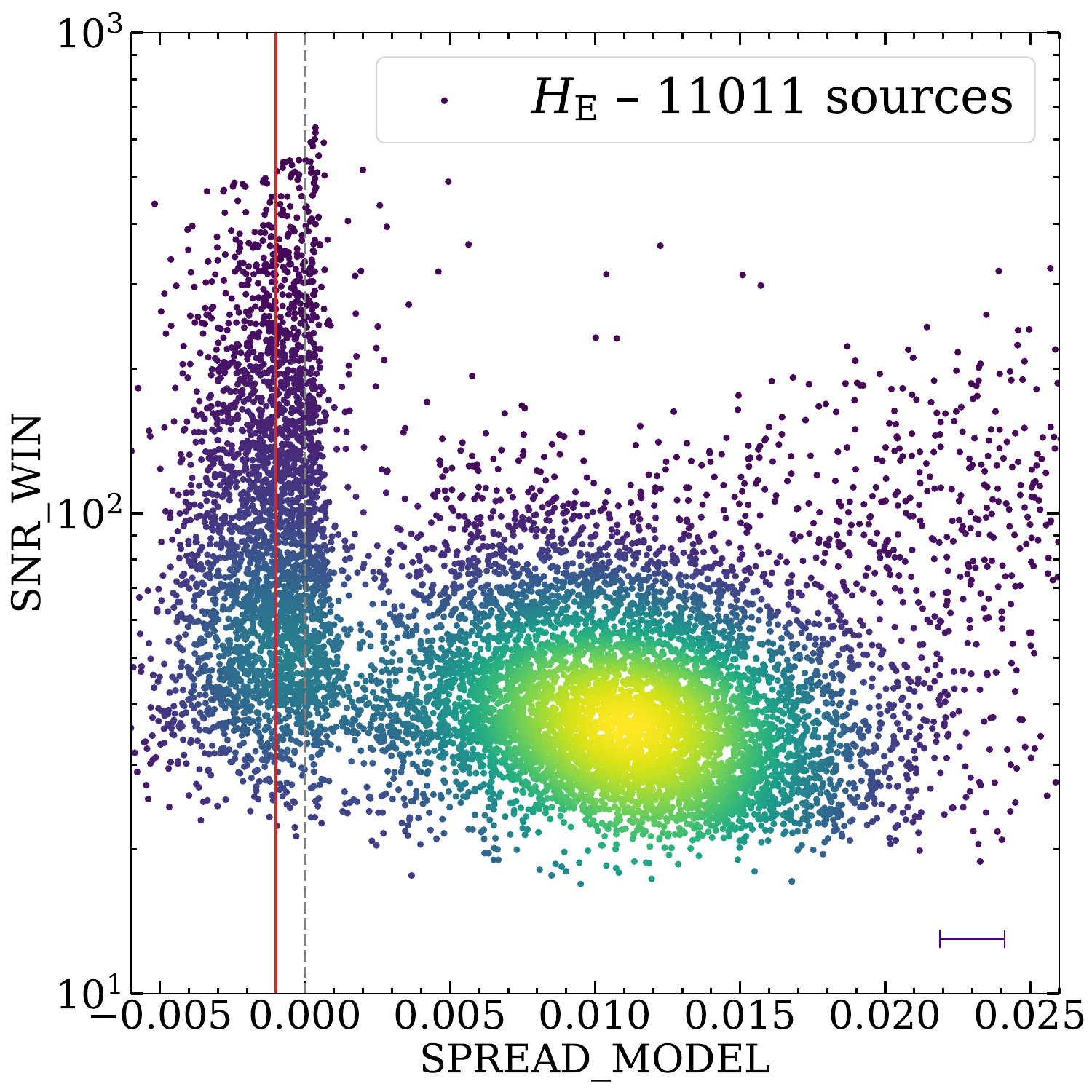}}
\end{minipage} \\
\begin{minipage}[t]{\textwidth}
\hspace*{18pt}
\resizebox{0.95\hsize}{!}{\includegraphics{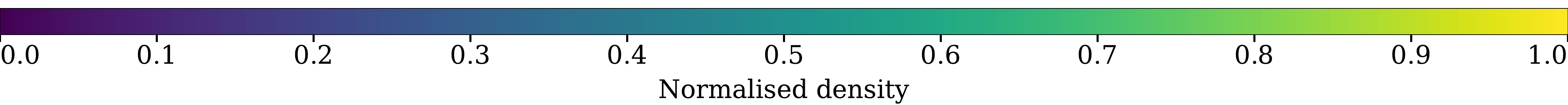}}
\end{minipage}
\caption{Signal-to-noise ratio {\tt SNR\_WIN} as a function of the morphological classifier {\tt SPREAD\_MODEL}, both derived by the {\tt SourceExtractor} code. The panels show sources extracted from the 16 detectors of single-dither exposures (Observation ID 6185). The solid red line indicates the median calculated for sources with {\tt SPREAD\_MODEL} $<$\,0.002.}
\label{figure:spread_models}
\end{figure*}

\begin{figure}
\centering
\begin{minipage}[t]{0.49\textwidth}
\centering
\resizebox{\hsize}{!}{\includegraphics{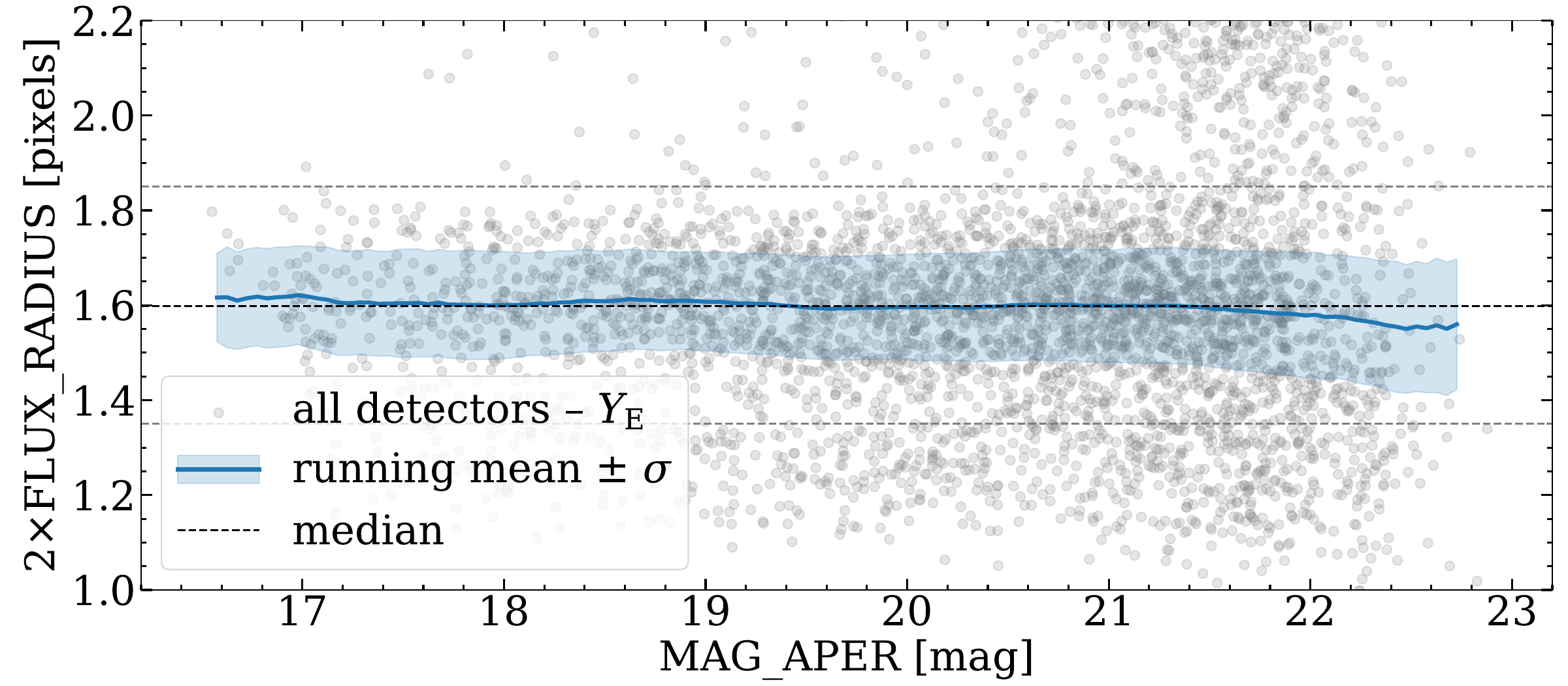}}
\end{minipage} \\
\begin{minipage}[t]{0.49\textwidth}
\centering
\resizebox{\hsize}{!}{\includegraphics{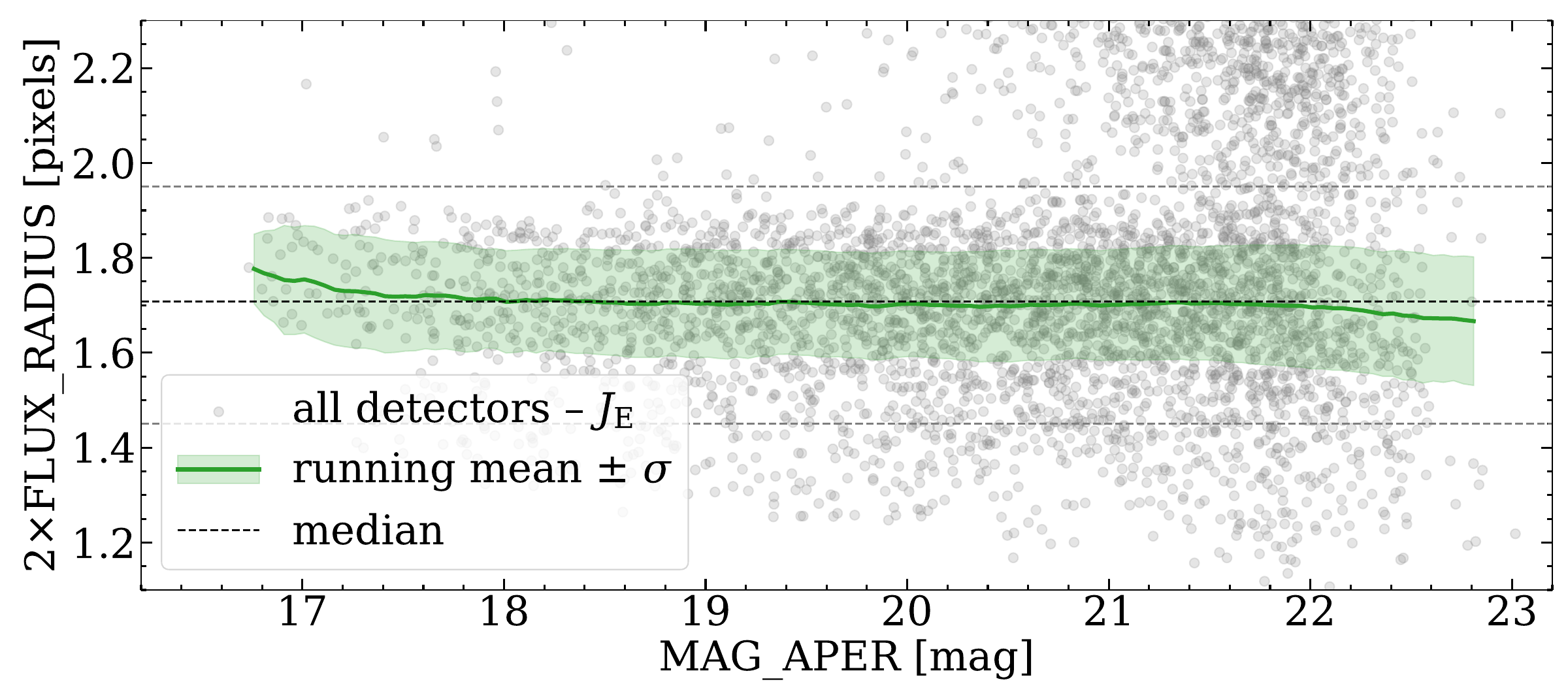}}
\end{minipage} \\
\begin{minipage}[t]{0.49\textwidth}
\centering
\resizebox{\hsize}{!}{\includegraphics{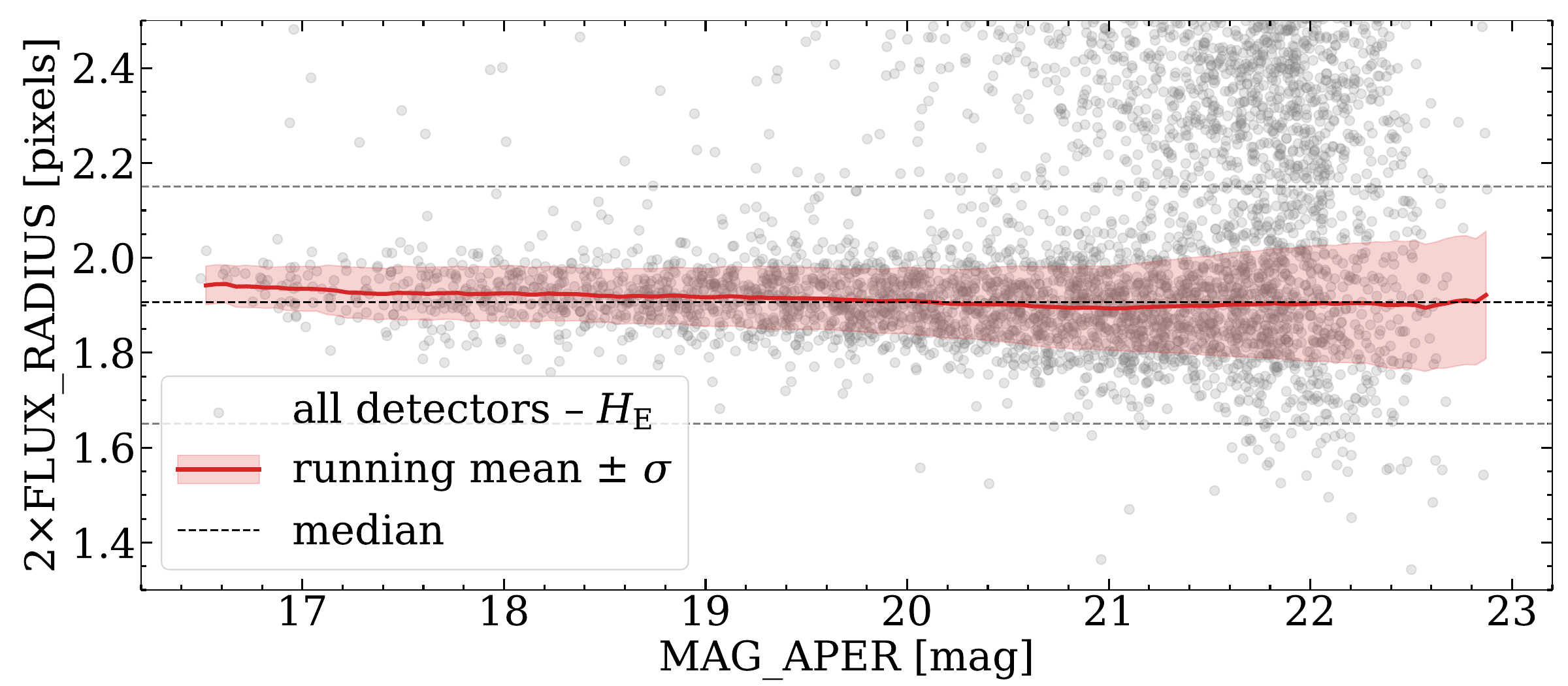}}
\end{minipage}
\caption{Half-light diameter as a function of fixed-aperture photometry for single-dither exposures (Observation ID 6185). The panels show the median and the running mean for the 16 detectors (sampling width of 1~mag) calculated for points within the two dashed grey lines.}
\label{figure:flux_radius_mag_aper}
\end{figure}

\begin{figure*}
\centering
\begin{minipage}[t]{0.49\textwidth}
\centering
\resizebox{\hsize}{!}{\includegraphics{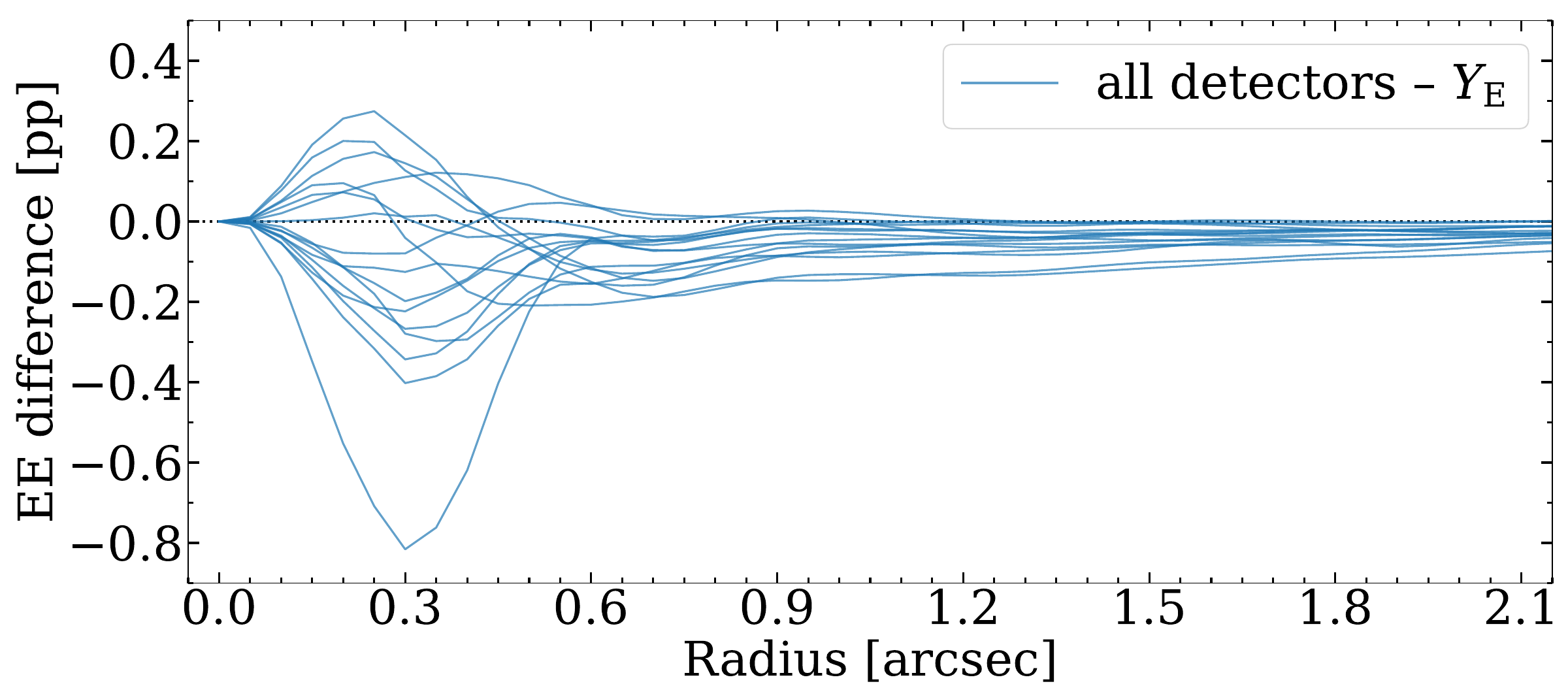}}
\end{minipage}
\begin{minipage}[t]{0.49\textwidth}
\centering
\resizebox{\hsize}{!}{\includegraphics{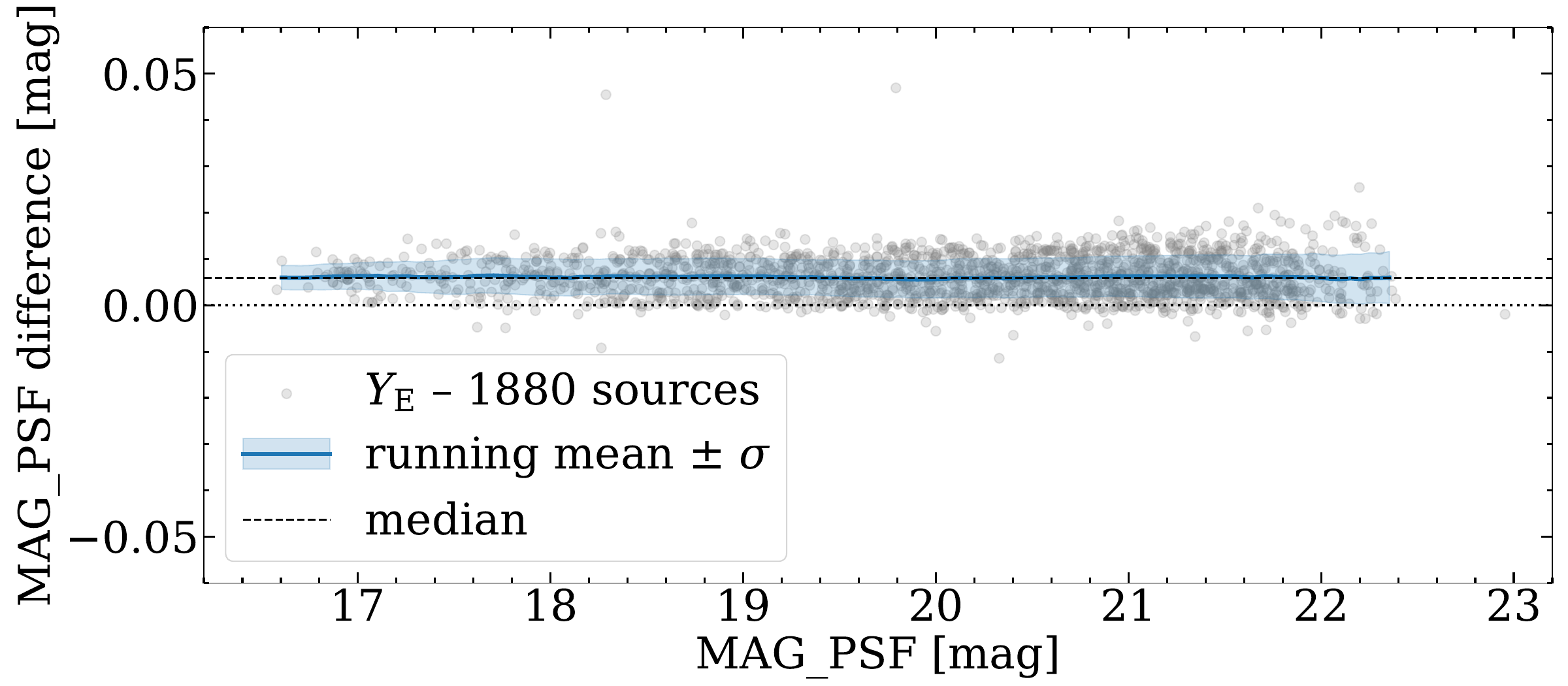}}
\end{minipage} \\
\begin{minipage}[t]{0.49\textwidth}
\centering
\resizebox{\hsize}{!}{\includegraphics{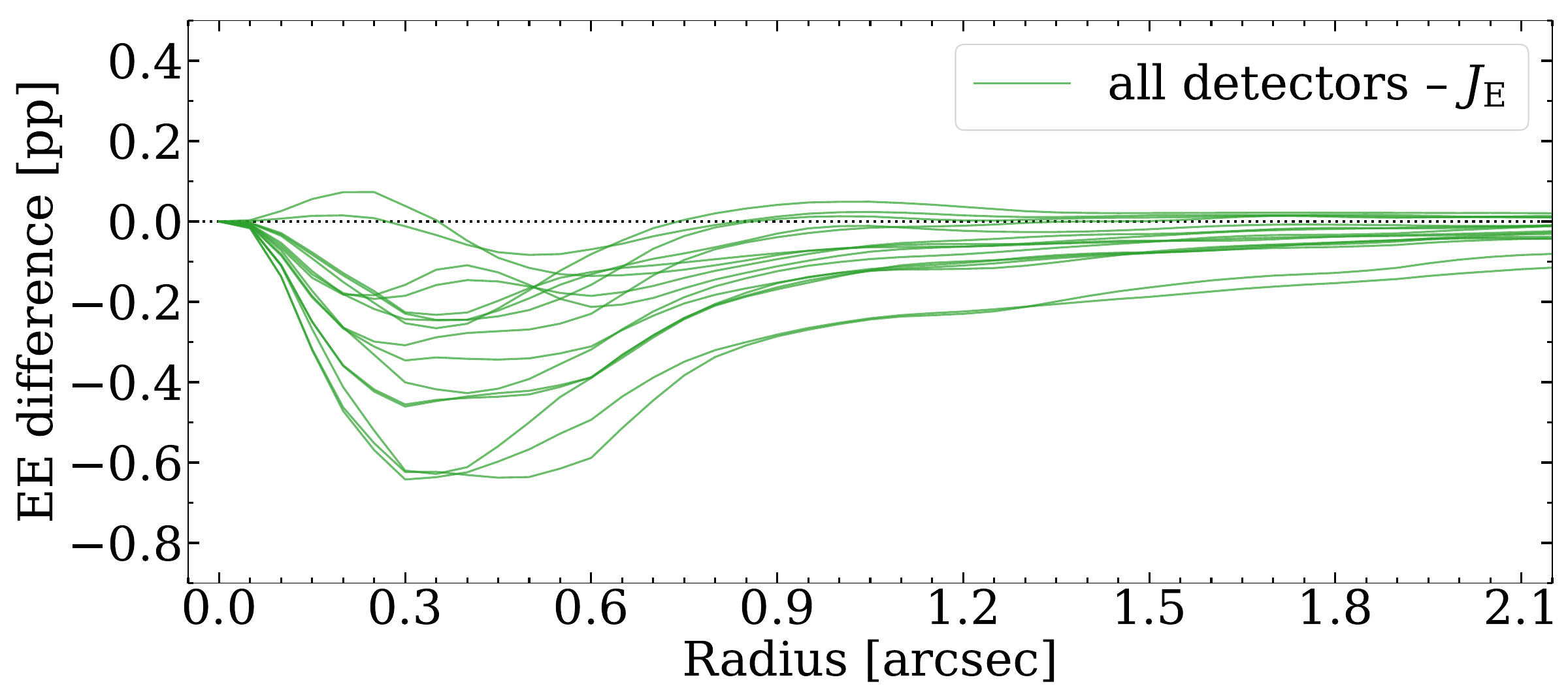}}
\end{minipage}
\begin{minipage}[t]{0.49\textwidth}
\centering
\resizebox{\hsize}{!}{\includegraphics{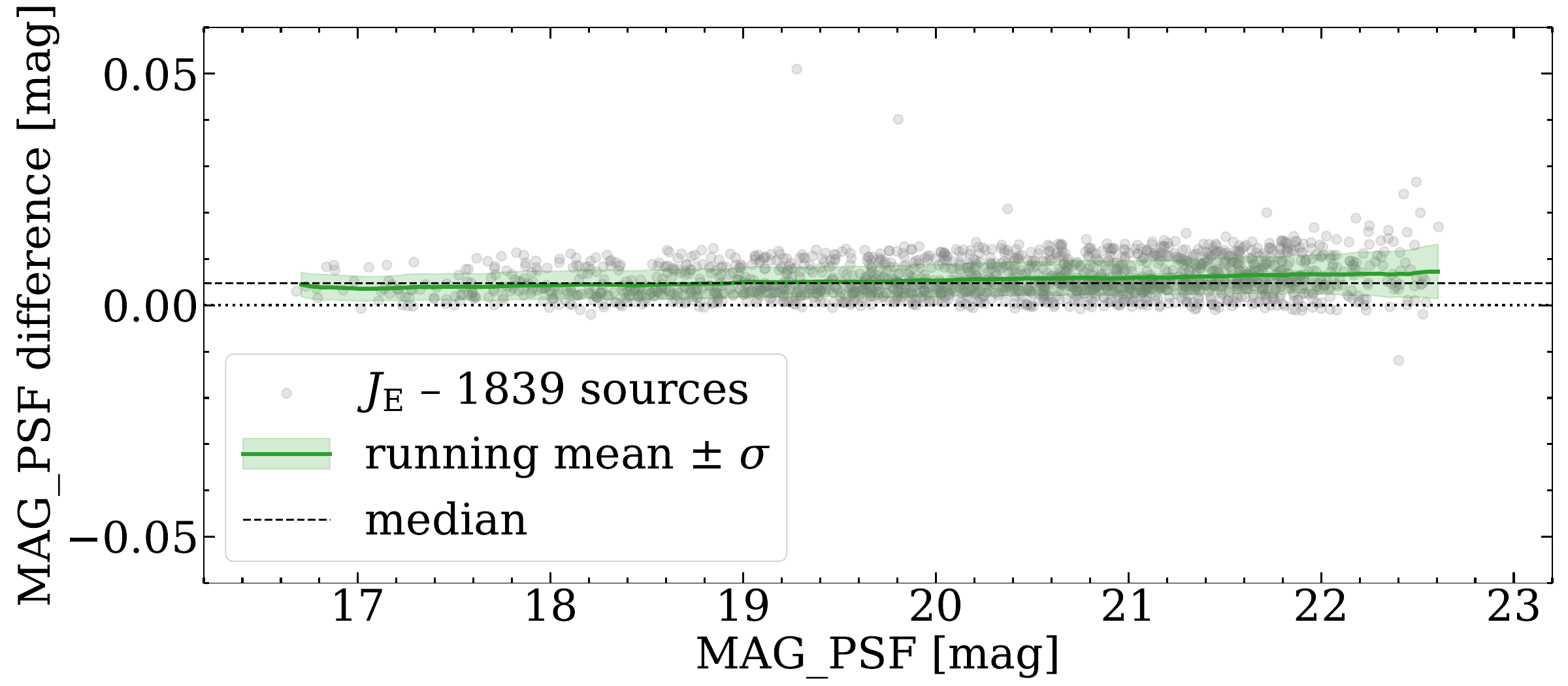}}
\end{minipage} \\
\begin{minipage}[t]{0.49\textwidth}
\centering
\resizebox{\hsize}{!}{\includegraphics{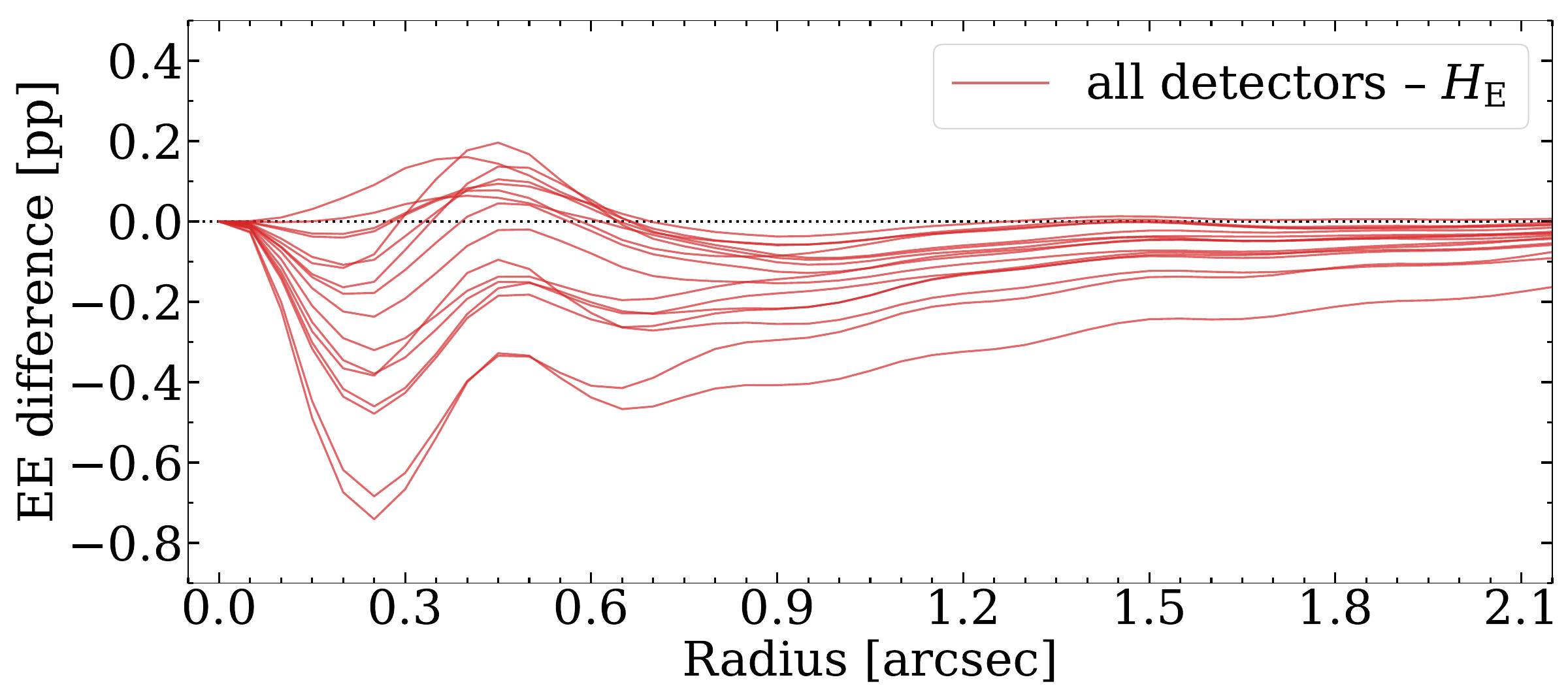}}
\end{minipage}
\begin{minipage}[t]{0.49\textwidth}
\centering
\resizebox{\hsize}{!}{\includegraphics{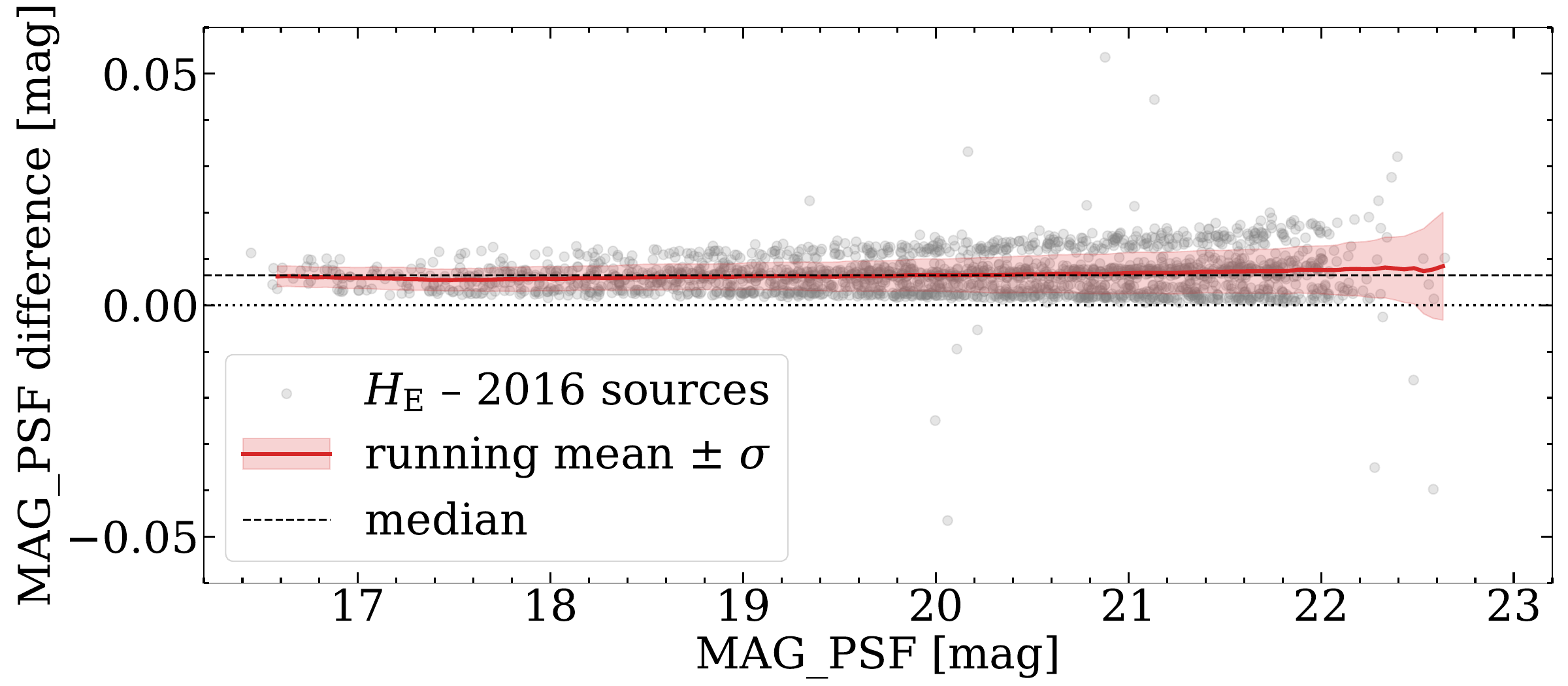}}
\end{minipage}
\caption{Comparison between extended PSF models derived using data from Observation IDs 2174 and 519 processed in January and April 2025, respectively. {\it Left panels}: differences in encircled energy (pp = percentage points) as a function of the distance to the peak of the PSF profile. {\it Right panels}: Differences in {\tt MAG\_PSF} photometry derived from PSF profile fitting.}
\label{figure:ee_mag_psf_differences}
\end{figure*}

For true point-like sources, the value of {\tt SPREAD\_MODEL} should be independent of their flux or their signal-to-noise ratio. Checking the relation between {\tt SPREAD\_MODEL} and {\tt SNR\_WIN} (also provided by {\tt SourceExtractor}) is therefore a powerful diagnostic not only for spotting spurious detections and performing a star/galaxy classification, but also for assessing the quality of the input PSF model. Such plots are shown in Fig.~\ref{figure:spread_models} for sources extracted from the 16 detectors of single-dither exposures taken in the \YE, \JE, and \HE passbands. According to these figures, compact sources would be distributed along the vertical area where $\abs{{\tt SPREAD\_MODEL}}$\,$<$\,\mbox{0.002--0.003}, while the concentration of the points where {\tt SPREAD\_MODEL}\,$>$\,\mbox{0.002--0.003} would be associated with extended galaxies. The region of points with negative values is sparsely populated because most of these sources were previously flagged as cosmic rays or flagged for other reasons and, therefore, removed from the catalogues.

From a more detailed inspection of Fig.~\ref{figure:spread_models}, we see that the point-like sources do not perfectly follow the expected vertical distribution around {\tt SPREAD\_MODEL}\,=\,0, especially for the \YE and \HE filters. For \JE, even if the median value calculated for these sources is close to zero, the distribution is not sufficiently symmetric. This is a clear indication that the derived PSF models still have some room for improvement. The models shown and discussed in the current work are based on a single visit of the SCF, from which data were processed using a certain version of the NIR processing tasks. The NIR pipelines are under continuous evolution, with the aim of better dealing with the flagging of spurious detections, the correction of small- and large-scale detector response and illumination, or the masking of unwanted detector effects, such as the persistence signal \citep[e.g.][]{DR1-TP038}. All these improvements might help to increase the number of reliable sources available for accurately modelling the PSF profile. A new tuning of the {\tt PSFEx} parameters is also required, based on the data processed more recently. In addition, a post-processing of the data would help as well, since it offers the possibility to combine data from different visits of the SCF, or data collected using different filters, thus accounting for possible wavelength dependence of the models.

The lack of a vertical distribution of the point-like sources plotted in Fig.~\ref{figure:spread_models} may also be related to the BFE, in which the width of the PSF profile is observed to be larger for brighter sources. This is more clearly seen in Fig.~\ref{figure:flux_radius_mag_aper} for the three \mbox{NISP-P} passbands, even for \YE and \JE despite the larger scatter of the points. The three panels show the running mean of the half-light diameter as a function of magnitude, individually calculated for the 16 detectors, where the half-light radius is the {\tt FLUX\_RADIUS} parameter provided by {\tt SourceExtractor}. The running mean is a convolution of the data with a fixed width Gaussian function ($\sigma$\,=\,1~mag), where the central value of the Gaussian varies in different magnitude bins. In the current version of the NIR PF, the BFE is not corrected as is done in the VIS PF \citep{Q1-TP002}. If not corrected, bright sources may have the flux slightly overestimated.

\subsection{\label{sec:psf_variab_valid} PSF variation and validation}

In the general scenario, we expect that the shape of the \mbox{NISP-P} PSF will depend on the position within the focal plane, the wavelength, and the time of observations. As stated in Sect.~\ref{sec:psf_task_inflight}, we set the {\tt PSFEx} parameters to produce models that are variable among the detectors but remain constant within each. This is because, when searching for the PSF model that could best represent the source profiles in each detector, we found no significant improvement in the models if derived with a certain dependence on position within the detectors.

The variation of the PSF profiles from detector to detector is small but detectable, as seen from the encircled energy curves plotted in the bottom panels of Fig.~\ref{figure:psf_profile_ee}. The differences are slightly larger, and therefore clearly seen when comparing the models produced from data observed at different times, i.e., at different visits of the SCF. This is what is shown in the left panels of Fig.~\ref{figure:ee_mag_psf_differences} for the 16 detectors of each \mbox{NISP-P} passband, where we compare the encircled energy curves of the PSF models derived using data processed in January and April 2025 (Table~\ref{table:data}). The right panels of the same figure highlight the differences between the two visits by comparing the PSF-fitting photometry generated by {\tt SourceExtractor}. To achieve this comparison, the code was executed on the same dither exposure but using as input the two distinct models.

Regarding the dependence on time of the PSF profiles, any change in shape is monitored by checking the encircled energy radii at 50\% and 80\%, the FWHM in both $x$ and $y$ directions, and the corresponding ellipticity $e$, which we defined as
\be
 e = 1 - \frac{{\rm FWHM}_{\rm minor}}{{\rm FWHM}_{\rm major}}\,,
\ee
where ${\rm FWHM}_{\rm minor}$ and ${\rm FWHM}_{\rm major}$ are the minimum and maximum values of FWHM estimated in the $x$ and $y$ directions in the focal plane, respectively. For this purpose, instead of using the extended models provided by the PSF calibration pipeline, we adopted the models with 61/6$\times$61/6 detector pixels derived from individual exposures using the NIR scientific pipeline. Although not accurately representing the diffraction patterns seen in the source profile of the true observations, these simplified models are good enough to provide a reliable estimate of the encircled energy radii and the FWHM. Moreover, with these models, we can also have access to the number of sources used in the reconstruction of the PSF profile. Therefore, since these parameters are calculated for each detector and for each dither exposure, they can be used to keep track of the quality and variation of the PSF model over time. Indeed, this is the aim of the DQC pipeline (Appendix~\ref{sec:appendix_dqc}) that is used by the NIR Organisation Unit (OU-NIR) to validate the \mbox{NISP-P} observations.

Figure~\ref{figure:psf_ellip_hist} shows how the ellipticity changes across the array of detectors for different \mbox{NISP-P} passbands, i.e., it shows the dependence of the ellipticity on the position in the FoV and on wavelength. The histograms include ellipticity values (estimated either from the PSF models or from point-like sources) for a set of EWS observations of three Patch IDs (Table~\ref{table:data}) that spanned more than 180~days. This means that these distributions also provide evidence for the temporal stability of the ellipticity, which remains within 0.05 for most of the PSF models and sources of the three filters. This is an example of how the fourth requirement in Sect.~\ref{sec:requirem} is verified and validated by OU-NIR.

\section{\label{sec:summary} Summary and final remarks}

In the present work, we have described the procedure adopted to model the PSF profile of point-like sources detected in NISP images observed in the photometric mode. Starting from models derived on the basis of Q1 data, we introduce the improvements that have been implemented in the PSF task of the NIR PF currently used for DR1. The key findings and conclusions of this paper are listed below.

\begin{description}

\item[(i)] The PSF models derived in Inflight mode from single-dither exposures clearly reproduce some diffraction patterns and optical aberrations commonly seen in astronomical images, such as rings, spikes, coma, and trefoil. For stacked frames, the PSF models are currently created with a smaller size and need to be updated for future data releases.

\item[(ii)] For single-dither exposures, the radii that enclose 50\% and 80\% of the energy (Table~\ref{table:psf_params}), as estimated from the PSF models derived in the \YE, \JE, and \HE passbands, are safely within the requirements presented in Sect.~\ref{sec:requirem}. These radii are also consistent with the experimental findings discussed in \citet{EuclidSkyNISP}, in which the values of rEE50 and rEE80 are compared with a diffraction-limited image delivered by the optics of the NISP instrument. For stacked frames, the PSF profiles are a bit broader, which is expected considering that they are a combination of multiple dithered exposures. As discussed in Sect.~\ref{sec:ee}, the PSF models derived from stacked images are still not optimal, and some refinement is required. Nevertheless, the values of rEE50 and rEE80 from stacked PSF models also comfortably satisfy the requirements in each \mbox{NISP-P} passband.

\item[(iii)] We adopted a PSF profile-fitting procedure compared to fixed- and automatic-aperture photometry to evaluate the quality of the derived PSF models. A new version of the catalogue extraction code combined with more detailed PSF models reduced the scatter in the comparison of PSF-fitting and aperture photometry. An offset is still present, which requires the application of a post-processing correction to the magnitudes or a modification of the PSF-fitting procedure. The mission requirement related to the centroid positions estimated during the PSF fitting is comfortably satisfied. As a provisional measure to account for the observed offset, we recommend applying a correction to the {\tt MAG\_PSF} values based on the results from fixed-aperture photometry.

\item[(iv)] The {\tt SPREAD\_MODEL} morphological classifier provided by {\tt SourceExtractor} also indicates that the models currently derived using the PSF calibration pipeline can be further improved. A new tuning of the {\tt PSFEx} parameters is required, using data processed with the most recent version of the NIR pipelines. A correction of the BFE \citep[e.g.][]{Q1-TP002} added to a post-processing of the data, based on multiple visits to the SCF and the combination of different filters, could also contribute to improving the current results.

\item[(v)] The DR1 PSF models are constant within each detector but variable among different detectors. Larger differences are seen when comparing the profile of models obtained from the data observed at different times. In compliance with mission requirements, the changes in position on the FoV, wavelength, and time of the encircled energy curves, FWHM, ellipticity, and number of sources are continuously evaluated by the NIR PF to monitor the variation and quality of the derived models. Figure~\ref{figure:psf_ellip_hist} shows how the distribution of ellipticity (estimated from PSF models or from point-like sources) changes with wavelength and with different positions in the FoV. Histograms were created using EWS observations of three Patch IDs (Table~\ref{table:data}) that spanned more than 180~days. Therefore, they also provide evidence for the temporal stability of the ellipticity.

\end{description}

Several improvements to the NIR pipelines are planned for upcoming data releases. These include a new tuning of {\tt PSFEx} parameters, particularly for PSF models derived from stacked images, alongside the modelling and correction of sub-pixel magnitude shifts. Additionally, the impact of the BFE on the NIR photometry should be investigated. Lastly, accounting for the wavelength dependence of the PSF profile by employing distinct models across different spectral types remains a recommended direction for future work.


\begin{acknowledgements}
We thank the internal reviewers for helpful comments on the manuscript.
\AckQone
\AckDRone
\AckEC
\end{acknowledgements}


%
%

\bibliography{Euclid, Q1, my} 


\begin{appendix}

\section{\label{sec:appendix_tables} Complementary tables}

This section includes additional tables that provide more detailed information on the data and parameters used in this study. Table~\ref{table:data} specifies which PSF models, images, and catalogues were used to present the current results for single and stacked cases. Table~\ref{table:psfex_tuning} lists the {\tt PSFEx} parameters provided by the tuning procedure. Table~\ref{table:sextractor_conf} lists the {\tt SourceExtractor} parameters that are different from the default values. Finally, Table~\ref{table:sextractor_psfex_params} gives a description of the configuration parameters and the catalogue keywords discussed throughout the paper.

\section{\label{sec:appendix_ips_ipc} IPS and IPC kernels}

The NIR scientific pipeline executed in MDB mode requires as input the IPS and IPC kernels to be convolved with the library of MDB PSF models (Sect.~\ref{sec:psf_task_mdb}). The IPS kernel was defined using R distributions from the {\tt SciPy} library\footnote{\url{https://docs.scipy.org/doc/scipy/reference/generated/scipy.stats.rdist.html}}. We adopted a function with a parabolic shape ($c$\,=\,4) to construct a 6$\times$6 grid of sub-pixels, which is normalised in such a way as to have mean\,=\,1. The standard deviation is about 0.0325 and the function drop is about 8.5\%. These values were set on the basis of some tests that compared the PSF profiles derived from the images with the MDB PSF profiles after the convolution. The IPC kernel was estimated by the NISP instrument team and stored in a file as a grid of 3$\times$3 detector pixels. Figure~\ref{figure:ips_ipc_kernels} shows the adopted kernels with the IPS response function (left) and the IPC coupling matrix (right).

\section{\label{appendix:psfex_tuning} Tuning of {\tt PSFEx} parameters}

The shape of the derived PSF models is particularly sensitive to some of the {\tt PSFEx} parameters. Some of them are kept fixed, whereas others require a tuning in order to set their optimal values. The PSF profile is modelled as a linear combination of basis vectors, also known as polar shapelets \citep{MasseyRefregier2005}. The corresponding coefficients are estimated by fitting the image of every selected source with the PSF model through a global minimisation of $\chi^2$. The width of the PSF profile and the number of basis vectors are iteratively changed, and the reconstruction of the PSF core and wings is continuously evaluated. The authors calculate the $\chi^2$ values from the difference between the derived models and an observed image. In the case of NISP images, given the strong undersampling of the PSF \citep{EuclidSkyNISP}, the reconstruction process must rely on a large number of undersampled image stamps at different positions on the FPA. Therefore, to evaluate the goodness of the derived PSF models, we adopted a different approach in which we compared the profile fitting and aperture photometry provided by {\tt SourceExtractor}. During the tuning process, our aim is to minimise the scatter and slope of {\tt MAG\_PSF} minus {\tt MAG\_APER} as a function of {\tt MAG\_APER} (Fig.~\ref{figure:hist_mag_psf_aper}) by changing the values of {\tt BASIS\_SCALE} (to set the width of the PSF profile) and {\tt BASIS\_NUMBER} (to set the number of basis vectors).

The slope of the linear regression in the plot of {\tt MAG\_PSF} versus {\tt MAG\_APER} is particularly affected by the value of the {\tt PSF\_ACCURACY} parameter, which is used to set the expected accuracy of the stamp pixels. Therefore, during tuning, we preferred to keep {\tt PSF\_ACCURACY} as a free parameter until no slope was observed. Moreover, this parameter also has an impact on the offset between {\tt MAG\_PSF} and {\tt MAG\_APER} photometry, since it affects the normalisation of the PSF models, which in turn are used during the profile-fitting procedure (see Sect.~\ref{sec:phot} for a more detailed discussion regarding the offset).

Table~\ref{table:psfex_tuning} lists the parameters of {\tt PSFEx} adopted to derive the PSF models in the NIR scientific pipeline (for single-dither exposures and stacked frames) and in the PSF calibration pipeline (for single-dither exposures). {\tt PSF\_SAMPLING} is set to 1/6 to account for the oversampling of the derived models, while {\tt PSF\_SIZE} is set to produce 61/6$\times$61/6 and 141/6$\times$141/6 detector pixel stamps for the scientific and calibration pipelines, respectively. Together with {\tt BASIS\_NUMBER}, this table also shows the number of basis vectors used by {\tt PSFEx} to model the PSF profile, which is calculated using
\be
\label{eq:n_max}
N_{\rm max} = \frac{(n_{\rm max}+1)\,(n_{\rm max}+2)}{2}\,,
\ee
where $n_{\rm max}$ is the {\tt BASIS\_NUMBER} parameter. The larger $N_{\rm max}$, the larger the level of detail of the models derived. However, we note that modelling the PSF profile using a large number of basis vectors requires a large number of sources. This is the reason why {\tt BASIS\_NUMBER}, or $N_{\rm max}$, has large values in Table~\ref{table:psfex_tuning} only for the PSF calibration pipeline, which is executed using data from SCF observations. In this case, about 50 sources are normally available for each detector and for each of the 76 dither exposures, resulting in thousands of available sources observed at different positions on the focal plane. For comparison, in the case of the NIR scientific pipeline, normally executed using data from EWS observations, the number of good (unflagged) point-like sources from the four dither exposures together is rarely greater than 100.

The selection of point-like sources is performed using the ranges in {\tt SAMPLE\_FWHMRANGE} and is based on the distribution of the {\tt FLUX\_RADIUS} parameter. This is defined as the radius that encompasses a certain fraction of the total flux. To obtain the half-light radius that contains half of the total flux, we need to set {\tt PHOT\_FLUXFRAC}\,=\,0.5. The left panels of Fig.~\ref{figure:hist_mag_psf_aper} show, for the three \mbox{NISP-P} passbands, the distributions of $2\times${\tt FLUX\_RADIUS}, which is a good estimate for {\tt SAMPLE\_FWHMRANGE}. A first estimate of these ranges was adopted to derive the PSF models in the NIR scientific pipeline, and they were later revised to be used in the PSF calibration pipeline. For the stacked case, we also used similar histograms to set those ranges, which are based on sources extracted from the stacked frames. The {\tt SAMPLE\_MAXELLIP} parameter is also used to select point-like objects. It is set to 0.05 for single exposures, while we kept the default value of 0.2 for stacked frames, helping to exclude very elliptical sources.

Regarding the flagging of spurious detections, {\tt PSFEx} uses the parameters {\tt SAMPLE\_FLAGMASK} and {\tt SAMPLE\_IMAFLAGMASK}, which are set to bitmasks 0$\times$00fe and 0$\times$0001, respectively. The former is used to exclude sources internally flagged by {\tt SourceExtractor} (truncated objects, blends, etc.), whereas the latter accounts for the exclusion of detections flagged as {\tt INVALID} by the NIR pipelines.

Finally, it is worth mentioning that the central positions of the sources used for PSF modelling are set through the keyword {\tt CENTER\_KEYS}. For this, we adopted the parameters {\tt XWIN\_IMAGE} and {\tt YWIN\_IMAGE}, which provide a more precise estimate of the source centroids compared to the default values. For other parameters used by {\tt PSFEx} that are not listed in Table~\ref{table:psfex_tuning}, or are not mentioned in the text, we adopted the values predefined internally by the code.

\section{\label{sec:appendix_extended_psf} Extended PSF models and their application to PSF-fitting photometry}

Additional diagnostic plots are provided here to illustrate further results from the PSF-fitting photometry. Figure~\ref{figure:psf_models} shows two-dimensional representations of the \mbox{NISP-P} PSF models derived for each detector of a dither exposure from SCF data observed in the three passbands. Figure~\ref{figure:mag_psf_aper_offset} illustrates how the magnitude changes for sources observed in different dither exposures, i.e., having centroids in different sub-pixel positions. The figure compares the results for {\tt MAG\_APER}, which remains quite constant across the surface of the pixel, with {\tt MAG\_PSF}, for which the differences increase when going from the centre to the edges of a pixel. This means that {\tt MAG\_PSF} becomes underestimated (i.e. the sources are brighter than expected) compared to {\tt MAG\_APER}, which might explain the offset discussed in Sect.~\ref{sec:phot_offset} and also seen in the left panels of Fig.~\ref{figure:hist_mag_psf_aper}.

\section{\label{sec:appendix_dqc} The NIR DQC pipeline}

As mentioned in Sect.~\ref{sec:psf_variab_valid}, the validation of the images acquired in photometric mode by the NISP instrument is performed using the NIR DQC pipeline. Several parameters, such as FWHM, ellipticity, and encircled energy radii, can be extracted from the data products to examine their variation over position, wavelength, and time. Figure~\ref{figure:psf_ellip_hist} is an example that shows the distribution of the ellipticity in different NISP detectors for the three passbands. In each panel, the histograms compare the ellipticity estimated using two different approaches: in one case, the ellipticity is calculated from the simplified PSF models by measuring the semi-minor and semi-major axes of the profile; in the other case, the ellipticity is provided by {\tt SourceExtractor}, which internally measures the profiles of the extracted sources (only point-like sources are considered).

\clearpage

\renewcommand{\thesection}{A}
\setcounter{table}{0}
\renewcommand{\thetable}{A.\arabic{table}}

\begin{table*}
\caption{DR1 observations used in the present paper.}
\centering
\smallskip
\label{table:data}
\smallskip
{\small
\begin{tabular}{lcccl}
\hline\hline\noalign{\vskip 2pt}
Calblock ID & Processing date & Patch ID & Observation ID & Data product \\
\noalign{\vskip 2pt}\hline\noalign{\vskip 2pt}
CALBLOCK-F-001 & 2025-01-24 & 41  & 2174        & extended PSF models \\
CALBLOCK-F-001 & 2025-02-04 & 131 & 6185        & single-dither exposures and catalogues \\
CALBLOCK-F-001 & 2025-04-12 & 8   & 519         & extended PSF models \\
CALBLOCK-F-001 & 2025-11-09 & 255 & 10614       & single-dither catalogues \\
\noalign{\vskip 5pt}
CALBLOCK-F-005 & 2025-04-23 & 56  & 2955        & stacked frames and catalogues \\
CALBLOCK-F-005 & 2025-04-23 & 56  & 2955...2971 & single-dither exposures \\
CALBLOCK-F-005 & 2025-04-23 & 98  & 4876...4983 & single-dither exposures \\
CALBLOCK-F-005 & 2025-04-25 & 129 & 6125...6141 & single-dither exposures \\
\hline
\end{tabular}}
\tablefoot{Some observations were not included in our analysis because they were not available in the \Euclid science archive. The extended PSF models were derived using the PSF calibration pipeline applied to SCF observations.}
\end{table*}

\begin{table*}
\caption{Parameters in the {\tt PSFEx} configuration file used by the PSF task in the NIR scientific and PSF calibration pipelines.}
\centering
\smallskip
\label{table:psfex_tuning}
\smallskip
{\small
\begin{tabular}{l@{\hskip 2em} ccc@{\hskip 2em} ccc@{\hskip 2em} ccc}
\hline\hline\noalign{\vskip 2pt}
 & \multicolumn{6}{c@{\hskip 2em}}{NIR scientific pipeline}\hskip 1em & \multicolumn{3}{c}{NIR PSF calibration pipeline} \\
\noalign{\vskip 2pt}\hline\noalign{\vskip 2pt}
 & \multicolumn{3}{c@{\hskip 2em}}{Single exposure} & \multicolumn{3}{c@{\hskip 2em}}{Stacked frame} & \multicolumn{3}{c}{Single exposure} \\
\noalign{\vskip 2pt}\hline\noalign{\vskip 2pt}
{\tt PSFEx} parameter & \YE & \JE & \HE & \YE & \JE & \HE & \YE & \JE & \HE \\
\noalign{\vskip 2pt}\hline\noalign{\vskip 2pt}
{\tt BASIS\_NUMBER}     ($N_{\rm max}$) & 5 (21)    & 9 (55)    & 5 (21)    & 9 (55)    & 9 (55)    & 7 (36)    & 36 (703)  & 40 (861)  & 40 (861)  \\
{\tt BASIS\_SCALE}                      & 1.40      & 1.38      & 1.70      & 2.10      & 2.05      & 2.10      & 1.35      & 1.40      & 1.57      \\
{\tt PSF\_ACCURACY}                     & 0.09      & 0.14      & 0.12      & 0.01      & 0.01      & 0.01      & 0.20      & 0.20      & 0.17      \\
{\tt SAMPLE\_FWHMRANGE}                 & 1.45,1.85 & 1.55,1.95 & 1.75,2.05 & 1.65,1.90 & 1.75,2.05 & 1.95,2.20 & 1.45,1.70 & 1.55,1.80 & 1.82,1.98 \\
{\tt SAMPLE\_MINSN}                     & 50        & 40        & 30        & 30        & 30        & 30        & 30        & 30        & 40        \\
\hline
\end{tabular}}
\tablefoot{During the tuning procedure, {\tt BASIS\_NUMBER}, {\tt BASIS\_SCALE}, and {\tt PSF\_ACCURACY} are free parameters while the other two are kept fixed. $N_{\rm max}$ is related to {\tt BASIS\_NUMBER} according to Eq.~(\ref{eq:n_max}) and represents the number of basis vectors used to model the PSF profile. {\tt SAMPLE\_FWHMRANGE} must be written as string, with the values in pixels separated by a comma. A description of these parameters is available in Table~\ref{table:sextractor_psfex_params}.}
\end{table*}

\begin{table*}
\caption{Parameters in the {\tt SourceExtractor} configuration file used by the PSF task in the NIR pipelines.}
\centering
\smallskip
\label{table:sextractor_conf}
\smallskip
{\small
\begin{tabular}{l@{\hskip 2em} cc}
\hline\hline\noalign{\vskip 2pt}
{\tt SourceExtractor} parameter & Value & Unit \\
\noalign{\vskip 2pt}\hline\noalign{\vskip 2pt}
{\tt CATALOG\_TYPE}    & FITS\_LDAC    &  \\
{\tt DETECT\_MINAREA}  & 4             & pixel \\
{\tt DETECT\_MAXAREA}  & 0 (unlimited) & pixel \\
{\tt DETECT\_THRESH}   & 3             & $\sigma$ \\
{\tt ANALYSIS\_THRESH} & 1.5           & $\sigma$ \\
{\tt RESCALE\_WEIGHTS} & N             &  \\
{\tt WEIGHT\_GAIN}     & N             &  \\
{\tt BACK\_TYPE}       & AUTO          &  \\
{\tt BACKPHOTO\_TYPE}  & LOCAL         &  \\
{\tt PHOT\_APERTURES}  & 6             & pixel \\
{\tt SATUR\_LEVEL}     & 64511         & ADU \\
{\tt SATUR\_KEY}       & NONE          &  \\
{\tt GAIN}             & 2             & ${\rm e}^-{\rm ADU}^{-1}$ \\
{\tt GAIN\_KEY}        & NONE          &  \\
\hline
\end{tabular}}
\tablefoot{These values are valid for the three \mbox{NISP-P} passbands. For the parameters not listed in this table, the default values are adopted. A description of these parameters is available in Table~\ref{table:sextractor_psfex_params}.}
\end{table*}

\begin{table*}
\caption{Description of the {\tt SourceExtractor} and {\tt PSFEx} parameters discussed in the present work.}
\centering
\smallskip
\label{table:sextractor_psfex_params}
\smallskip
{\small
\begin{tabular}{ll}
\hline\hline\noalign{\vskip 2pt}
Parameter & Definition \\
\noalign{\vskip 2pt}\hline\noalign{\vskip 2pt}
{\tt ANALYSIS\_THRESH}    & Surface brightness threshold used for source detection and classification \\
{\tt BACK\_TYPE}          & Type of subtracted background: internal map (AUTO) or user-supplied value (MANUAL) \\
{\tt BACKPHOTO\_TYPE}     & Type of computed background: from the internal map (GLOBAL) or recomputed around each object (LOCAL) \\
{\tt CATALOG\_TYPE}       & Format of output catalogue (ASCII, ASCII\_HEAD, FITS\_LDAC, etc.) \\
{\tt DETECT\_MAXAREA}     & Maximum number of pixels above threshold triggering detection \\
{\tt DETECT\_MINAREA}     & Minimum number of pixels above threshold triggering detection \\
{\tt DETECT\_THRESH}      & Detection threshold \\
{\tt GAIN}                & Detector conversion factor [e-/ADU] \\
{\tt GAIN\_KEY}           & Keyword for detector conversion factor [e-/ADU] \\
{\tt PHOT\_APERTURES}     & Diameter(s) for aperture photometry [pixel] \\
{\tt PHOT\_FLUXFRAC}      & Flux fraction(s) used for {\tt FLUX\_RADIUS} \\
{\tt RESCALE\_WEIGHTS}    & Set if input weights/variances should be rescaled (Y/N) \\
{\tt SATUR\_KEY}          & Keyword for saturation level [ADU] \\
{\tt SATUR\_LEVEL}        & Value above which a pixel is considered saturated [ADU] \\
{\tt WEIGHT\_GAIN}        & Set if gain should be modulated with weights (Y/N) \\
\noalign{\vskip 5pt}
{\tt BASIS\_NUMBER}       & Number of basis vectors used to model the PSF profile \\
{\tt BASIS\_SCALE}        & The $\beta$ parameter of the Gauss-Laguerre function \\
{\tt CENTER\_KEYS}        & Catalogue parameters for source pre-centring \\
{\tt PSF\_ACCURACY}       & Expected accuracy of the pixel values \\
{\tt PSF\_SAMPLING}       & Sampling step of the PSF models [pixel] \\
{\tt PSF\_SIZE}           & Image size of the PSF models \\
{\tt PSFVAR\_DEGREES}     & Degree of the polynomial function used to model the PSF variations across the FoV \\
{\tt SAMPLE\_FLAGMASK}    & Bit mask applied to the {\tt SourceExtractor} {\tt FLAGS} parameter \\
{\tt SAMPLE\_FWHMRANGE}   & Allowed FWHM range [pixel] \\
{\tt SAMPLE\_IMAFLAGMASK} & Bit mask applied to the {\tt SourceExtractor} {\tt IMAFLAGS\_ISO} parameter \\
{\tt SAMPLE\_MAXELLIP}    & Maximum ellipticity for a source to be used \\
{\tt SAMPLE\_MINSN}       & Minimum signal-to-noise ratio for a source to be used \\
\noalign{\vskip 5pt}
{\tt FLAGS}               & Internal extraction flags \\
{\tt IMAFLAGS\_ISO}       & External image flags \\
{\tt MAG\_APER}           & Fixed-aperture magnitude vector [mag] \\
{\tt MAG\_PSF}            &	Magnitude from PSF-fitting [mag] \\
{\tt MAGERR\_PSF}         & RMS magnitude error from PSF-fitting [mag] \\
{\tt FLUX\_APER}          & Flux vector within fixed circular aperture(s) [count] \\
{\tt FLUX\_RADIUS}        & Radius containing the fraction of light set by {\tt PHOT\_FLUXFRAC} [pixel] \\
{\tt SNR\_WIN}            & Gaussian-weighted signal-to-noise ratio \\
{\tt SPREAD\_MODEL}       & Spread parameter from model-fitting \\
{\tt VIGNET}              & Size of the image stamps [pixel] \\
{\tt XWIN\_IMAGE}         & Windowed position estimate along the $x$-axis [pixel] \\
{\tt YWIN\_IMAGE}         & Windowed position estimate along the $y$-axis [pixel] \\
\hline
\end{tabular}}
\tablefoot{The first two set of rows refer to parameters set in the {\tt SourceExtractor}\footnote{User's manual available at \url{https://astromatic.github.io/sextractor/}} and {\tt PSFEx}\footnote{User's manual available at \url{http://psfex.readthedocs.io/}} configuration files, respectively. The last set of rows represent the name of the columns in the {\tt SourceExtractor} output catalogue.}
\end{table*}

\clearpage

\renewcommand{\thesection}{B}
\setcounter{figure}{0}
\renewcommand{\thefigure}{B.\arabic{figure}}

\begin{figure*}
\sidecaption
\begin{minipage}[t]{0.33\textwidth}
\centering
\resizebox{\hsize}{!}{\includegraphics{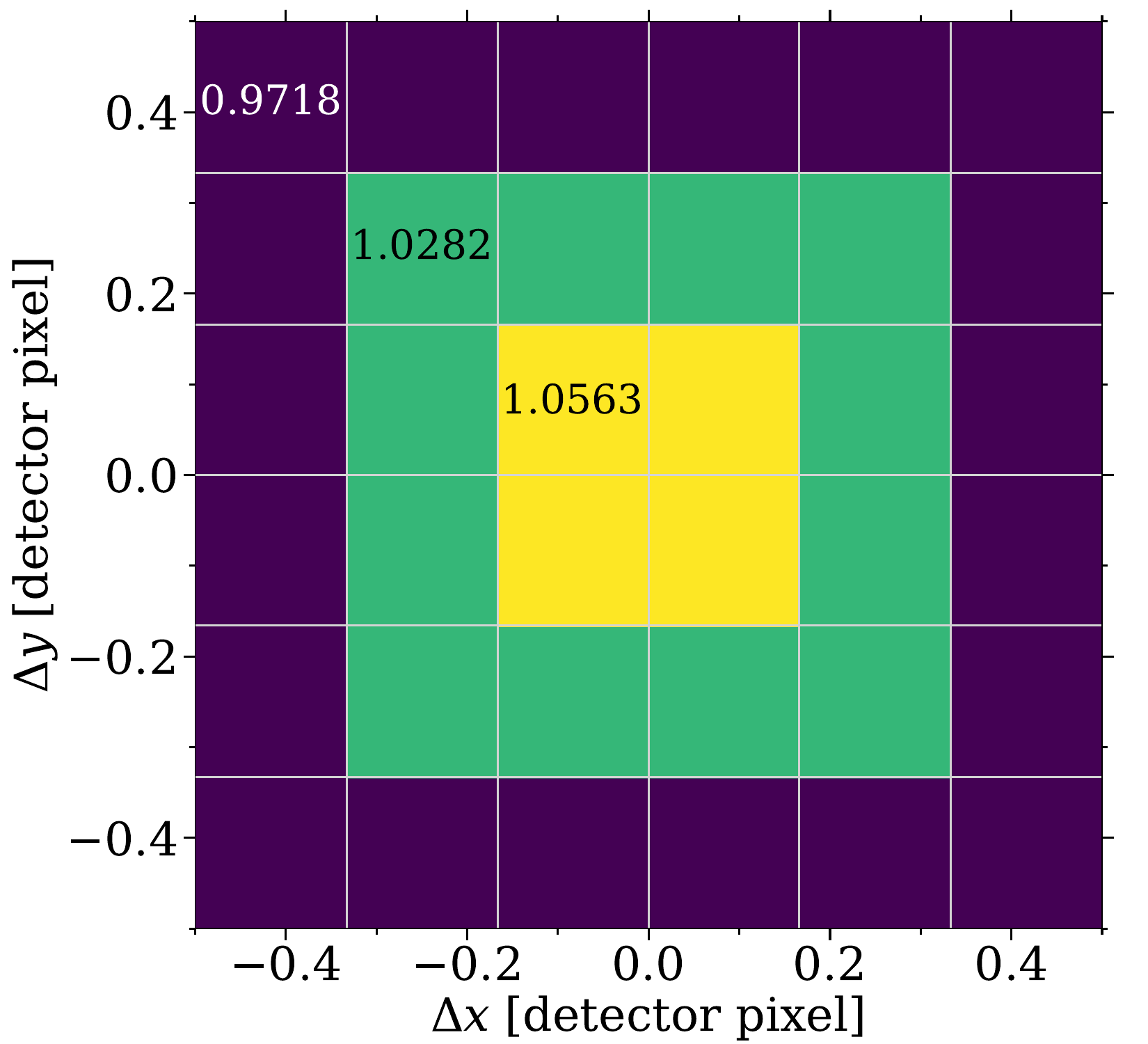}}
\end{minipage}
\begin{minipage}[t]{0.33\textwidth}
\centering
\resizebox{\hsize}{!}{\includegraphics{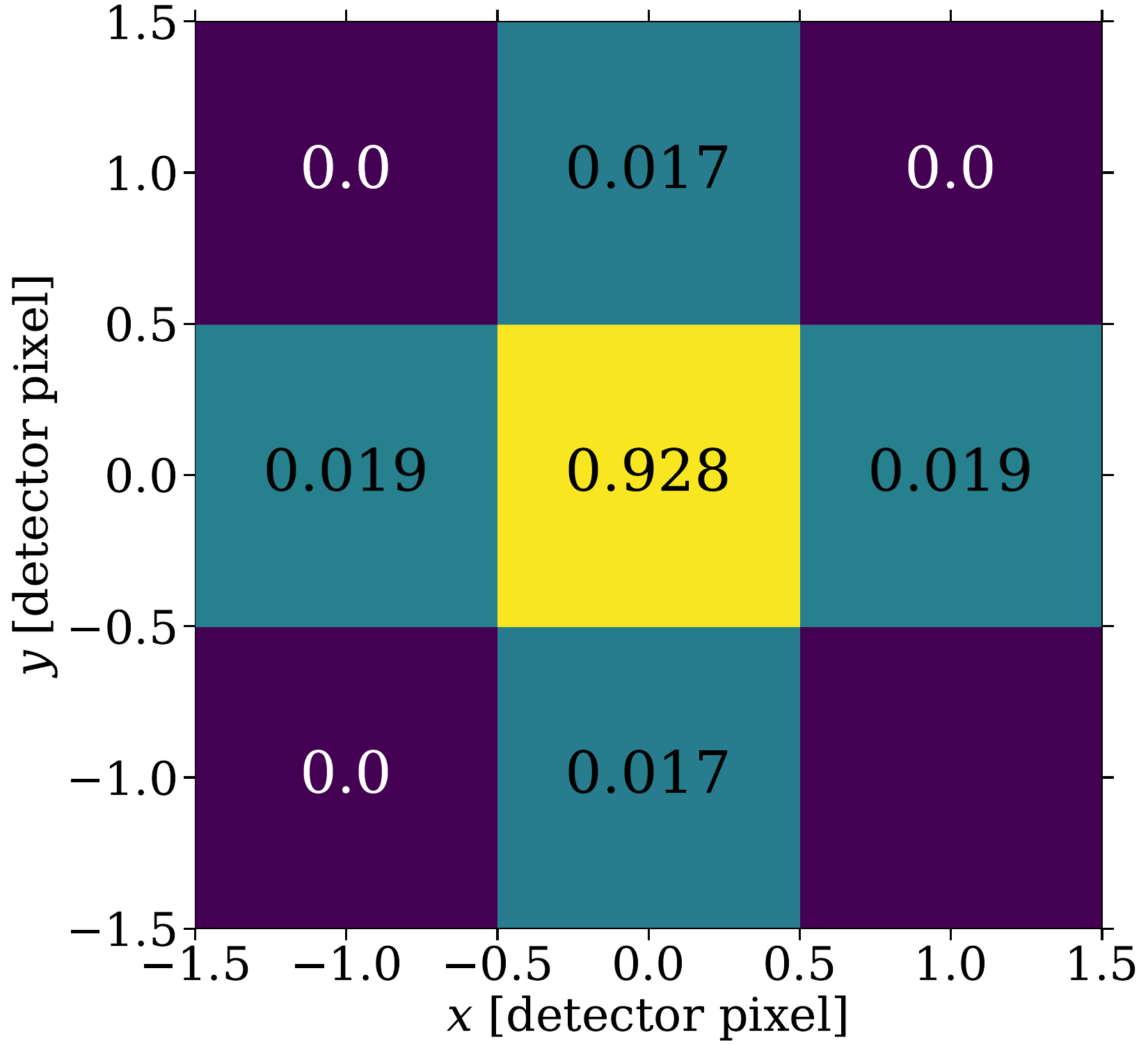}}
\end{minipage}
\caption{IPS (\emph{left}) and IPC (\emph{right}) kernels adopted when executing the NIR scientific pipeline in MDB mode. The values on the left panel indicate the sub-pixel response function (mean\,=\,1) while those on the right represent the fraction of the signal that reaches the central and adjacent pixels (sum\,=\,1).}
\label{figure:ips_ipc_kernels}
\end{figure*}

\renewcommand{\thesection}{C}
\setcounter{figure}{0}
\renewcommand{\thefigure}{C.\arabic{figure}}

\begin{figure*}
\centering
\begin{minipage}[t]{\textwidth}
\centering
\resizebox{\hsize}{!}{\includegraphics{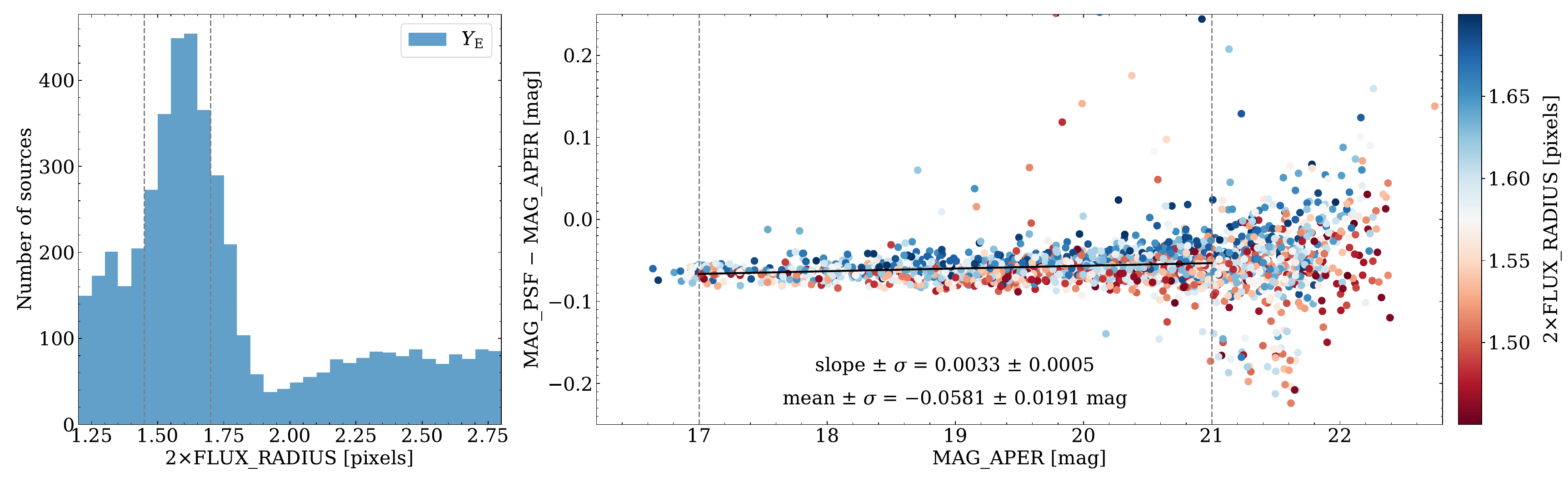}}
\end{minipage} \\
\begin{minipage}[t]{\textwidth}
\centering
\resizebox{\hsize}{!}{\includegraphics{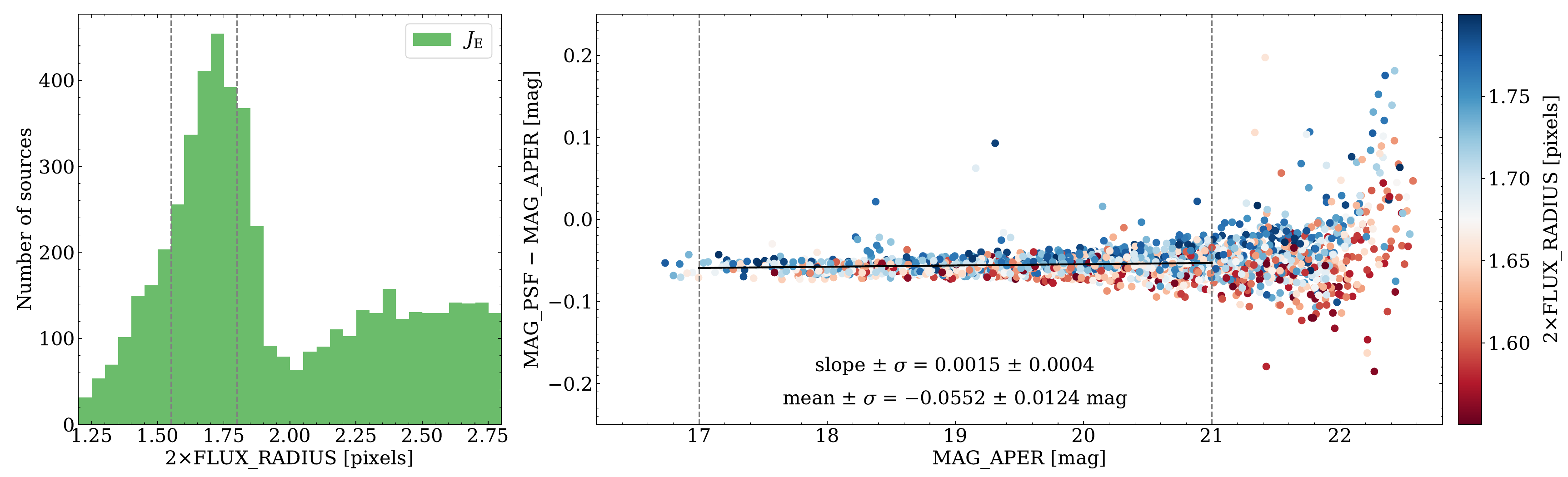}}
\end{minipage} \\
\begin{minipage}[t]{\textwidth}
\centering
\resizebox{\hsize}{!}{\includegraphics{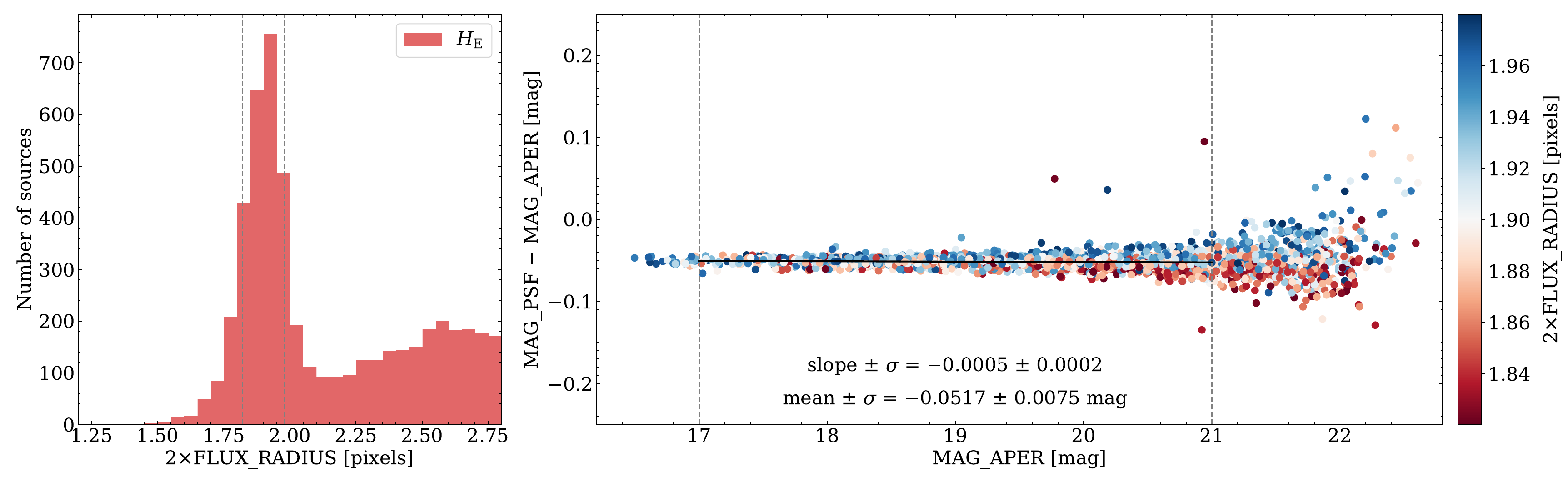}}
\end{minipage}
\caption{Comparison between PSF-fitting and aperture photometry for the three \mbox{NISP-P} passbands (Observation ID 6185). {\it Left column:} Distribution of the half-light diameter, indicating the ranges (between the dashed lines) of point-like sources used by {\tt PSFEx} for PSF modelling. {\it Right column:} Difference between {\tt MAG\_PSF} and {\tt MAG\_APER}, for point-like sources only, colour-coded according to the half-light diameter. The mean difference and the slope of the linear regression (black line) for magnitudes between 17 and 21 are also shown.}
\label{figure:hist_mag_psf_aper}
\end{figure*}

\renewcommand{\thesection}{D}
\setcounter{figure}{0}
\renewcommand{\thefigure}{D.\arabic{figure}}

\begin{figure*}
\centering
\begin{minipage}[t]{0.33\textwidth}
\centering
\resizebox{\hsize}{!}{\includegraphics{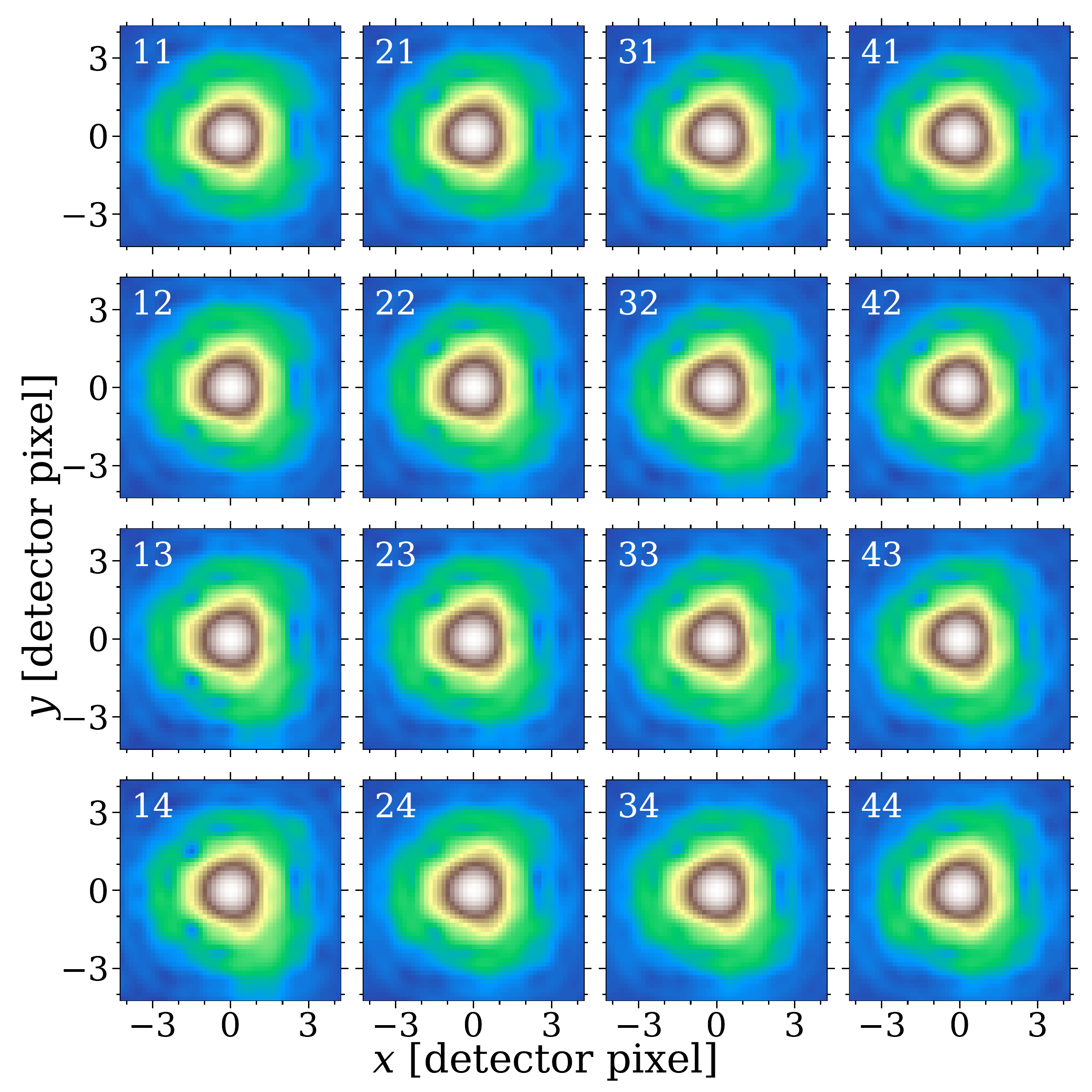}}
\end{minipage}
\begin{minipage}[t]{0.33\textwidth}
\centering
\resizebox{\hsize}{!}{\includegraphics{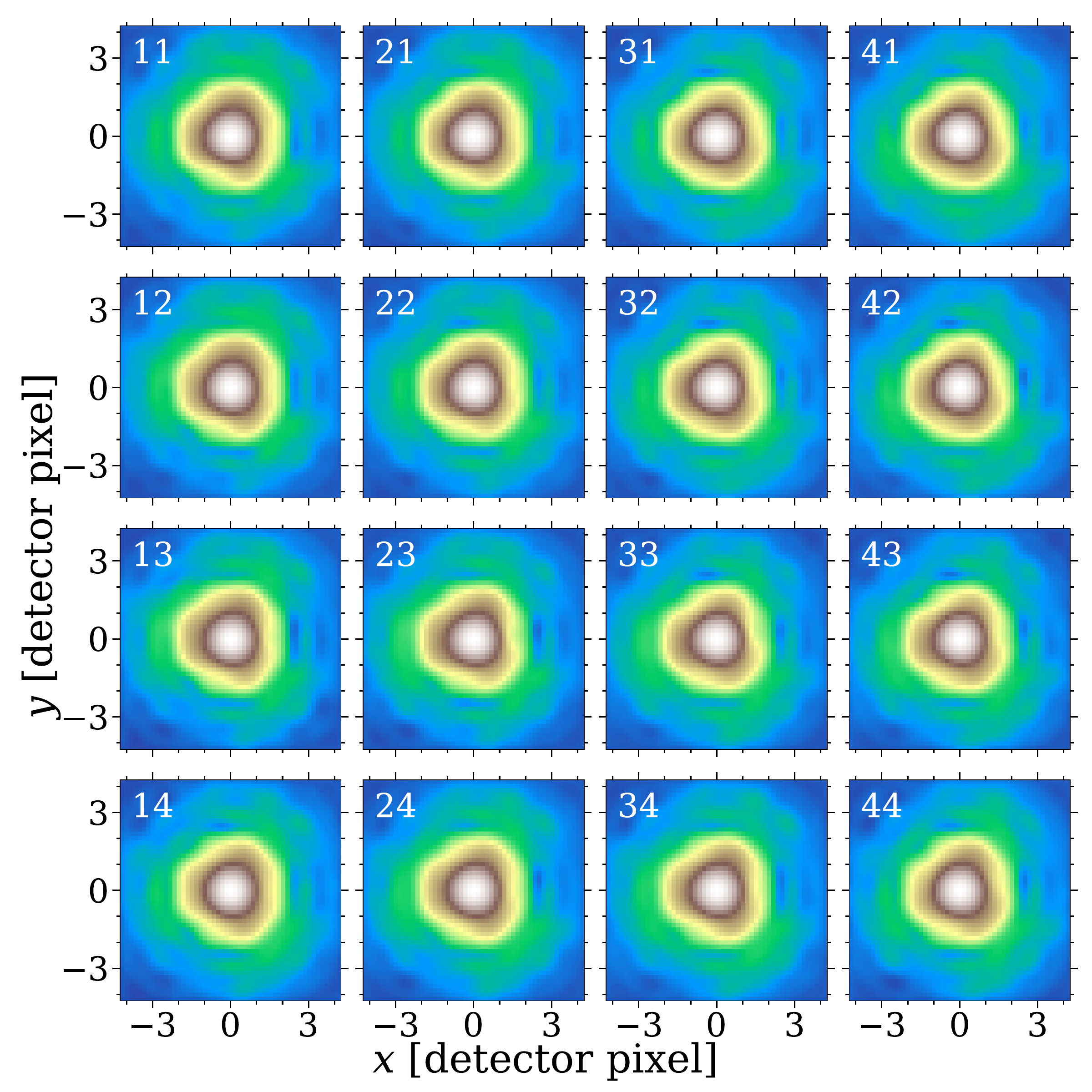}}
\end{minipage}
\begin{minipage}[t]{0.33\textwidth}
\centering
\resizebox{\hsize}{!}{\includegraphics{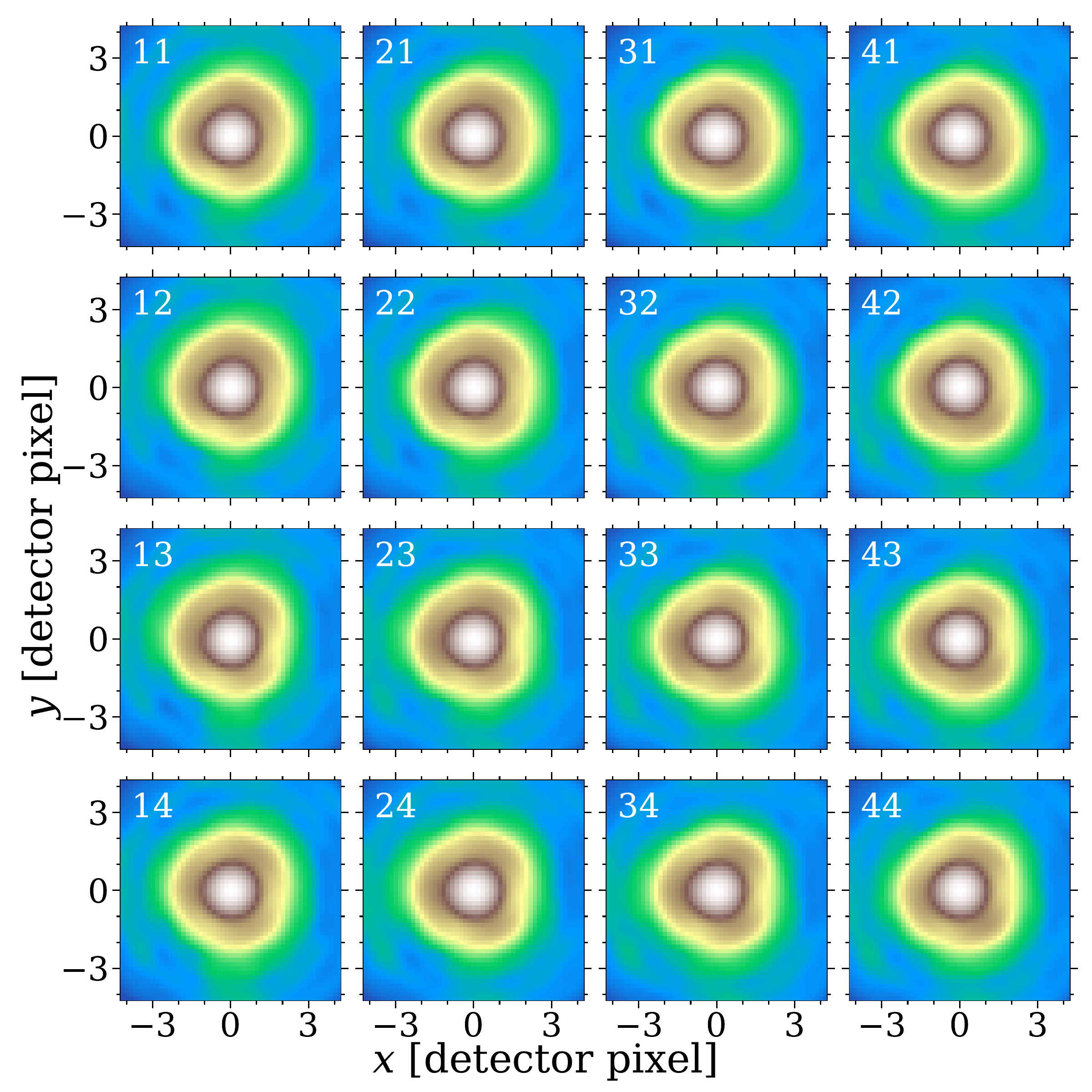}}
\end{minipage} \\
\begin{minipage}[t]{\textwidth}
\hspace*{16pt}
\resizebox{0.96\hsize}{!}{\includegraphics{figures/psf_model_image_vignet_colorbar.pdf}}
\end{minipage}
\caption{Two-dimensional representations of \mbox{NISP-P} extended PSF models derived from SCF data (Observation ID 6185) for each of the 16 NISP detectors. The different panels are for the \YE (\emph{left}), \JE (\emph{middle}), and \HE (\emph{right}) passbands. These are cut-outs (51/6$\times$51/6 detector pixels) of the same models shown in Fig.~\ref{figure:psf_models_vignets} (\emph{top panels}) for Detector ID {\tt 11} but also showing the other detectors.}
\label{figure:psf_models}
\end{figure*}

\begin{figure*}
\centering
\begin{minipage}[t]{0.49\textwidth}
\centering
\resizebox{\hsize}{!}{\includegraphics{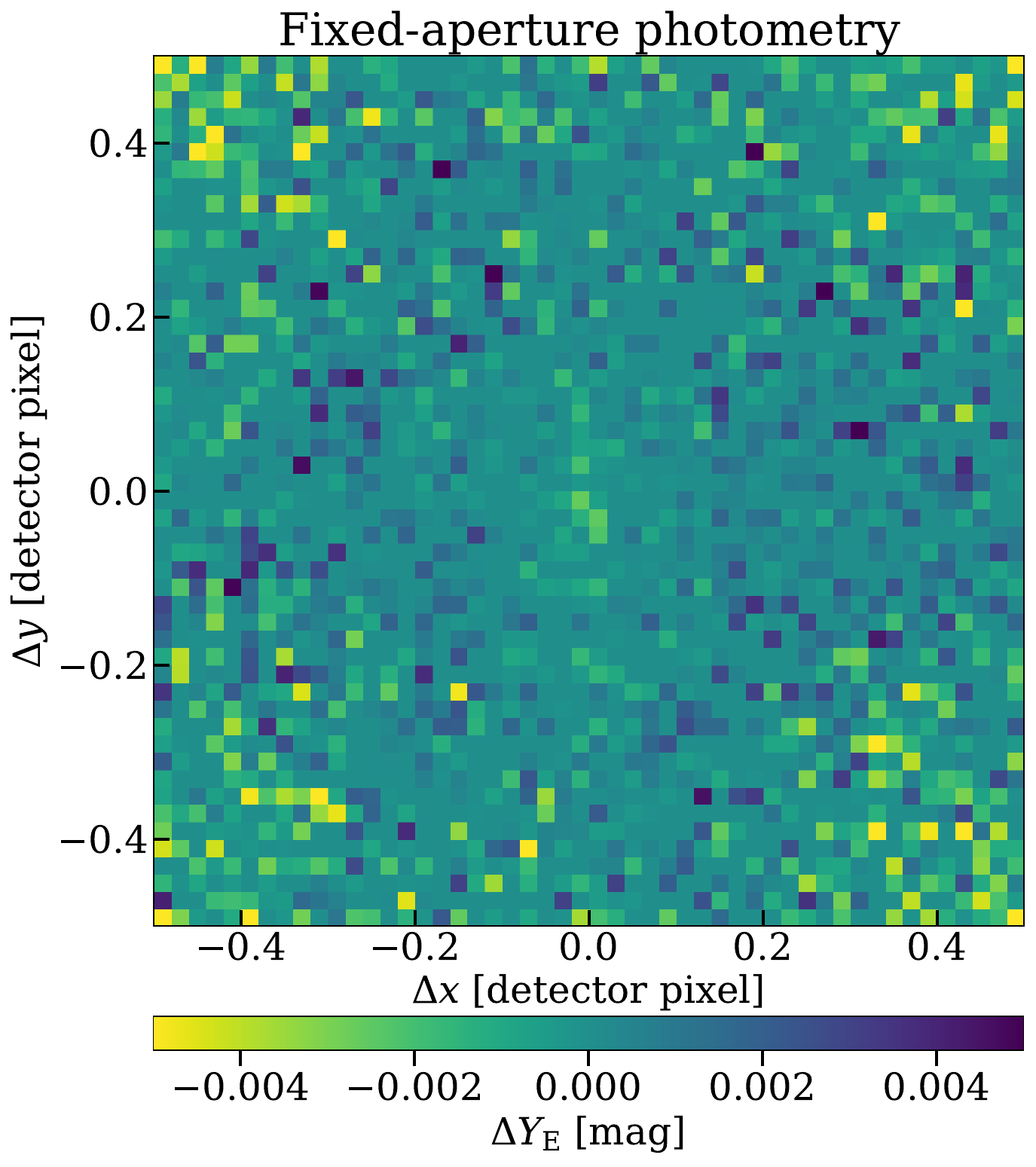}}
\end{minipage}
\begin{minipage}[t]{0.49\textwidth}
\centering
\resizebox{\hsize}{!}{\includegraphics{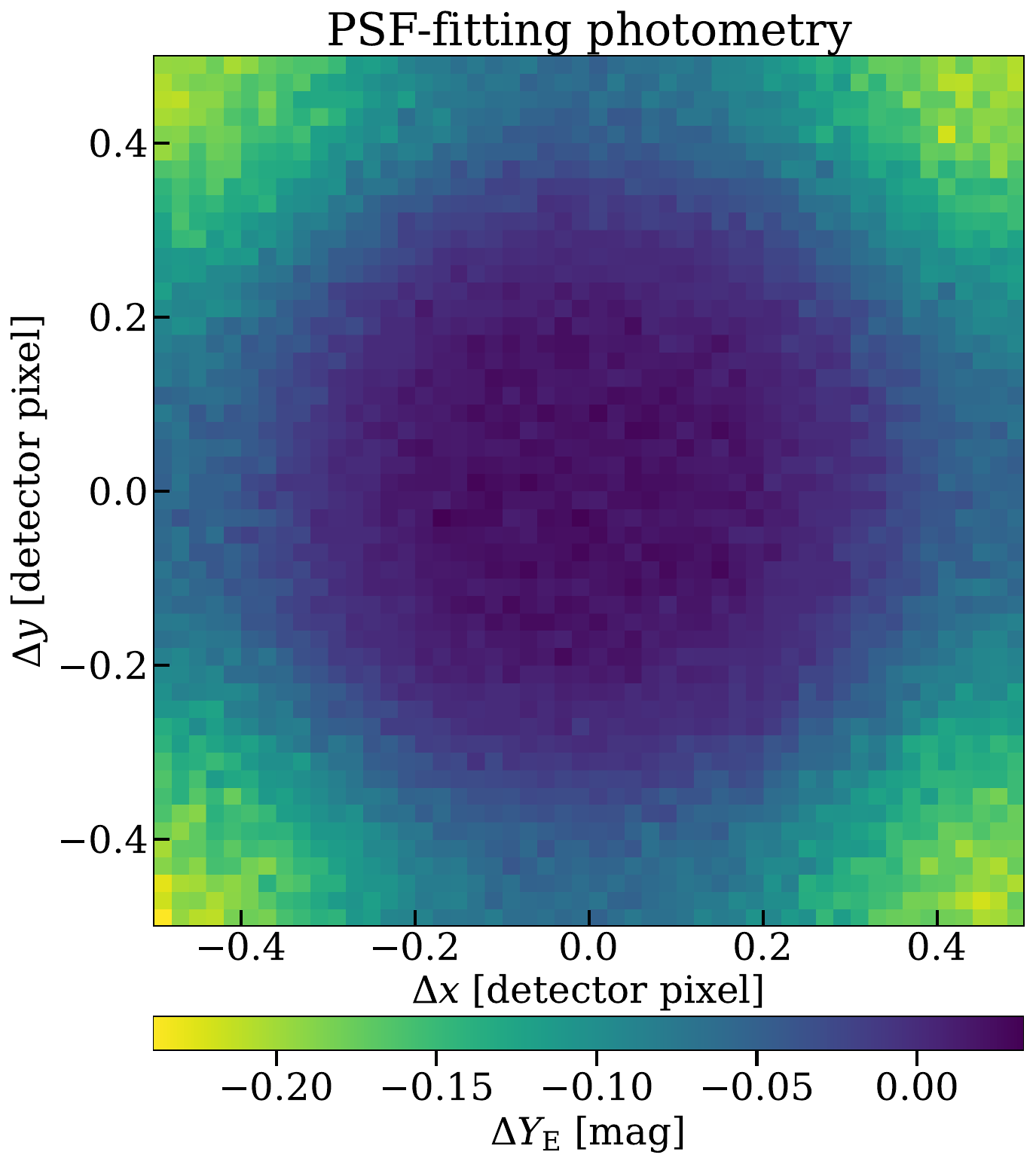}}
\end{minipage}
\caption{Magnitude shift as a function of $x$ and $y$ sub-pixel positions. The panels show the results for {\tt MAG\_APER} (\emph{left}) and {\tt MAG\_PSF} (\emph{right}) photometry for sources observed in at least ten different dither exposures of the SCF (Observation ID 10614). For each source in each dither, $\Delta$\YE represents the difference in magnitude with respect to the median value calculated for the same source.}
\label{figure:mag_psf_aper_offset}
\end{figure*}

\renewcommand{\thesection}{E}
\setcounter{figure}{0}
\renewcommand{\thefigure}{E.\arabic{figure}}

\begin{figure*}
\centering
\begin{minipage}[t]{0.33\textwidth}
\centering
\resizebox{\hsize}{!}{\includegraphics{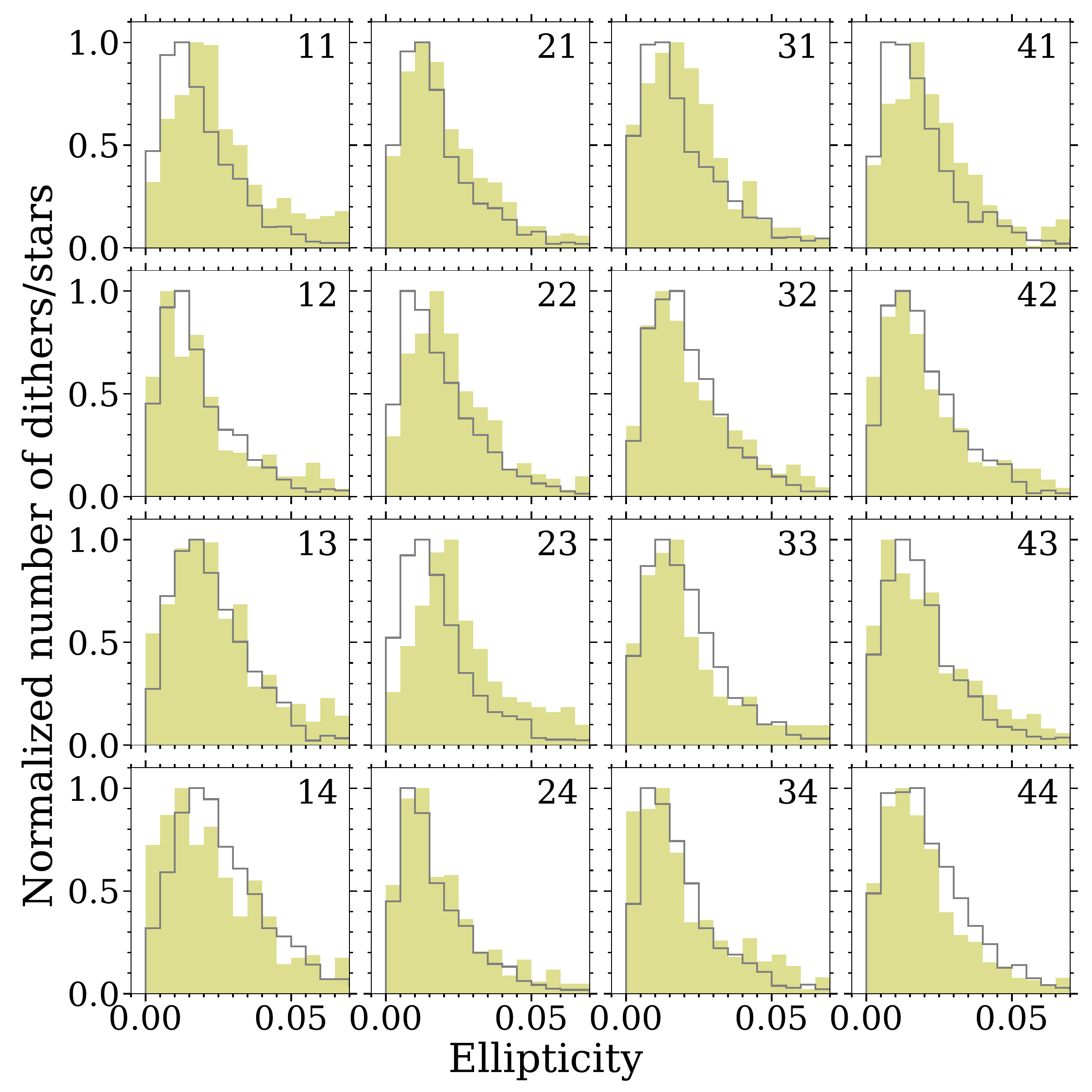}}
\end{minipage}
\begin{minipage}[t]{0.33\textwidth}
\centering
\resizebox{\hsize}{!}{\includegraphics{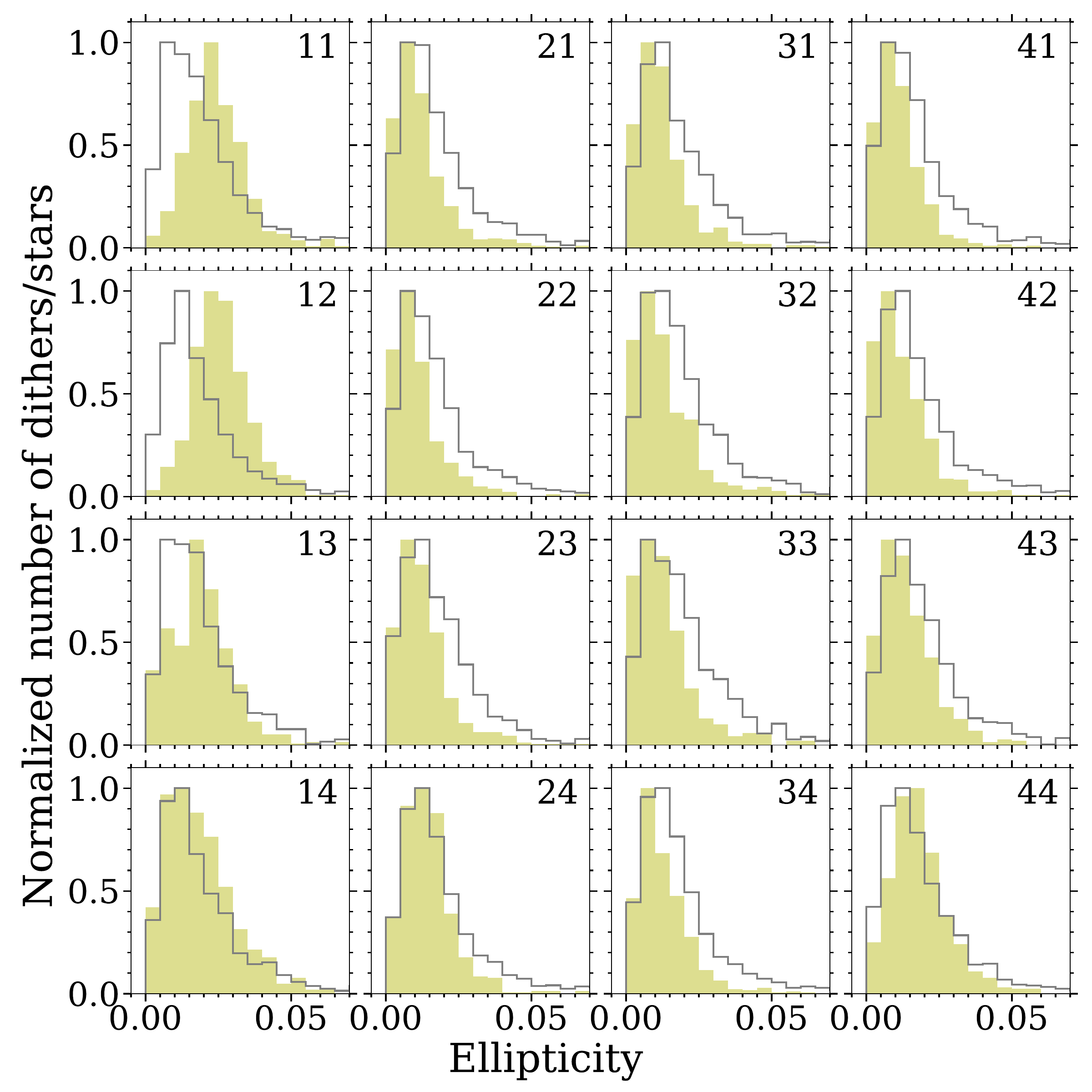}}
\end{minipage}
\begin{minipage}[t]{0.33\textwidth}
\centering
\resizebox{\hsize}{!}{\includegraphics{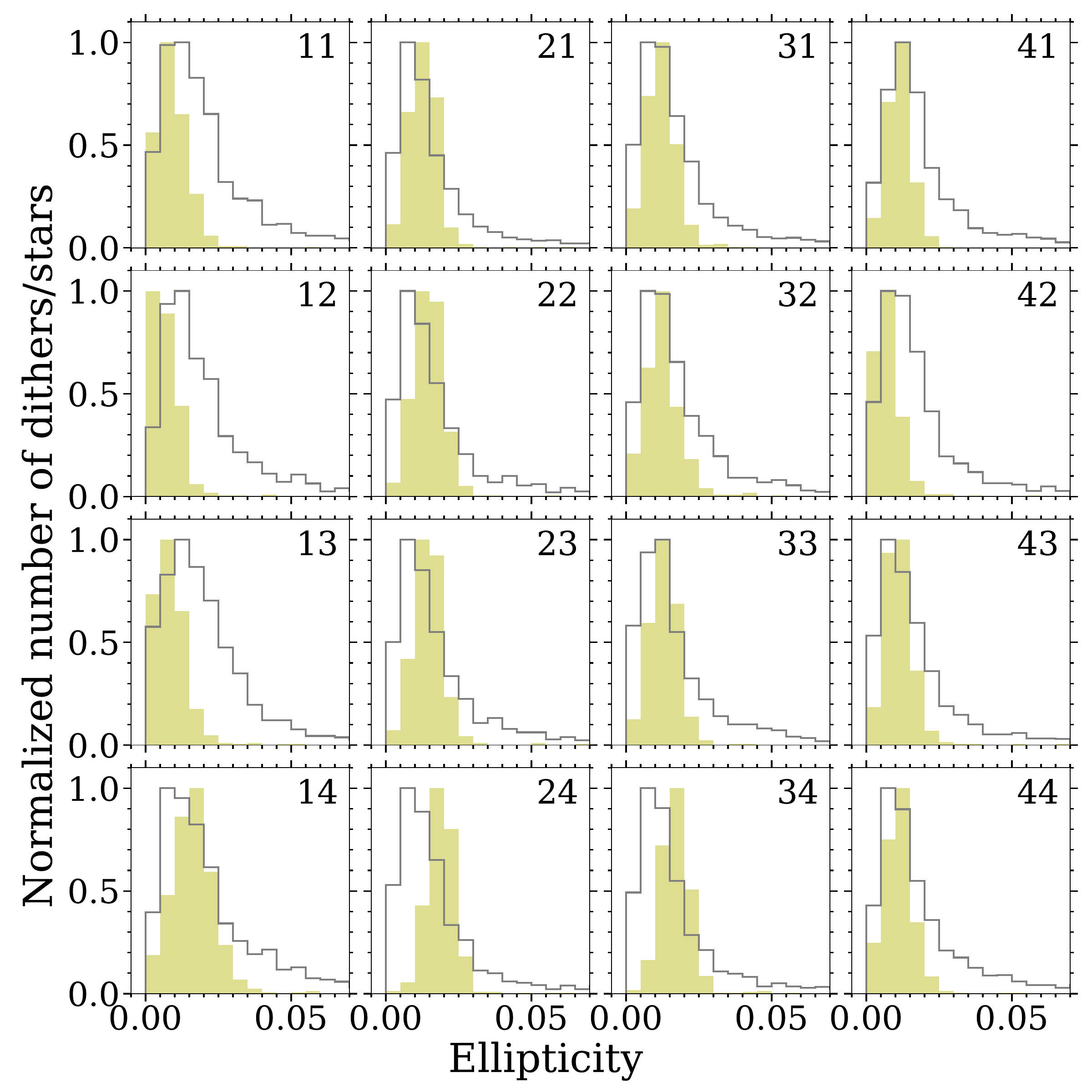}}
\end{minipage}
\caption{Distribution of the ellipticity in the 16 NISP detectors for the \YE (\emph{left}), \JE (\emph{middle}), and \HE (\emph{right}) passbands. The histograms compare the ellipticity calculated from the simplified PSF models (filled histograms, Patch IDs 56, 98, and 129) with the values measured by {\tt SourceExtractor} from point-like sources (unfilled histograms, Observation ID 6185).}
\label{figure:psf_ellip_hist}
\end{figure*}

\end{appendix}


\end{document}